\documentclass[11pt,draftclsnofoot,onecolumn]{IEEEtran}

\usepackage{amsbsy,amscd,amsfonts,amsgen,amsmath,amsopn,amssymb,amstext,amsthm,amsxtra,url,citesort,graphicx,algorithmic,algorithm,epsfig,verbatim,epic,graphicx}
\usepackage{subfigure}
\usepackage[usenames]{color}
\renewcommand{\baselinestretch}{1.2}
\newcommand{\eg}{{\em e.g., }}

\newcommand{\sinc}{\operatorname{sinc}}

\newcommand{\C}{{\mathcal{C}}}

\newcommand{\I}{{\mathcal{I}}}

\newcommand{\R}{{\mathcal{R}}}
\newcommand{\A}{{\mathcal{A}}}
\newcommand{\U}{{\mathcal{U}}}

\newcommand{\bbm}{{{\bf M}}}
\newcommand{\bbw}{{{\bf W}}}

\newcommand{\bbu}{{{\bf U}}}
\newcommand{\bbv}{{{\bf V}}}

\newcommand{\bbd}{{{\bf D}}}

\newcommand{\bbx}{{{\bf X}}}
\newcommand{\bby}{{{\bf Y}}}
\newcommand{\bbi}{{{\bf I}}}

\newcommand{\bbq}{{{\bf Q}}}
\newcommand{\bbs}{{{\bf S}}}
\newcommand{\bbn}{{{\bf N}}}
\newcommand{\bc}{{{\bf c}}}
\newcommand{\bd}{{{\bf d}}}
\newcommand{\ba}{{{\bf a}}}

\newcommand{\bx}{{{\bf x}}}
\newcommand{\by}{{{\bf y}}}

\newcommand{\bh}{{{\bf h}}}
\newcommand{\bs}{{{\bf s}}}

\newcommand{\bl}{\left(}
\newcommand{\br}{\right)}

\newcommand{\ZZ}{{\mathbb{Z}}}

\newcommand{\RR}{{\mathbb{R}}}

\newcommand{\inner}[2]{{\langle#1,#2\rangle}}

\DeclareMathOperator*{\supp}{supp}
\DeclareMathOperator*{\rank}{rank}
\DeclareMathOperator*{\spark}{spark}
\def\range{\mathcal{R}}
\def\R{\mathcal{R}}
\newcommand{\mixed}{{2,\I}}
\newcommand{\mub}{\mu_{\operatorname{B}}}

\newcommand{\bPhi}{\mbox{\boldmath{$\Phi$}}}
\newcommand{\bbg}{{{\bf G}}}
\newcommand{\bw}{{{\bf w}}}
\def\fmax{f_\textrm{max}}
\newcommand{\fnyq}{{f_\textrm{NYQ}}}
\newcommand{\Tnyq}{{T_\textrm{NYQ}}}

\def \b {b}
\def \d {d}
\def \e {e}
\def \n {n} 
\def \r {r}
\def \x {x}
\def \y {y}
\def \N {N} 
\def \M {M} 

\def \E {\mathbb{E}}
\def \P {\mathbb{P}}

\def \bR {{\bf R}}
\def \bU {{\bf U}}

\def \bV {{\bf V}}
\def \bX {{\bf X}}
\def \bY {{\bf Y}}
\def \bI {{\bf I}}
\def \bS {{\bf S}}
\def \bM {{\bf M}}
\def \bG {{\bf G}}
\def \pinv {^\dag} 
\def \real { \mathbb{R} }
\def \ident {{\bf I}}
\def \thetahat {\widehat{\theta}}
\def \thresh {\mathcal{T}}
\def \trans {^T} 
\def \xhat {\widehat{\x}}

\def \xbar {\overline{\x}}
\def \Phibar {\overline{\Phi}}
\def \Psibar {\overline{\Psi}}
\def \thetabar {\overline{\theta}}
\def \aalg {\mathbb{M}}

\def \bXhat {\widehat{\bX}}

\definecolor{gray}{rgb}{0.8,0.8,0.8}

\newcommand{\bigo}[1]{\mathcal{O}\left(#1\right)}
\usepackage{signal}

\theoremstyle{plain}
\newtheorem{theorem}{Theorem}
\newtheorem{lemma}{Lemma}
\newtheorem{proposition}{Proposition}
\newtheorem{corollary}[theorem]{Corollary}

\theoremstyle{definition}
\newtheorem{definition}{Definition}

\title{Structured Compressed Sensing: \\
From Theory to Applications}

\author{Marco~F.~Duarte~\IEEEmembership{Member,~IEEE,}
        and~Yonina~C.~Eldar,~\IEEEmembership{Senior Member,~IEEE}
\thanks{Manuscript submitted September 1, 2010; revised February 18, 2011 and June 21, 2011; accepted July 3, 2011.}%
\thanks{Copyright (c) 2011 IEEE. Personal use of this material is permitted. However, permission to use this material for any other purposes must be obtained from the IEEE by sending a request to pubs-permissions@ieee.org.}%
\thanks{MFD was supported in part by NSF Supplemental 
Funding DMS-0439872 to UCLA-IPAM, P.I.\ R.\ Caflisch. 
YCE was supported in part by the Israel Science Foundation under Grant no. 170/10, and by a Magneton grant from the Israel Ministry of Industry and Trade.}%
\thanks{MFD is with the Department of Electrical and Computer Engineering, University of Massachusetts, Amherst, MA 01003. 
Email: mduarte@ecs.umass.edu}%
\thanks{YCE is with the Department of Electrical Engineering,
Technion---Israel Institute of Technology, Haifa 32000, Israel. 
She is also a Visiting Professor at Stanford University, Stanford, CA, USA.
E-mail: yonina@ee.technion.ac.il}}

\begin{document}
\maketitle

\begin{abstract}
Compressed sensing (CS) is an emerging field that
has attracted considerable research interest over the past few
years. Previous review articles in CS limit their scope to standard 
discrete-to-discrete measurement architectures using matrices of 
randomized nature and signal models based on standard sparsity. 
In recent years, CS has worked its way into several new 
application areas. This, in turn, necessitates a fresh look on many 
of the basics of CS. The random matrix measurement operator must 
be replaced by more structured sensing architectures that 
correspond to the characteristics of feasible acquisition hardware. 
The standard sparsity prior has to be extended to include a much 
richer class of signals and to encode broader data models, 
including continuous-time signals. In our overview, the theme is 
exploiting signal and measurement structure in compressive sensing. 
The prime focus is bridging theory and practice; that is, to pinpoint the 
potential of structured CS strategies to emerge from the math to the 
hardware. 
Our summary highlights new directions as well as relations to 
more traditional CS, with the hope of serving both as a review to practitioners
wanting to join this emerging field, and as a reference for researchers 
that attempts to put some of the existing ideas in perspective of practical applications.
\end{abstract}

\section{Introduction and Motivation}

Compressed sensing (CS) is an emerging field that has
attracted considerable research interest in the signal processing
community. Since its introduction only several years
ago~\cite{DonohoCS,CandesUES}, thousands of papers have
appeared in this area, and hundreds of conferences, workshops, and
special sessions have been dedicated to this growing research field.

Due to the vast interest in this topic, there exist several excellent review
articles on the basics of CS~\cite{BaraniukCS,CandesCS,CandesWakinCS}.
These articles focused on the first CS efforts: the use of standard
discrete-to-discrete measurement architectures using matrices of randomized
nature, where no structure beyond sparsity is assumed on the signal or in its
representation. This basic formulation already required the use of
sophisticated mathematical tools and rich theory in order to analyze
recovery approaches and provide performance guarantees.  It was
therefore essential to confine attention to this simplified setting in the
early stages of development of the CS framework.

Essentially all analog-to-digital converters (ADCs) to date follow the
celebrated Shannon-Nyquist theorem which requires the sampling rate
to be at least twice the bandwidth of the signal. This basic
principle underlies the majority of digital signal processing (DSP)
applications such as audio, video, radio receivers, radar applications,
medical devices and more. The ever growing demand for data, as well
as advances in radio frequency (RF) technology, have promoted the
use of high-bandwidth signals, for which the rates dictated by the
Shannon-Nyquist theorem impose severe challenges both on the
acquisition hardware and on the subsequent storage and DSP
processors. CS was motivated in part by the desire to sample wideband
signals at rates far lower than the Shannon-Nyquist rate, while still
maintaining the essential information encoded in the underlying signal.
In practice, however, much of the work to date on CS has focused on
acquiring finite-dimensional sparse vectors using random measurements.
This precludes the important case of continuous-time (i.e., analog)
input signals, as well as practical hardware architectures which inevitably
are structured. Achieving the holy grail of compressive ADCs and increased
resolution requires a broader framework which can treat more general signal
models, including continuous-time signals with various types of structure, as
well as practical measurement schemes.

In recent years, the area of CS has branched out to many new fronts and
has worked its way into several application areas. This, in turn,
necessitates a fresh look on many of the basics of CS. The random matrix
measurement operator, fundamental in all early presentations of CS, must
be replaced by more structured measurement operators that correspond
to the application of interest, such as wireless channels, analog sampling
hardware, sensor networks and optical imaging. The standard sparsity
prior that has characterized early work in CS has to be extended to
include a much richer class of signals: signals that have underlying
low-dimensional structure, not necessarily represented by standard
sparsity, and signals that can have arbitrary dimensions, not only
finite-dimensional vectors.

A significant part of the recent work on CS from the signal processing
community can be classified into two major contribution areas. The
first group consists of theory and applications related to CS matrices
that are not completely random and that often exhibit considerable
structure. This largely follows from efforts to model the way the
samples are acquired in practice, which leads to sensing matrices
that inherent their structure from the real world. The second group
includes signal representations that exhibit structure beyond sparsity
and broader classes of signals, such as continuous-time signals with
infinite-dimensional representations. For many types of signals, such
structure allows for a higher degree of signal compression when
leveraged on top of sparsity. Additionally, infinite-dimensional signal
representations provide an important example of richer structure which
clearly cannot be described using standard sparsity. Since reducing the
sampling rate in analog signals was one of the driving forces behind CS,
building a theory that can accommodate low-dimensional signals in
arbitrary Hilbert spaces is clearly an essential part of the CS framework.
Both of these categories are motivated by real-world CS
involving actual hardware implementations.

In our review, the theme is exploiting signal and measurement
structure in CS. The prime focus is bridging
theory and practice -- that is, to pinpoint the potential of
CS strategies to emerge from the math to the
hardware by generalizing the underlying theory where needed.
We believe that this is an essential ingredient in taking CS to the
next step: incorporating this fast growing field into real-world
applications. Considerable efforts have been devoted in recent
years by many researchers to adapt the theory of CS to better
solve real-world signal acquisition challenges. This has also led
to parallel low-rate sampling schemes that combine the principles
of CS with the rich theory of sampling such as the finite rate of
innovation (FRI)~\cite{VMB02,DVB07,BDVMC08} and Xampling
frameworks~\cite{MEDS09,MEE10}. There are already dozens
of papers dealing with these broad ideas. In this review we have
strived to provide a coherent summary, highlighting new
directions and relations to more traditional CS. This material can
both serve as a review to those wanting to join this emerging field,
as well as a reference that attempts to summarize some of
the existing results in the framework of practical applications.
Our hope is that this presentation will
attract the interest of both mathematicians and engineers in the
desire to promote the CS premise into practical applications, and
encourage further research into this new frontier.

This review paper is organized as follows. Section~\ref{sec:bg}
provides background motivating the formulation of CS and the
layout of the review. A primer on standard CS theory is presented
in Section~\ref{sec:csbasics}. This material serves as a basis for
the later developments. Section~\ref{sec:csmatrices} reviews
alternative constructions for structured CS matrices beyond those
generated completely at random. In Section~\ref{sec:finitemodels}
we introduce finite-dimensional signal models that exhibit additional
signal structure. This leads to the more general union-of-subspaces
framework, which will play an important role in the context of structured
infinite-dimensional representations as well. Section~\ref{sec:analog}
extends the concepts of CS to infinite-dimensional signal models and
introduces recent compressive ADCs which have been developed
based on the Xampling and FRI frameworks. For each of the matrices
and models introduced, we summarize the details of the theoretical
and algorithmic frameworks and provide example applications where
the structure is applicable.

\section{Background}
\label{sec:bg}

We live in a digital world. Telecommunication, entertainment,
medical devices, gadgets, business -- all revolve around digital media.
Miniature sophisticated black-boxes process streams of bits
accurately at high speeds. Nowadays, electronic consumers feel
natural that a media player shows their favorite movie, or that
their surround system synthesizes pure acoustics, as if sitting in
the orchestra instead of the living room. The digital world plays
a fundamental role in our everyday routine, to such a point that
we almost forget that we cannot ``hear'' or ``watch'' these streams
of bits, running behind the scenes.

Analog to digital conversion lies at the heart of this
revolution. ADC devices translate the physical information into a
stream of numbers, enabling digital processing by sophisticated
software algorithms. The ADC task is inherently intricate: its
hardware must hold a snapshot of a fast-varying input signal
steady while acquiring measurements. Since these measurements are
spaced in time, the values between consecutive snapshots are lost.
In general, therefore, there is no way to recover the analog
signal unless some prior on its structure is incorporated.

After sampling, the numbers or bits retained must be stored and later
processed. This requires ample storage devices and sufficient
processing power. As technology advances, so does the requirement
for ever-increasing amounts of data, imposing unprecedented strains
on both the ADC devices and the subsequent DSP and storage media.
How then does consumer electronics keep up with these high
demands? Fortunately, most of the data we acquire can be discarded
without much perceptual loss. This is evident in essentially all
compression techniques used to date.
However, this paradigm of high-rate sampling followed by compression
does not alleviate the large strains on the acquisition device and on the
DSP. In his seminal work on CS \cite{DonohoCS}, Donoho
posed the ultimate goal of merging compression and sampling: ``why go
to so much effort to acquire all the data when most of what we get will
be thrown away? Can't we just directly measure the part that won't end
up being thrown away?''.

\subsection{Shannon-Nyquist Theorem}

ADCs provide the interface between an analog signal being recorded
and a suitable discrete representation. A common approach in
engineering is to assume that the signal is bandlimited, meaning that
the spectral contents are confined to a maximal frequency $B$.
Bandlimited signals have limited time variation, and can therefore be
perfectly reconstructed from equispaced samples with rate at least
$2B$, termed the Nyquist rate. This fundamental result is often
attributed in the engineering community to
Shannon-Nyquist~\cite{N28,S49}, although it dates back to earlier
works by Whittaker~\cite{W15} and Kotel\'{n}ikov~\cite{K33}.
\begin{theorem}~\cite{S49}
If a function $x(t)$ contains no frequencies higher than $B$ hertz, then
it is completely determined by giving its ordinates at a series of points
spaced $1/(2B)$ seconds apart.
\label{th:snk}
\end{theorem}

A fundamental reason for processing at the Nyquist rate is the clear
relation between the spectrum of $x(t)$ and that of its samples $x(nT)$,
so that digital operations can be easily substituted for their
analog counterparts. Digital filtering is an example where
this relation is successfully exploited. Since the power spectral
densities of analog and discrete random processes are
associated in a similar manner, estimation and detection of
parameters of analog signals can be performed by DSP. In contrast,
compression is carried out by a series of algorithmic steps,
which, in general, exhibit a nonlinear complicated relationship
between the samples $x(nT)$ and the stored data.

While this framework has driven the development of signal acquisition
devices for the last half century, the increasing complexity of emerging
applications dictates increasingly higher sampling rates that cannot
always be met using available hardware. Advances in related fields
such as wideband communication and RF technology open a
considerable gap with ADC devices. Conversion speeds which are
twice the signal's maximal frequency component have become more
and more difficult to obtain. Consequently, alternatives to high rate
sampling are drawing considerable attention in both academia and
industry.

Structured analog signals can often be processed far more efficiently
than what is dictated by the Shannon-Nyquist theorem, which does
not take any structure into account. For example, many wideband
communication signals are comprised of several narrow transmissions
modulated at high carrier frequencies. A common practice in
engineering is demodulation in which the  input signal is multiplied by
the carrier frequency of a band of interest, in order to shift the contents
of the narrowband transmission from the high frequencies to the origin.
Then, commercial ADC devices at low rates are utilized. Demodulation,
however, requires knowing the exact carrier frequency. In this review
we focus on structured models in which the exact parameters defining
the structure are unknown. In the context of multiband communications,
for example, the carrier frequencies may not be known, or may be
changing over time. The goal then is to build a compressed sampler
which does not depend on the carrier frequencies, but can nonetheless
acquire and process such signals at rates below Nyquist.

\subsection{Compressed Sensing and Beyond}

A holy grail of CS is to build acquisition devices that exploit signal
structure in order to reduce the sampling rate, and
subsequent demands on storage and DSP. In such an approach, the
actual information contents dictate the sampling rate, rather than the
dimensions of the ambient space in which the signal resides.
The challenges in achieving this task both theoretically and in terms of
hardware design can be reduced substantially when considering
finite-dimensional problems in which the signal to be measured
can be represented as a discrete finite-length vector.
This has spurred a surge of research on various mathematical and
algorithmic aspects of sensing sparse signals, which were mainly studied
for discrete finite vectors.

At its core, CS is a mathematical framework that studies accurate
recovery of a signal represented by a vector of length $N$ from $M \ll N$
measurements, effectively performing
compression during signal acquisition. The measurement paradigm
consists of linear projections, or inner products, of the signal vector
into a set of carefully chosen projection vectors that act as a
multitude of probes on the information contained in the signal. In the
first part of this review (Sections~\ref{sec:csbasics}
and~\ref{sec:csmatrices}) we survey the fundamentals of CS and
show how the ideas can be extended to allow for more elaborate
measurement schemes that incorporate structure into the
measurement process. When considering real-world acquisition
schemes, the choices of possible measurement
matrices are dictated by the constraints of the application. Thus,
we must deviate from the general randomized constructions and
apply structure within the projection vectors that can be easily
implemented by the acquisition hardware. Section~\ref{sec:csmatrices}
focuses on such alternatives; we survey both existing theory and
applications for several classes of structured CS matrices.
In certain applications, there exist hardware designs that measure 
analog signals at a sub-Nyquist rate, obtaining measurements for 
finite-dimensional signal representations via such structured CS matrices.

In the second part of this review (Sections~\ref{sec:finitemodels}
and~\ref{sec:analog}) we expand the theory of CS to signal models
tailored to express structure beyond standard sparsity.
A recent emerging theoretical framework that allows a broader class of
signal models to be acquired efficiently is the union of subspaces
model~\cite{LD08,BD09a,EM09a,E09,GE10,BCDH10}. We introduce
this framework and some of its applications in a finite-dimensional
context in Section~\ref{sec:finitemodels}, which include more general
notions of structure and sparsity. Combining the principles and insights
from the previous sections, in Section~\ref{sec:analog} we extend
the notion of CS to analog signals with infinite-dimensional
representations. This new framework, referred to as
Xampling~\cite{MEDS09,MEE10}, relies on more general signal 
models -- union of subspaces and FRI signals -- together with guidelines 
on how to exploit these mathematical structures in order to build sensing 
devices that can directly acquire analog signals at reduced rates.
We then survey several compressive ADCs that result from
this broader framework.

\section{Compressed Sensing Basics}
\label{sec:csbasics}

Compressed sensing (CS)~\cite{DonohoCS,CandesUES,CandesCS,BaraniukCS,CandesWakinCS}
offers a framework for simultaneous sensing and compression
of finite-dimensional vectors, that relies on linear dimensionality
reduction. Specifically, in CS we do not acquire $\x$ directly but rather
acquire $M<N$ linear measurements $\y=\Phi\x$ using an $M\times N$
{\em CS matrix} $\Phi$. We refer to $\y$ as the {\em measurement vector}.
Ideally, the matrix $\Phi$ is designed to reduce the number of measurements
$M$ as much as possible while allowing for recovery of a wide class
of signals $\x$ from their measurement vectors $\y$. However, the
fact that $M < N$ renders the matrix $\Phi$ {\em rank-defficient},
meaning that it has a nonempty nullspace; this, in turn, implies that
for any particular signal $x_0 \in \real^N$, an infinite number of
signals $\x$ will yield the same measurements
$\y_0 = \Phi x_0 = \Phi x$ for the chosen CS matrix $\Phi$.

The motivation behind the design of the matrix $\Phi$ is, therefore,
to allow for distinct signals $\x,\x'$ within a class of signals of interest
to be uniquely identifiable from their measurements
$\y = \Phi \x$, $\y' = \Phi x'$, even though $M \ll N$. We must therefore
make a choice on the class of signals that we aim to recover from
CS measurements.

\subsection{Sparsity}

Sparsity is the signal structure behind many compression algorithms
that employ transform coding, and is the most prevalent signal structure
used in CS. Sparsity also has a rich history of applications in signal
processing problems in the last century (particularly in imaging), including
denoising, deconvolution, restoration, and
inpainting~\cite{GR97,DonohoBP,BDE09}.

To introduce the notion of sparsity, we rely on a signal representation in a
given basis $\{\psi_i\}_{i=1}^N$ for $\real^N$. Every signal $\x\in\real^N$
is representable in terms of $N$ coefficients $\{\theta_i\}_{i=1}^N$
as $\x = \sum_{i=1}^N \psi_i\theta_i$; arranging the $\psi_i$ as columns
into the $N\times N$ matrix $\Psi$ and the coefficients $\theta_i$ into
the $N\times 1$ {\em coefficient vector} $\theta$, we can write succinctly
that $\x= \Psi \theta$, with $\theta \in \real^N$. Similarly, if we use a
frame\footnote{A matrix $\Psi$ is said to be a frame if there
exist constants $A$, $B$ such that $A\|x\|_2 \le \|\Psi x\|_2 \le B\|x\|_2$ for all
$x$.} $\Psi$ containing $N$ unit-norm column vectors of length $L$ with
$L < N$ (i.e., $\Psi \in \real^{L \times N}$), then for any vector
$\x \in \real^L$ there exist infinitely many decompositions
$\theta \in \real^N$ such that $\x = \Psi \theta$. In a general setting,
we refer to $\Psi$ as the sparsifying {\em dictionary}~\cite{MZ93}.
While our exposition is restricted to real-valued signals, the concepts are
extendable to complex signals as well~\cite{Foucart,FoucartGribonval}.

We say that a signal $\x$ is $K$-{\em sparse} in the basis or frame $\Psi$
if there exists a vector $\theta \in \real^N$ with only $K\ll N$ nonzero
entries such that $\x = \Psi \theta$.  We call the set of indices
corresponding to the nonzero entries the {\em support} of $\theta$
and denote it by $\supp(\theta)$. We also define the set $\Sigma_K$
that contains all signals $\x$ that are $K$-sparse.

A $K$-sparse signal can be efficiently compressed by preserving only the
values and locations of its nonzero coefficients, using $\mathcal{O}(K \log_2 N)$
bits: coding each of the $K$ nonzero coefficient's locations takes
$\log_2 N$ bits, while coding the magnitudes uses a constant amount of
bits that depends on the desired precision, and is independent of $N$.
This process is known as {\em transform coding}, and relies on the
existence of a suitable basis or frame $\Psi$ that renders signals of
interest sparse or approximately sparse.

For signals that are not exactly sparse, the amount of compression
depends on the number of coefficients of $\theta$ that we keep.
Consider a signal $\x$ whose coefficients $\theta$, when sorted in
order of decreasing magnitude, decay according to the power law
\begin{equation}
|\theta({\mathcal I}(n))| \le S \, n^{-1/r}, ~ n=1,\ldots,N,
\label{eq:comp}
\end{equation}
where ${\mathcal I}$ indexes the sorted coefficients. Thanks to the
rapid decay of their coefficients, such signals are
well-approximated by $K$-sparse signals.
The best $K$-term approximation error for such a signal obeys
\begin{equation}
\sigma_\Psi(\x,K) := \arg \min_{\x' \in \Sigma_K} \|\x-\x'\|_2 \le CS K^{-s}, \label{eq:kta}
\end{equation}
with $s = \frac{1}{r}-\frac{1}{2}$ and $C$ denoting a constant that does
not depend on $N$~\cite{CDD09}. That is, the signal's best
approximation error (in an $\ell_2$-norm sense) has a power-law decay
with exponent $s$ as $K$
increases. We dub such a signal {\em $s$-compressible}. When $\Psi$
is an orthonormal basis, the best sparse approximation of $x$ is obtained by
hard thresholding the signal's coefficients, so that only the $K$ coefficients
with largest magnitudes are preserved. The situation is decidedly more
complicated when $\Psi$ is a general frame, spurring the development of sparse
approximation methods, a subject that we will focus on in
Section~\ref{sec:csrecovery}.

When sparsity is used as the signal structure enabling CS, we aim to
recover $\x$ from $\y$ by exploiting its sparsity. In contrast with transform
coding, we do not operate on the signal $\x$ directly, but rather only have access to the CS
measurement vector $\y$. Our goal is to push $M$ as close as possible to
$K$ in order to perform as much signal ``compression'' during acquisition
as possible. In the sequel, we will assume that $\Psi$ is taken to be the
identity basis so that the signal $\x = \theta$ is itself sparse. In certain
cases we will explicitly define a different basis or frame $\Psi$ that arises
in a specific application of CS.

\subsection{Design of CS Matrices}

The main design criteria for the CS matrix $\Phi$ is to enable the unique
identification of a signal of interest $\x$ from its measurements $\y = \Phi \x$.
Clearly, when we consider the class of $K$-sparse signals $\Sigma_K$,
the number of measurements $M > K$ for any matrix design, since the
identification problem has $K$ unknowns even when the support
$\Omega = \supp(\x)$ of the signal $\x$ is provided. In this case,
we simply restrict the matrix $\Phi$ to its columns corresponding to the
indices in $\Omega$, denoted by $\Phi_\Omega$, and then use the
pseudoinverse to recover the nonzero coefficients of $\x$:
\begin{equation}
\x_\Omega = \Phi_\Omega\pinv \y.
\label{eq:pinv}
\end{equation}
Here $\x_\Omega$ is the restriction of the vector $\x$ to the set of indices
$\Omega$, and $\bM\pinv = (\bM\trans\bM)^{-1}\bM\trans$ denotes the
pseudoinverse of the matrix $\bM$. The implicit assumption in (\ref{eq:pinv}) is that
$\Phi_\Omega$ has full column-rank so that there is a unique solution to the equation
$y=\Phi_\Omega\x_\Omega$.

We begin by determining properties of
$\Phi$ that guarantee that distinct signals $\x, \x' \in \Sigma_K$, $\x \ne \x'$,
lead to different measurement vectors $\Phi \x \ne \Phi \x'$. In other words,
we want each vector $\y \in \real^M$ to be matched to at most one vector
$\x \in \Sigma_K$ such that $\y = \Phi \x$.
A key relevant property of the matrix in this context is its {\em spark}.
\begin{definition}
~\cite{DonohoOSR} The spark $\spark(\Phi)$ of a given matrix $\Phi$ is the smallest
number of columns of $\Phi$ that are linearly dependent.
\end{definition}
\noindent The spark is related to the {\em Kruskal Rank} 
from the tensor product literature; the matrix $\Phi$ has Kruskal rank $\spark(\Phi)-1$.
This definition allows us to pose the following straightforward guarantee.
\begin{theorem}
~\cite{DonohoOSR} If $\spark(\Phi) > 2K$, then for each measurement vector
$\y \in \real^M$ there exists at most one signal $\x \in \Sigma_K$
such that $\y = \Phi \x$.
\label{th:spark}
\end{theorem}

\noindent It is easy to see that $\spark(\Phi) \in [2, M+1]$, so that
Theorem~\ref{th:spark} yields the requirement $M \ge 2K$.

While Theorem~\ref{th:spark} guarantees
uniqueness of representation for $K$-sparse
signals, computing the spark of a general matrix $\Phi$ has combinatorial
computational complexity, since one must verify that all sets of columns
of a certain size are linearly independent. Thus, it is preferable to use
properties of $\Phi$ that are easily computable to provide recovery
guarantees. The {\em coherence} of a matrix is one such property.
\begin{definition}\cite{DH01,DonohoOSR,TroppGreed,GN03}
The coherence $\mu(\Phi)$ of a matrix $\Phi$ is the largest absolute
inner product between any two columns of $\Phi$:
\begin{align}
\mu(\Phi) = \max_{1 \le i \ne j \le N} \frac{|\langle\phi_i,\phi_j \rangle|}{\|\phi_i\|_2\|\phi_j\|_2}.
\end{align}
\end{definition}
It can be shown that
$\mu(\Phi) \in \left[\sqrt{\frac{N-M}{M(N-1)}},1\right]$; the lower bound is
known as the Welch bound~\cite{Welch,StrohmerHeath}. Note that when
$N \gg M$, the lower bound is approximately $\mu(\Phi) \ge 1/\sqrt{M}$.
One can tie the coherence and spark of a matrix by
employing the Gershgorin circle theorem.
\begin{theorem}
~\cite{Gershgorin}
The eigenvalues of an $m \times m$ matrix $\bM$ with entries $\bM_{i,j}$,
$1 \le i,j \le m$, lie in the union of $m$ discs $d_i = d_i(c_i,r_i)$,
$1 \le i \le m$, centered at $c_i = \bM_{i,i}$ with radius
$r_i = \sum_{j\ne i}|\bM_{i,j}|$.
\label{th:gershgorin}
\end{theorem}
Applying this theorem on the Gram matrix
$\bG = \Phi_\Omega^T\Phi_\Omega$ leads to the following result.
\begin{lemma}\cite{DonohoOSR}
For any matrix $\Phi$,
\begin{align}
\spark(\Phi) \ge 1 + \frac{1}{\mu(\Phi)}.
\end{align}
\label{lemma:mu}
\end{lemma}

By merging Theorem~\ref{th:spark} with Lemma~\ref{lemma:mu},
we can pose the following condition on $\Phi$ that guarantees uniqueness.
\begin{theorem}\cite{DonohoOSR,TroppGreed,GN03}
If
\begin{equation}
K < \frac{1}{2} \left(1+\frac{1}{\mu(\Phi)}\right),
\label{eq:maxK}
\end{equation}
then for each measurement vector $\y \in \real^M$ there exists at most
one signal $\x \in \Sigma_K$ such that $\y = \Phi \x$.
\label{th:mu}
\end{theorem}
Theorem~\ref{th:mu}, together with the Welch bound, provides an
upper bound on the level of sparsity $K$ that guarantees uniqueness
using coherence: $K = \mathcal{O}(\sqrt{M})$.

The prior properties of the CS matrix provide guarantees of uniqueness
when the measurement vector $\y$ is obtained without error. 
Hardware considerations introduce two main
sources of inaccuracies in the measurements: inaccuracies due to noise 
at the sensing stage (in the form of additive noise $\y = \Phi \x + \n$), and 
inaccuracies due to mismatches between the CS matrix used during 
recovery, $\Phi$, and that implemented during acquisition, 
$\Phi' = \Phi+\Delta$ (in the form of multiplicative 
noise~\cite{HS10,CSPC10}). Under these sources of error, it is no longer 
possible to guarantee uniqueness; however, it is desirable for the 
measurement process to be tolerant to both types of error.
To be more formal, we would like the
distance between the measurement vectors  for two sparse signals
$\y = \Phi \x$, $\y' = \Phi \x'$ to be proportional to the distance between
the original signal vectors $\x$ and $\x'$. Such a property allows us to
guarantee that, for small enough noise, two sparse vectors that are far
apart from each other cannot lead to the same (noisy) measurement
vector. This behavior has been formalized into the {\em restricted
isometry property} (RIP).
\begin{definition}\cite{CandesCS}
A matrix $\Phi$ has the {\em $(K,\delta)$-restricted isometry
property} ($(K,\delta)$-RIP) if, for all $x \in \Sigma_K$,
\begin{equation}
(1-\delta) \|x\|_2^2 \le \|\Phi x\|_2^2 \le (1+\delta)
\|x\|_2^2.
\label{eq:rip}
\end{equation}
\end{definition}
\noindent In words, the $(K,\delta)$-RIP ensures that all submatrices of 
$\Phi$ of size $M \times K$ are close to an isometry, and therefore 
distance-preserving.
We will show later that this property suffices to prove that the 
recovery is stable to presence of additive noise $n$. 
In certain settings, noise is introduced to the signal $x$ prior to 
measurement. Recovery is also stable for this case; however, there is 
a degradation in the distortion of the recovery by a factor of 
$N/M$~\cite{Treichler,Aeron,BenHaimMichaeliEldar,Arias}.
Furthermore, the 
RIP also leads to stability with respect to the multiplicative noise 
introduced by the CS matrix mismatch $\Delta$~\cite{HS10,CSPC10}.

The RIP can be connected to the coherence property by using, once 
again, the Gershgorin circle theorem (Theorem~\ref{th:gershgorin}).
\begin{lemma}~\cite{CaiXuZhang}
If $\Phi$ has unit-norm columns and coherence $\mu = \mu(\Phi)$, then
$\Phi$ has the $(K,\delta)$-RIP with $\delta \le (K-1)\mu$.
\end{lemma}
One can also easily connect RIP with the spark.
For each $K$-sparse vector to be uniquely identifiable by its
measurements, it suffices for the matrix $\Phi$ to have the $(2K,\delta)$-RIP
with $\delta > 0$, as this implies that all sets of $2K$ columns of $\Phi$
are linearly independent, i.e., $\spark(\Phi) > 2K$
(cf. Theorems~\ref{th:spark} and~\ref{th:mu}).
We will see later that the RIP enables recovery guarantees that are much
stronger than those based on spark and coherence. However, checking
whether a CS matrix $\Phi$ satisfies the $(K,\delta)$-RIP
has combinatorial computational complexity.

Now that we have defined relevant properties of a CS matrix $\Phi$,
we discuss specific matrix constructions that are suitable for CS.
An $M \times N$ Vandermonde matrix $\bV$ constructed from
$N$ distinct scalars has $\spark(\bV) = M+1$~\cite{CDD09}.
Unfortunately, these matrices are poorly conditioned for large values of
$N$, rendering the recovery problem numerically unstable.
Similarly, there are known matrices $\Phi$ of size $M \times M^2$ that
achieve the coherence lower bound 
\begin{equation}
\mu(\Phi) = 1/\sqrt{M}
\label{eq:mubound}
\end{equation}
such as the Gabor frame generated from the Alltop sequence~\cite{herman:tsp09}
and more general equiangular tight frames~\cite{StrohmerHeath}.
It is also possible to construct deterministic CS matrices of size $M \times N$
that have the $(K,\delta)$-RIP for
$K = \mathcal{O}(\sqrt{M} \log M / \log(N/M))$~\cite{DeVoreDeterministic}.
These constructions restrict the number of measurements needed to recover
a $K$-sparse signal to be $M = \mathcal{O}(K^2\log N)$, which is undesirable for
real-world values of $N$ and $K$.

Fortunately, these bottlenecks can be defeated by randomizing the matrix
construction. For example, random matrices $\Phi$ of size $\M \times \N$
whose entries are independent and identically distributed (i.i.d.) with
continuous distributions have $\spark(\Phi) = M+1$ with high probability.
It can also be shown that when the distribution used has zero mean
and finite variance, then in the asymptotic regime (as $M$ and $N$
grow) the coherence converges to
$\mu(\Phi) = 2 \sqrt{\log N/M}$~\cite{DonohoL1L0,CandesPlan}.
Similarly, random matrices from Gaussian, Rademacher, or more generally
a subgaussian distribution\footnote{A Rademacher distribution gives
probability $1/2$ to the values $\pm 1$. A random variable $X$ is called
subgaussian if there exists $c>0$ such that $\mathbb{E}\left(e^{Xt}\right) \le
e^{c^2t^2/2}$ for all $t \in \real$. Examples include the Gaussian, Bernoulli,
and Rademacher random variables, as well as any bounded random
variable.} have the $(K,\delta)$-RIP with high probability if~\cite{BDDW08}
\begin{equation}
M=\mathcal{O}(K \log(N/K)/\delta^2).
\label{eq:ripbound}
\end{equation}
Finally, we point out that
while the set of RIP-fulfilling matrices provided above might seem limited,
emerging numerical results have shown that a variety of classes of matrices
$\Phi$ are suitable for CS recovery at regimes similar to those of the matrices
advocated in this section, including subsampled Fourier and Hadamard
transforms~\cite{TD09,DPF10}.

\subsection{CS Recovery Algorithms}
\label{sec:csrecovery}

We now focus on solving the CS recovery problem: given $\y$ and $\Phi$,
find a signal $\x$ within the class of interest such that $\y = \Phi \x$ exactly
or approximately.

When we consider sparse signals, the CS recovery process consists of a
search for the sparsest signal $\x$ that yields the measurements $\y$.
By defining the $\ell_0$ ``norm'' of a vector $\|\x\|_0$ as the number of
nonzero entries in $\x$, the simplest way to pose a recovery algorithm
is using the optimization
\begin{equation}
\xhat = \arg \min_{x \in \real^N} \|\x\|_0~\textrm{subject to}~\y = \Phi\x.
\label{eq:L0}
\end{equation}
Solving (\ref{eq:L0}) relies on an exhaustive search and is successful
for all $x \in \Sigma_K$
when the matrix $\Phi$ has the sparse solution uniqueness property
(i.e., for $M$ as small as $2K$, see Theorems~\ref{th:spark}
and~\ref{th:mu}). However, this algorithm has combinatorial
computational complexity, since we must check whether the
measurement vector $\y$ belongs to the span of each set of $K$
columns of $\Phi$, $K=1,2,\ldots,N$. Our goal, therefore, is to find
computationally feasible algorithms that can successfully recover
a sparse vector $\x$ from the measurement vector $\y$ for the
smallest possible number of measurements $M$.

An alternative to the $\ell_0$ ``norm'' used in (\ref{eq:L0}) is to use the
$\ell_1$ norm, defined as $\|\x\|_1 = \sum_{n=1}^N |\x(n)|$. The
resulting adaptation of (\ref{eq:L0}), known as {\em basis pursuit}
(BP)~\cite{DonohoBP}, is formally defined as
\begin{equation}
\xhat = \arg \min_{\x \in \real^N} \|\x\|_1~\textrm{subject to}~\y = \Phi \x.
\label{eq:L1}
\end{equation}
Since the $\ell_1$ norm is convex, (\ref{eq:L1}) can be seen as a convex
relaxation of~(\ref{eq:L0}).
Thanks to the convexity, this algorithm can be implemented
as a linear program, making its computational complexity polynomial in
the signal length~\cite{LP}.\footnote{A similar set of recovery algorithms,
known as total variation minimizers, operate on the gradient of an image,
which exhibits sparsity for piecewise smooth images~\cite{ROF92}.}

The optimization (\ref{eq:L1}) can be modified to allow for noise in the
measurements $\y = \Phi \x+n$; we simply change the constraint on the
solution to
\begin{equation}
\xhat = \arg \min_{\x \in \real^N} \|\x\|_1~\textrm{subject to}~\|\y - \Phi \x\|_2 \le \epsilon,
\label{eq:bpic}
\end{equation}
where $\epsilon \ge \|n\|_2$ is an appropriately chosen bound on the noise
magnitude. This modified optimization is known as basis pursuit with
inequality constraints (BPIC) and is
a quadratic program with polynomial complexity solvers~\cite{LP}.
The Lagrangian relaxation of this quadratic program is written as
\begin{equation}
\xhat = \arg \min_{\x \in \real^N} \|\x\|_1+\lambda \|\y - \Phi \x\|_2,
\label{eq:BPDN}
\end{equation}
and is known as basis pursuit denoising (BPDN). There exist many
efficient solvers to find BP, BPIC, and BPDN solutions; for an overview, 
see~\cite{TroppWright}.

Oftentimes, a bounded-norm noise model is overly pessimistic, and it
may be reasonable instead to assume that the noise is random. For
example, additive white Gaussian noise
$n \sim \mathcal{N}(0,\sigma^2{\bf I})$ is a common choice.
Approaches designed to address stochastic noise include
complexity-based regularizaton~\cite{HauptNowak} and Bayesian
estimation~\cite{JXC08}. These methods pose probabilistic or
complexity-based priors, respectively, on the set of observable signals.
The particular prior is then leveraged together with the noise
probability distribution during signal recovery. Optimization-based
approaches can also be formulated in this case; one of the most popular
techniques is the Dantzig selector~\cite{CandesDS}:
\begin{equation}
\xhat = \arg \min_{\x \in \real^N} \|\x\|_1~\textrm{s. t.}~\|\Phi^T(\y-\Phi\x)\|_\infty \le \lambda\sqrt{\log N}\sigma,
\label{eq:dantzig}
\end{equation}
where $\|\cdot\|_\infty$ denotes the $\ell_\infty$-norm, which provides
the largest-magnitude entry in a vector and $\lambda$ is a constant
parameter that controls the probability of successful recovery.

An alternative to optimization-based approaches, are {\em greedy
algorithms} for sparse signal recovery. These methods are iterative in
nature and select columns of $\Phi$ according to their correlation with
the measurements $\y$ determined by an appropriate inner product.
For example, the matching pursuit and orthogonal matching pursuit algorithms 
(OMP)~\cite{MZ93,PatiOMP} proceed
by finding the column of $\Phi$ most correlated to the signal residual, which
is obtained by subtracting the contribution of a partial estimate of the signal from
$\y$. The OMP method is formally defined as
Algorithm~\ref{alg:OMP}, where $\thresh(\x,K)$ denotes a {\em thresholding}
operator on $\x$ that sets all but the $K$ entries of $\x$ with the largest
magnitudes to zero, and $\bx|_\Omega$ denotes the restriction of $\bx$
to the entries indexed by $\Omega$. The convergence criterion used to find
sparse representations consists of checking whether $\y = \Phi \x$ exactly or
approximately; note that due to its design, the algorithm cannot run for more
than $M$ iterations, as $\Phi$ has $M$ rows. Other greedy techniques
that have a similar flavor to OMP include CoSaMP~\cite{cosamp}, detailed as
Algorithm~\ref{alg:cosamp}, and Subspace Pursuit
(SP)~\cite{DM09}. A simpler variant is known as iterative hard thresholding
(IHT)~\cite{IHT}:
starting from an initial signal estimate $\xhat_0 = 0$,
the algorithm iterates a gradient descent step followed by hard thresholding, i.e.,
\begin{equation}
\xhat_i = \thresh(\xhat_{i-1}+\Phi^T(\y-\Phi\xhat_{i-1}),K),
\label{eq:iht}
\end{equation}
until a convergence criterion is met.
\begin{algorithm}[!t]
\caption{Orthogonal Matching Pursuit \label{alg:OMP}}
\begin{algorithmic}
\STATE Input: CS matrix $\Phi$, measurement vector $\y$
\STATE Output: Sparse representation $\xhat$
\STATE Initialize: $\xhat_0=0$, $\r = \y$, $\Omega = \emptyset$, $i = 0$
\WHILE{halting criterion false}
\STATE $i \leftarrow i+1$
\STATE $\b \leftarrow \Phi\trans \r$
\COMMENT{form residual signal estimate}
\STATE $\Omega \leftarrow \Omega \cup \supp(\thresh(\b,1))$
\COMMENT{update support with residual}
\STATE $\xhat_i\vert_\Omega \leftarrow \Phi_\Omega\pinv \y$, $\xhat_i\vert_{\Omega^C} \leftarrow 0$
\COMMENT{update signal estimate}
\STATE $\r \leftarrow \y - \Phi \xhat_i$
\COMMENT{update measurement residual}
\ENDWHILE
\STATE return $\xhat \leftarrow \xhat_i$
\end{algorithmic}
\end{algorithm}
\begin{algorithm}[t]
\caption{CoSaMP} \label{alg:cosamp}
\begin{algorithmic}
\STATE Input: CS matrix $\Phi$, measurement vector $\y$, sparsity $K$
\STATE Output: $K$-sparse approximation $\xhat$ to true signal $x$
\STATE Initialize: $\xhat_0=0$, $\r = \y$, $i = 0$
\WHILE{halting criterion false}
\STATE $i \leftarrow i+1$
\STATE $\e \leftarrow \Phi\trans \r$
\COMMENT{form residual signal estimate}
\STATE $\Omega \leftarrow \mathrm{supp}(\thresh(\e,2K))$
\COMMENT{prune residual}
\STATE $T \leftarrow \Omega \cup \mathrm{supp}(\xhat_{i-1})$
\COMMENT{merge supports}
\STATE $\b\vert_T \leftarrow \Phi_T\pinv y$, $\b\vert_{T^C} \leftarrow 0$
\COMMENT{form signal estimate}
\STATE $\xhat_i \leftarrow \thresh(\b,K)$
\COMMENT{prune signal using model}
\STATE $\r \leftarrow \y - \Phi \xhat_i$
\COMMENT{update measurement residual}
\ENDWHILE
\STATE return $\xhat \leftarrow \xhat_i$
\end{algorithmic}
\end{algorithm}

\subsection{CS Recovery Guarantees}

Many of the CS recovery algorithms above come with guarantees
on their performance. We group these results according to the matrix
metric used to obtain the guarantee.

First, we review reuslts that rely on coherence. As a first example,
BP and OMP recover a $K$-sparse vector from noiseless measurements
when the matrix $\Phi$ satisfies
(\ref{eq:maxK})~\cite{DonohoOSR,GN03,TroppGreed}.
There also exist coherence-based guarantees designed for meausrements
corrupted with arbitrary noise.
\begin{theorem}~\cite{DonohoBPIC}
Let the signal $\x \in \Sigma_K$ and write $y = \Phi x + n$. Denote
$\gamma = \|\n\|_2$. Suppose that $K \le (1/\mu(\Phi)+1)/4$ and 
$\epsilon \ge \gamma$ in (\ref{eq:bpic}).
Then the output $\xhat$ of (\ref{eq:bpic}) has error bounded by
\begin{align}
\|\x-\xhat\|_2 \le \frac{\gamma+\epsilon}{\sqrt{1-\mu(\Phi)(4K-1)}},
\end{align}
while the output $\xhat$ of the OMP algorithm with halting criterion $\|r\|_2 \le \gamma$ has error bounded by
\begin{align}
\|\x-\xhat\|_2 \le \frac{\gamma}{\sqrt{1-\mu(\Phi)(K-1)}},
\end{align}
provided that $\gamma \le A(1-\mu(\Phi)(2K-1))/2$ for OMP, with $A$
being a positive lower bound on the magnitude of the nonzero entries of $\x$.
\label{theo:bpic}
\end{theorem}
\noindent Note here that BPIC must be aware of the noise magnitude $\gamma$ 
to make $\epsilon \ge \gamma$, while OMP must be aware of the noise 
magnitude $\gamma$ to set an appropriate convergence criterion.
Additionally, the error in Theorem~\ref{theo:bpic} is proportional to
the noise magnitude $\gamma$. This is because the only assumption
on the noise is its magnitude, so that $n$ might be aligned to maximally
harm the estimation process. 

In the random noise case, bounds on $\|x-\xhat\|_2$ can only be stated in high 
probability, since there is always a small probability that the noise will be very 
large and completely overpower the signal. For example, under additive white
Gaussian noise (AWGN), the guarantees for BPIC
in Theorems~\ref{theo:bpic}  hold with high probability when the parameter
$\epsilon = \sigma\sqrt{M+\eta\sqrt{2M}}$, with $\eta$ denoting an
adjustable parameter to control the probability of $\|n\|_2$ being
too large~\cite{CandesSSR}. A second example gives a related result 
for the BPDN algorithm.
\begin{theorem}
\cite{BenHaimEldarElad}
Let the signal $\x \in \Sigma_K$ and write $y = \Phi x + n$, where
$n \sim \mathcal{N}(0, \sigma^2 {\bf I})$. Suppose that $K < 1/3\mu(\Phi)$ and consider the
BPDN optimization problem (\ref{eq:BPDN}) with
$\lambda = \sqrt{16 \sigma^2 \log M}$. Then, with probability on the order of
$1 - 1/M^2$, the solution $\xhat$ of (\ref{eq:BPDN}) is unique, its error is 
bounded by
\begin{equation}
\|x-\xhat\|_2 \le C \sigma \sqrt{K \log M},
\end{equation}
and its support is a subset of the true $K$-element support of $x$.
\label{theo:BPDN}
\end{theorem}
\noindent Under AWGN, the value of $\epsilon$ one would need to choose in 
Theorem~\ref{theo:bpic} is $\mathcal{O}(\sigma\sqrt{M})$, giving a bound 
much larger than Theorem~\ref{theo:BPDN}, which is
$\mathcal{O}(\sigma \sqrt{K\log M})$. This demonstrates the noise reduction
achievable due to the adoption of the random noise model. These
guarantees come close to the Cram\'er--Rao bound, which is given by
$\sigma\sqrt{K}$ \cite{BenHaimEldar}.
\noindent We finish the study of coherence-based guarantees for the AWGN setting with a result for OMP.
\begin{theorem}~\cite{BenHaimEldarElad}
Let the signal $\x \in \Sigma_K$ and write $y = \Phi x + n$, where
$n \sim \mathcal{N}(0, \sigma^2 {\bf I})$. Suppose that $K \le (1/\mu(\Phi)+1)/4$
and
\begin{align}
\min_{1\le n\le N} |x(n)| \ge \frac{2\sigma\sqrt{2(1+\alpha)\log N}}{1-(2K-1)\mu(\Phi)}
\end{align}
for some constant $\alpha > 0$.
Then, with probability at least $1-(N^\alpha\sqrt{\pi(1+\alpha) \log N})^{-1}$,
the output $\xhat$ of OMP after $K$ iterations has error bounded by
\begin{align}
\|\x-\xhat\|_2 \le C\sigma\sqrt{(1+\alpha)K\log N},
\end{align}
and its support matches the true $K$-element support of $x$.
\label{theo:ompgauss}
\end{theorem}
\noindent The greedy nature of OMP poses the requirement on the minimum
absolute-valued entry of $\x$ in order for the support to be correctly detected, 
in contrast to BPIC and BPDN.

A second class of  guarantees are based on the RIP.
The following result for OMP provides an interesting viewpoint of greedy
algorithms.
\begin{theorem}~\cite{WakinDavenport}
Let the signal $\x \in \Sigma_K$ and write $y = \Phi x$.
Suppose that $\Phi$ has the $(K+1,\delta)$-RIP with $\delta < \frac{1}{3\sqrt{K}}$.
Then OMP can recover a $K$-sparse signal $\x$ exactly
in $K$ iterations.
\end{theorem}
\noindent Guarantees also exist for noisy measurement settings, albeit with significantly 
more stringent RIP conditions on the CS matrix.
\begin{theorem}~\cite{ZhangOMP}
Let the signal $\x \in \Sigma_K$ and write $y = \Phi x+n$.
Suppose that $\Phi$ has the $(31K,\delta)$-RIP with $\delta < 1/3$.
Then the output of OMP after $30K$ iterations has error bounded by
\begin{align}
\|\x-\xhat\|_2 \le C\|n\|_2.
\end{align}
\end{theorem}

The next result extends guarantees from sparse to more general signals
measured under noise. We collect a set of independent statements in a
single theorem.
\begin{theorem}~\cite{CandesCS,cosamp,DM09,IHT}
Let the signal $\x \in \Sigma_K$ and write $y = \Phi x + n$.
The outputs $\xhat$ of the CoSaMP, SP, IHT, and BPIC algorithms,
with $\Phi$ having the $(cK,\delta)$-RIP, obey
\begin{equation}
\| x - \xhat \|_2 \leq C_1 \|x-x_K\|_2 + C_2\frac{1}{\sqrt{K}}
\|x-x_K\|_1 + C_3 \|\n\|_2,
\label{eq:global}
\end{equation}
where $\x_K = \arg \min_{\x' \in \Sigma_K}\|x-x'\|_2$ is the best $K$-sparse
approximation of the vector $\x$ when measured in the $\ell_2$ norm.
The requirements on the parameters $c$, $\delta$ of the RIP and
the values of $C_1$, $C_2$, and $C_3$ are specific to each algorithm.
For example, for the BPIC algorithm, $c=2$ and $\delta=\sqrt{2}-1$ suffice
to obtain the guarantee (\ref{eq:global}).
\label{theo:instanceoptimal}
\end{theorem}
The type of guarantee given in Theorem~\ref{theo:instanceoptimal}
is known as {\em uniform instance optimality}, in the sense that the CS recovery
error is proportional to that of the best $K$-sparse approximation to
the signal $x$ for any signal $x \in \real^N$. In fact, the formulation
of the CoSaMP, SP and IHT algorithms was driven by the goal of instance optimality,
which has not been shown for older greedy algorithms like MP and OMP.
Theorem~\ref{theo:instanceoptimal} can also be adapted to recovery of 
exactly sparse signals from noiseless measurements. 
\begin{corollary}
Let the signal $x \in \Sigma_K$ and write $y = \Phi x$. The CoSaMP, SP, IHT, 
and BP algorithms can exactly recover $x$ from $y$ if $\Phi$ has the $(cK,\delta)$-RIP, where the parameters $c$, $\delta$ of the RIP are specific to each algorithm.
\end{corollary}
Similarly to Theorem~\ref{theo:bpic}, the error in 
Theorem~\ref{theo:instanceoptimal} is proportional to
the noise magnitude $\|n\|_2$, and the bounds can be tailored to 
random noise with high probability. The Dantzig selector improves the scaling
of the error in the AWGN setting.
\begin{theorem}
\cite{CandesDS}
Let the signal $\x \in \Sigma_K$ and write $y = \Phi x + n$, where
$n \sim \mathcal{N}(0, \sigma^2 {\bf I})$. Suppose that $\lambda = \sqrt{2}(1+1/t)$ in
(\ref{eq:dantzig}) and that $\Phi$ has the $(2K,\delta_{2K})$ and
$(3K,\delta_{3K})$-RIPs with $\delta_{2K}+\delta_{3K} < 1$. Then,
with probability at least $1-N^t/\sqrt{\pi \log N}$, we have
\begin{align}
\|\xhat - \x\|_2 \le C(1+1/t)^2K \sigma^2\log N.
\end{align}
\label{theo:dantzig}
\end{theorem}
\noindent Similar results under AWGN have been shown for the OMP and thresholding algorithms \cite{BenHaimEldarElad}.

A third class of guarantees relies on metrics additional to coherence and RIP.
This class has a {\em non-uniform} flavor in the sense that the results
apply only for a certain subset of sufficiently sparse signals. Such
flexibility allows for significant relaxations on the properties required from
the matrix $\Phi$.
The next example has a probabilistic flavor and relies on the coherence property.
\begin{theorem}~\cite{T08}
Let the signal $\x \in \Sigma_K$ with support drawn uniformly at random from the
available $\binom{N}{K}$ possibilities and entries drawn independently from a
distribution $P(X)$ so that $P(X > 0) = P(X < 0)$. Write $\y = \Phi \x$
and fix $s \ge 1$ and $0 < \delta < 1/2$.
If $K \le \frac{\log(N/\delta)}{8\mu^2(\Phi)}$ and
\begin{align}
\sqrt{18 \log\frac{N}{\delta}\log\left(\frac{K}{2}+1\right)s} + \frac{2K}{N}\|\Phi\|_2^2 \le e^{-1/4}\left(1-2^{-1/2}\right), \nonumber
\end{align}
then $\x$ is the unique solution to BP (\ref{eq:L1})
with probability at least $1-2\delta-(K/2)^{-s}$.
\label{th:randomsubdic}
\end{theorem}
In words, the theorem says that as long as the coherence $\mu(\Phi)$ and the
spectral norm $\|\Phi\|_2$ of the CS matrix are small enough, we will be able
to recover the majority of $K$-sparse signals $\x$ from their measurements $y$.
Probabilistic results that rely on coherence can also be obtained for the
BPDN algorithm (\ref{eq:BPDN})~\cite{CandesPlan}.

The main difference between the guarantees that rely solely on coherence 
and those that rely on the RIP and probabilistic sparse signal models 
is the scaling of the number of measurements $M$ needed for 
successful recovery of $K$-sparse signals. According to the bounds 
(\ref{eq:mubound}) and (\ref{eq:ripbound}), the sparsity level that allows for 
recovery with high probability in Theorems~\ref{theo:instanceoptimal}, 
\ref{theo:dantzig}, and \ref{th:randomsubdic} is $K = \mathcal{O}(M)$, 
compared with $K = \mathcal{O}(\sqrt{M})$ for the deterministic guarantees 
provided by Theorems~\ref{theo:bpic}, \ref{theo:BPDN}, and~\ref{theo:ompgauss}. 
This so-called {\em square root bottleneck}~\cite{T08} gives an additional 
reason for the popularity of randomized CS matrices and sparse signal models.

\section{Structure in CS Matrices}
\label{sec:csmatrices}

While most initial work in CS has emphasized the use of randomized
CS matrices whose entries are obtained independently
from a standard probability distribution, such matrices are often not
feasible for real-world applications due to the cost of
multiplying arbitrary matrices with signal vectors of high
dimension. In fact, very often the physics of the sensing modality and
the capabilities of sensing devices limit the types of CS matrices that
can be implemented in a specific application. Furthermore, in the context of
analog sampling, one of the prime motivations for CS is to build analog
samplers that lead to sub-Nyquist sampling rates. These involve actual
hardware and therefore structured sensing devices. 
Hardware considerations require more elaborate signal models to 
reduce the number of measurements needed for recovery as much as 
possible. In this section, we review available alternatives for structured CS
matrices; in each case, we provide known performance guarantees, as
well as application areas where the structure arises. In
Section~\ref{sec:analog} we extend the CS framework to allow for analog
sampling, and introduce further structure into the measurement process.
This results in new hardware implementations for reduced rate samplers
based on extended CS principles.
Note that the survey of CS devices given in this section is by no
means exhaustive~\cite{GJBWS07,NK07,WJWB08}; our focus is on CS
matrices that have been investigated from both a theoretical and an
implementational point of view.

\subsection{Subsampled Incoherent Bases}

The key concept of a frame's coherence can be extended to pairs of
orthonormal bases. This enables a new choice for CS matrices: one
simply selects an orthonormal basis that is incoherent with the sparsity
basis, and obtains CS measurements by selecting a subset of the
coefficients of the signal in the chosen basis~\cite{CandesRomberg}.
We note that some degree of randomness remains in this
scheme, due to the choice of coefficients selected to represent the
signal as CS measurements.

\subsubsection{Formulation}

Formally, we assume that a basis $\Phi \in \real^{N \times N}$ is
provided for measurement purposes, where each column of
$\Phi = [\phi_1~\phi_2~\ldots~\phi_N]$ corresponds to a different basis
element.
Let $\Phibar$ be an
$N \times M$ column submatrix of $\Phi$ that preserves the basis vectors
with indices $\Gamma$ and set $\y = \Phibar^T \x$.
Under this setup, a different metric arises to evaluate the performance of CS.
\begin{definition}\cite{DonohoOSR,CandesRomberg}
The {\em mutual coherence} of the $N$-dimensional orthonormal bases
$\Phi$ and $\Psi$ is the maximum absolute value of the inner product
between elements of the two bases:
\begin{align}
\mu(\Phi,\Psi) = \max_{1\leq i,j \leq N} \left|\left<\phi_i,\psi_j\right>\right|,
\end{align}
where $\psi_j$ denotes the $j^{th}$ column, or element, of the basis $\Psi$.
\end{definition}
The mutual coherence $\mu(\Phi,\Psi)$ has values
in the range $\left[N^{-1/2},1\right]$. For example, $\mu(\Phi,\Psi) = N^{-1/2}$ when
$\Phi$ is the discrete Fourier transform basis, or Fourier matrix, and $\Psi$
is the canonical basis, or identity matrix, and $\mu(\Phi,\Psi) = 1$ when
both bases share at least one element or column. Note also that the concepts
of coherence and mutual coherence are connected by the equality
$\mu(\Phi,\Psi) = \mu([\Phi~\Psi])$.
The definition of mutual coherence can also be extended to infinite-dimensional
representations of continuous-time (analog) signals~\cite{StrohmerHeath,EldarUncertainty}.

\subsubsection{Theoretical guarantees}

The following theorem provides a recovery guarantee based on
mutual coherence.
\begin{theorem}
~\cite{CandesRomberg} Let $\x = \Psi \theta$ be a $K$-sparse
signal in $\Psi$ with support $\Omega \subset \{1,\ldots,N\}$,
$|\Omega| = K$, and with entries having signs chosen
uniformly at random. Choose a subset
$\Gamma \subseteq \{1,\ldots,N\}$ uniformly at random for the
set of observed measurements, with $M = |\Gamma|$. Suppose
that $M \ge CKN \mu^2(\Phi,\Psi) \log(N/\delta)$ and
$M \ge C' \log^2(N/\delta)$ for fixed values of $\delta<1,C,~C'$.
Then with probability at least $1-\delta$, $\theta$ is the
solution to (\ref{eq:L1}).\label{thm:coherence}
\end{theorem}
The number of measurements required by Theorem~\ref{thm:coherence}
ranges from $O(K\log N)$ to $O(N)$. It is possible to expand the
guarantee of Theorem~\ref{thm:coherence} to compressible signals by
adapting an argument of Rudelson and
Vershynin in~\cite{RudelsonVershynin} to link coherence and restricted
isometry constants.
\begin{theorem}
~\cite[Remark 3.5.3]{RudelsonVershynin}
Choose a subset $\Gamma \subseteq \{1,\ldots,N\}$ for the set of
observed measurements, with $M = |\Gamma|$. Suppose that
\begin{equation}
M \ge CK\sqrt{N}t \mu(\Phi,\Psi) \log(tK \log N)\log^2 K
\end{equation}
for a fixed value of $C$. Then with  probability at least $1-5e^{-t}$ the
matrix $\Phi^T\Psi$ has the RIP with constant
$\delta_{2K} \le 1/2$.\label{thm:coherenceRIP}
\end{theorem}
Using this theorem, we obtain the guarantee of
Theorem~\ref{theo:instanceoptimal} for compressible signals with the
number of measurements $M$ dictated by the coherence value
$\mu(\Phi,\Psi)$.

\subsubsection{Applications}
\label{sec:spc}

There are two main categories of applications where subsampled
incoherent bases are used. In the first category, the acquisition
hardware is limited by construction to measure directly in a transform
domain. The most relevant examples are magnetic resonance imaging
(MRI)~\cite{LustigDonohoSantosPauly} and tomographic
imaging~\cite{CandesRUP}, as well as optical
microscopy~\cite{GSES09,SGSES10}; in all of these cases, the measurements
obtained from the hardware correspond to coefficients of the image's
2-D continuous Fourier transform, albeit not typically selected in a
randomized fashion. Since the Fourier functions,
corresponding to sinusoids, will be incoherent with functions that have
localized support, this imaging approach works well in practice
for sparsity/compressibility transforms such as
wavelets~\cite{CandesRomberg}, total variation~\cite{CandesRUP}, and
the standard canonical representation~\cite{GSES09}.

In the case of optical microscopy, the Fourier coefficients that are
measured correspond to the lowpass regime. The highpass values are
completely lost. When the underlying signal $\x$ can change sign,
standard sparse recovery algorithms such as BP do not typically succeed
in recovering the true underlying vector. To treat the case of recovery from
lowpass coefficients, a special purpose sparse recovery method was
developed under the name of Nonlocal Hard Thresholding
(NLHT)~\cite{GSES09}. This technique attempts to allocate the off-support
of the sparse signal in an iterative fashion by performing a thresholding
step that depends on the values of the neighboring locations.

\begin{figure}
\begin{center}
\includegraphics[height=1.2in]{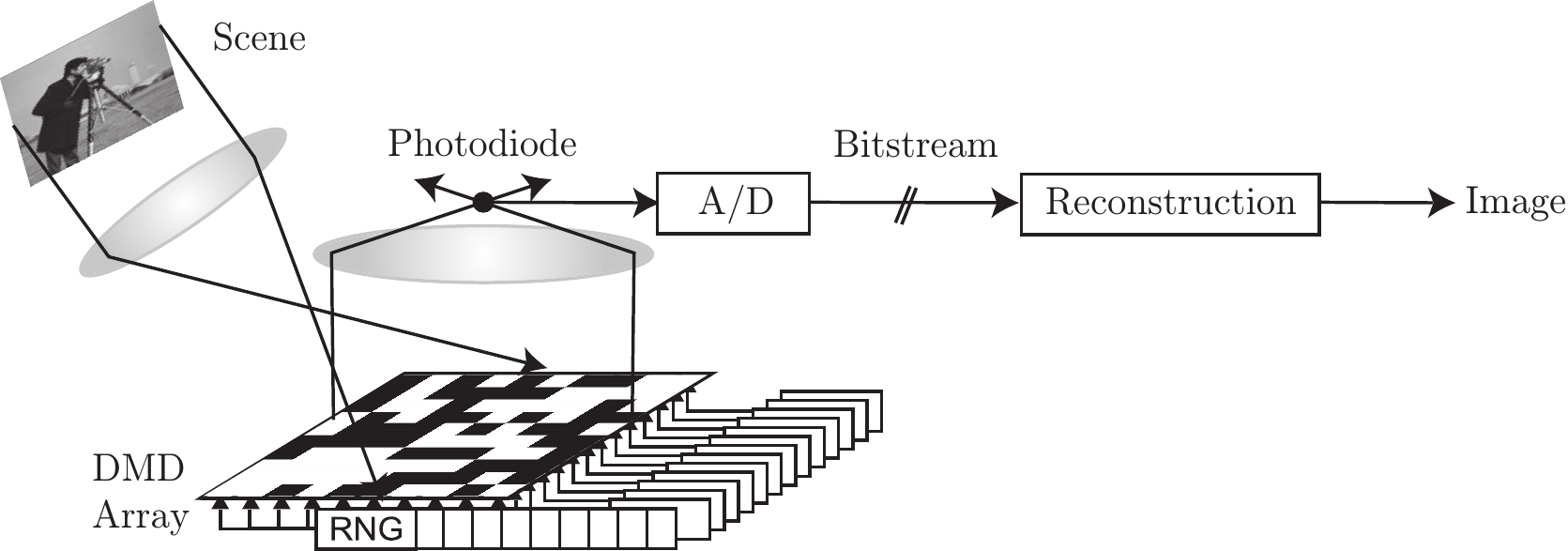}
\end{center}
\caption{{\sl Diagram of the single pixel camera.
The incident lightfield (corresponding to the desired image $x$) is
reflected off a digital micro-mirror device (DMD) array whose mirror
orientations are modulated in the pseudorandom pattern supplied by
the random number generator (RNG).  Each different mirror pattern
produces a voltage at the single photodiode that corresponds to
one measurement $y(m)$. The process is repeated with different patterns
$M$ times to obtain a full measurement vector $y$ (taken
from~\cite{DuarteDavenportTakharLaskaKellyBaraniuk}).}}
\label{fig:spc}
\end{figure}

The second category involves the design of new acquisition hardware
that can obtain projections of the signal against a class of vectors.
The goal of the matrix design step is to find a basis whose elements
belong to the class of vectors that can be implemented on the hardware.
For example, a class of single pixel imagers based on optical modulators~\cite{DuarteDavenportTakharLaskaKellyBaraniuk,CandesSSR}
(see an example in Fig.~\ref{fig:spc}) can obtain projections of an
image against vectors that have binary entries. Example bases that meet
this criterion include the Walsh-Hadamard and noiselet
bases~\cite{Noiselets}. The latter is particularly interesting for
imaging applications, as it is known to be maximally incoherent with the
Haar wavelet basis. In contrast, certain elements of the Walsh-Hadamard
basis are highly coherent with wavelet functions at coarse scales, due to
their large supports. Permuting the entries of the basis vectors (in a random or pseudorandom fashion) helps reduce the coherence between the
measurement basis and a wavelet basis.
Because the single pixel camera modulates the
light field through optical aggregation, it improves the signal-to-noise ratio
of the measurements obtained when compared to standard multisensor
schemes~\cite{Coifman}.
Similar imaging hardware architectures have been developed in~\cite{ChanCharanTakharKellyBaraniukMittleman,YeParedesArceWuChenPrather}.

\begin{figure}
\begin{center}
\includegraphics[height=2.5in]{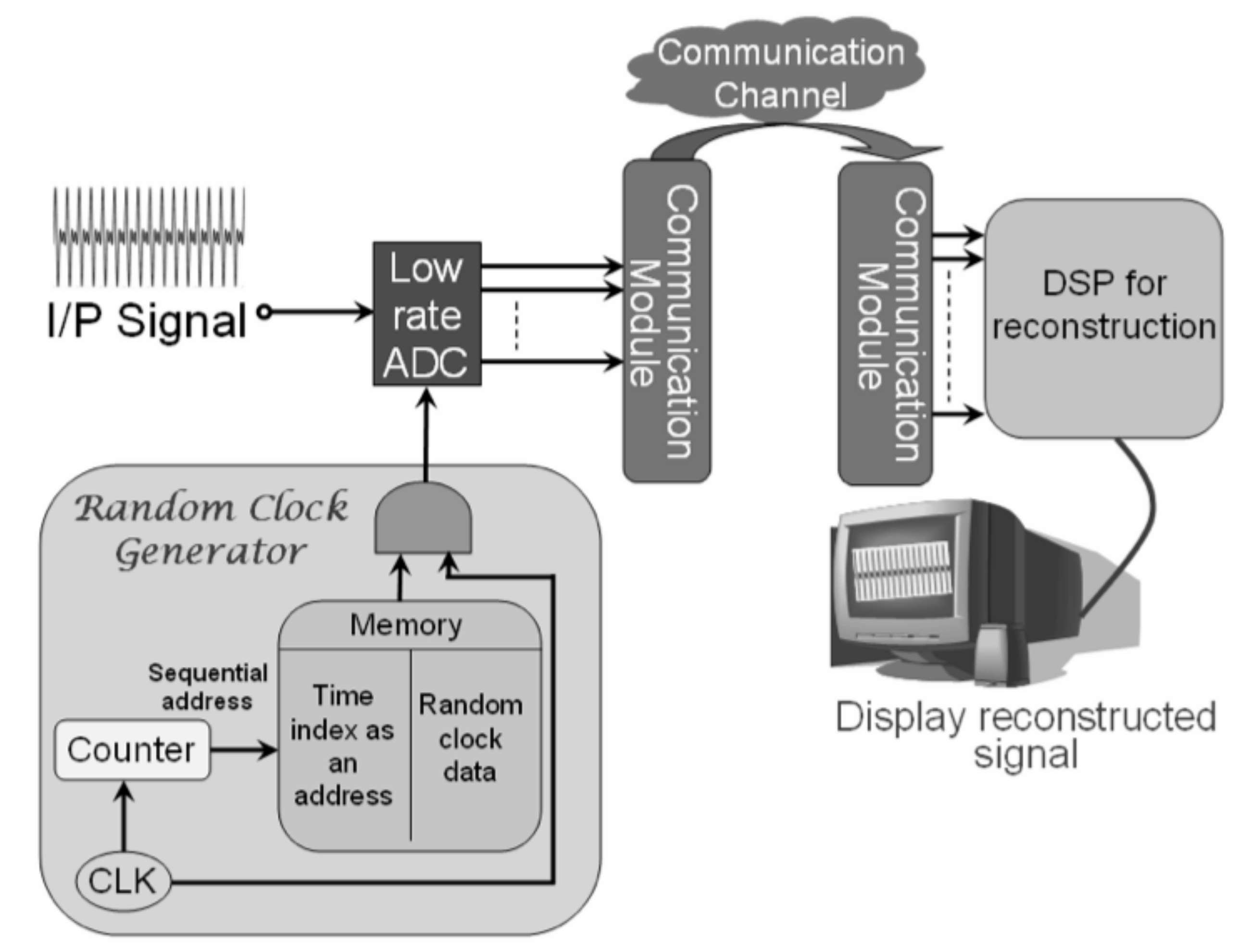}
\end{center}
\caption{{\sl Diagram of the random sampling ADC.
A pseudorandom clock generated by an FPGA drives a low-rate standard
ADC to sample the input signal at predetermined pseudorandom times.
The samples obtained are then fed into a CS recovery algorithm or a
sketching algorithm for Fourier sparse signals to estimate the frequencies
and amplitudes of the tones present in the signal (taken
from~\cite{PRLNGSBM}).}}
\label{fig:radc}
\end{figure}

An additional example of a configurable acquisition device is the
random sampling ADC~\cite{PRLNGSBM}, which is designed for acquisition of
periodic, multitone analog signals whose frequency components belong
to a uniform grid (similarly to the random demodulator of Section~\ref{sec:rd}).
The sparsity basis for the signal once again is chosen to be the discrete Fourier
transform basis. The measurement basis is chosen to be the identity basis,
so that the measurements correspond to standard signal
samples taken at randomized times. The hardware implementation employs
a randomized clock that drives a traditional low-rate ADC to sample
the input signal at specific times. As shown in Fig.~\ref{fig:radc}, the random
clock is implemented using an FPGA that outputs a predetermined
pseudorandom pulse sequence indicating the times at which the signal
is sampled. The patterns are timed according to a set of pseudorandom
arithmetic progressions. This process is repeated cyclically, establishing a
windowed acquisition for signals of arbitrary length. The measurement and
sparsity framework used by this randomized ADC is also compatible with
sketching algorithms designed for signals that are sparse in the frequency
domain~\cite{PRLNGSBM,GST08}.

\subsection{Structurally Subsampled Matrices}
\label{sec:subsampled}

In certain applications, the measurements obtained by the acquisition
hardware do not directly correspond to the sensed signal's coefficients
in a particular transform. Rather, the observations are a linear
combination of multiple coefficients of the signal.
The resulting CS matrix has been termed a structurally subsampled
matrix~\cite{BajwaSayeedNowak}.

\subsubsection{Formulation}

Consider a matrix of available measurement vectors that can be
described as the product $\Phi = \bR\bU$, where $\bR$ is a $P \times N$
mixing matrix and $\bU$ is a basis.
The CS matrix $\Phibar$ is obtained by selecting $M$
out of $P$ rows at random, and normalizing the columns of the resulting
subsampled matrix. There are two possible downsampling stages: first,
$\bR$ might offer only $P < N$ mixtures to be available as
measurements; second, we only preserve $M < P$ of the mixtures
available to represent the signal. This formulation includes the
use of subsampled incoherent bases simply by letting $P=N$ and
$\bR=\bI$, i.e., no coefficient mixing is performed.

\subsubsection{Theoretical guarantees}

To provide theoretical guarantees we place some additional constraints on
the mixing matrix $\bR$.
\begin{definition}
The $(P,N)$ integrator matrix $\bS$, for $P$ a divisor of $N$,
is defined as $\bS = [\bs_1^T~\bs_2^T~\ldots~\bs_P^T]^T$, where
the $p^{th}$ row of $\bS$ is defined as
$\bs_p = [{\bf 0}_{1 \times (p-1)L}~{\bf 1}_{1 \times L}~{\bf 0}_{1 \times (N-pL)}]$,
$1 \le p \le P$, with $L = P/N$.
\end{definition}
In words, using $\bS$ as a mixing matrix sums together intervals of $L$
adjacent coefficients of the signal under the transform $\bU$. We also use
a diagonal modulation matrix $\bM$ whose nonzero entries are
independently drawn from a Rademacher
distribution, and formulate our mixing matrix as $\bR = \bS \bM$.
\begin{theorem}
~\cite[Theorem 1]{BajwaSayeedNowak}
Let $\Phibar$ be a structurally subsampled matrix of size $M \times N$
obtained from the basis $\bU$ and the $P \times N$ mixing matrix
$\bR = \bS \bM$ via randomized subsampling. Then for each integer $K > 2$,
any $z > 1$ and any $\delta \in (0,1)$, there exist absolute positive
constants $c_1$, $c_2$ such that if
\begin{align}
M \ge c_1 z KN \mu^2(\bU,\Psi) \log^3 N \log^2 K,
\end{align}
then the matrix $\Phibar$ has the $(K,\delta)$-RIP with probability at least
$1-20 \max\{\exp(-c_2\delta^2z),N^{-1}\}$.
\end{theorem}
Similarly to the subsampled incoherent bases case, the 
possible values of $\mu(\bU,\Psi)$ provide us with a required number of
measurements $M$ ranging from $O(K\log^3 N)$ to $O(N)$.

\subsubsection{Applications}
\label{sec:rd}

Compressive ADCs are one promising application of CS, which we
discuss in detail in Section~\ref{sec:mwc} after introducing the
infinite-dimensional CS framework.
A first step in this direction is the architecture known as the random
demodulator (RD)~\cite{TroppLaskaDuarteRombergBaraniuk}. The RD
employs structurally subsampled matrices for the acquisition of
periodic, multitone analog signals whose frequency components
belong in a uniform grid. Such signals have a finite parametrization and
therefore fit the finite-dimensional CS setting.

\begin{figure}
\begin{center}
\includegraphics[height=1.4in]{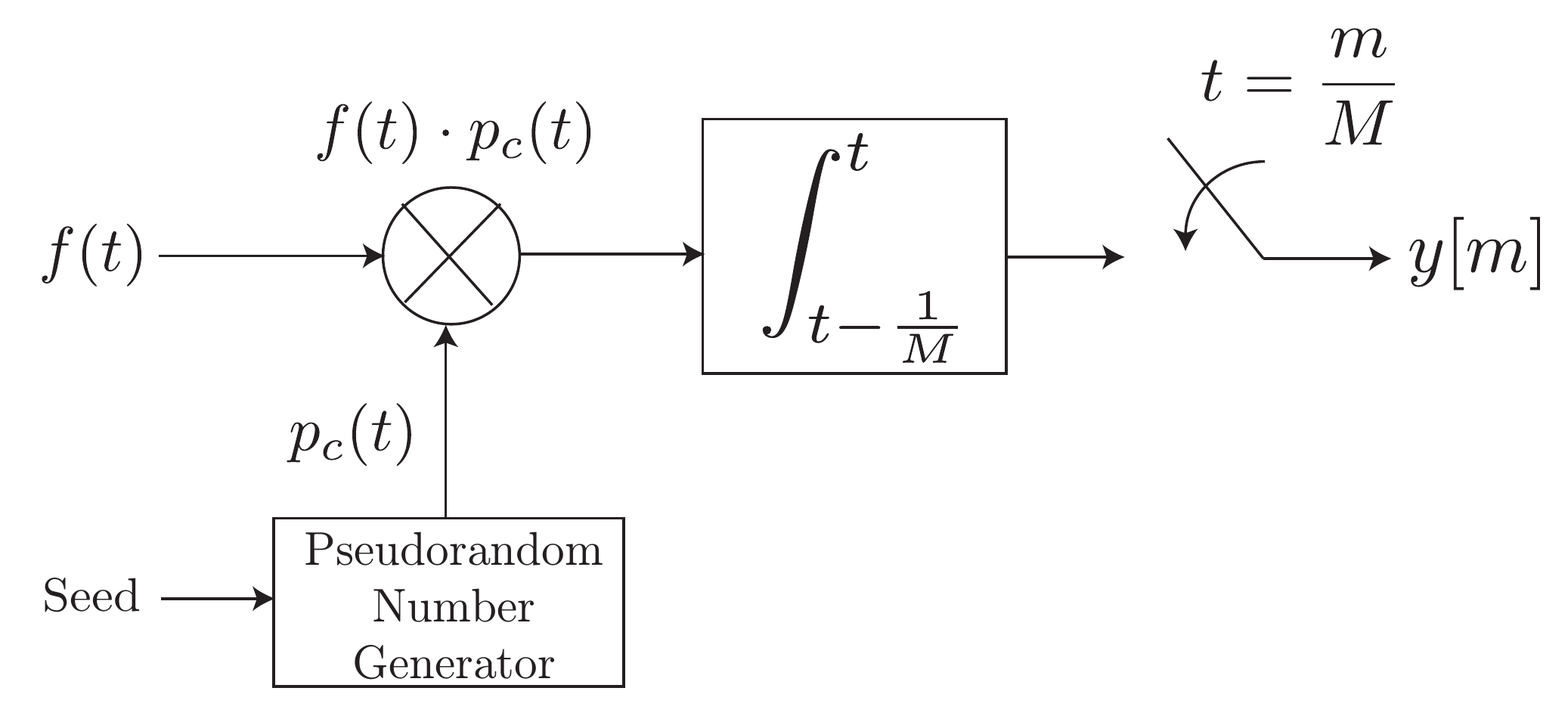}
\end{center}
\caption{{\sl Block diagram of the random demodulator
(taken from~\cite{TroppLaskaDuarteRombergBaraniuk}).}}
\label{fig:randdemod}
\end{figure}

To be more specific, our aim is to acquire a discrete uniform sampling
$x \in \real^N$ of a continuous-time signal $f(t)$ observed during an acquisition
interval of 1 sec where it is assumed that $f(t)$ is of the form
\begin{align}
f(t)=\sum_{k=1}^N x_k e^{j 2\pi k t}.
\end{align}
The vector $\x$ is sparse so that  only $K$ of its values are non-zero.
As shown in Fig.~\ref{fig:randdemod}, the sampling
hardware consists of a mixer element that multiplies the signal
$f(t)$ with a chipping sequence $p_c(t)$ at a rate $N$ chirps/second. This
chipping sequence is the output of a pseudorandom number generator.
The combination of these two units effectively performs the product of the
Nyquist-sampled discrete representation of the signal $x$ with the matrix
$\bM$ in the analog domain. The output from the
modulator is sampled by an ``integrate-and-dump'' sampler -- a combination
of an accumulator unit and a uniform sampler synchronized with each other --
at a rate of $M$ samples per interval. Such sampling effectively performs a
multiplication of the output of the modulator by the $(M,N)$ integrator matrix
$\bS$. To complete the setup, we set $\bU = \bI$ and $\Psi$ to be the discrete
Fourier transform basis; it is easy to see that these two bases are maximally
incoherent. In the RD architecture, all subsampling is performed at this
stage; there is no randomized subsampling of the output of the integrator.

A prototype implementation of this architecture is reported in~\cite{Ragheb};
similar designs are proposed in~\cite{YuHoyosSadler,CandesWakinCS}.
Note that the number of columns of the resulting CS matrix scales with the
maximal frequency in the representation of $f(t)$. Therefore, in practice, this
maximal frequency cannot be very large. For example, the implementations
reported above reach maximum frequency rates of 1MHz; the corresponding
signal representation therefore has one million coefficients. The
matrix multiplications can be implemented using algorithmic tools such as
the Fast Fourier Transform (FFT). In conjunction with certain optimization solvers
and greedy algorithms, this approach significantly reduces the complexity
of signal recovery. While the original formulation sets the
sparsity basis $\Psi$ to be the Fourier basis, limiting the set of recoverable
signals to periodic multitone signals, we can move beyond structurally
subsampled matrices by using redundant Fourier domain frames for sparsity.
These frames, together with modified recovery algorithms, can enable
acquisition of a wider class of frequency-sparse signals~\cite{DB10,VS10}.

While the RD is capable of implementing CS acquisition for a specific class
of analog signals having a finite parametrization, the CS framework can be
adapted to infinite-dimensional signal models, enabling more efficient analog
acquisition and digital processing architectures. We defer the details to
Section~\ref{sec:analog}.

\subsection{Subsampled Circulant Matrices}

The use of Toeplitz and circulant
structures~\cite{HauptBajwaRazNowak,RauhutStructured,RRT10}
as CS matrices was first inspired by applications in communications --
including channel estimation and multi-user detection -- where a sparse
prior is placed on the signal to be estimated, such as a channel response
or a multiuser activity pattern. When compared with generic CS matrices,
subsampled circulant matrices have a significantly smaller number of
degrees of freedom due to the repetition of the matrix entries along
the rows and columns.

\subsubsection{Formulation}
A circulant matrix $\bU$ is a square matrix where the entries in
each diagonal are all equal, and where the first entry of the second and
subsequent rows is equal to the last entry of the previous row.
Since this matrix is square, we perform random subsampling of the rows
(in a fashion similar to that described in Section~\ref{sec:subsampled}) to
obtain a CS matrix $\Phi = \bR \bU$, where $\bR$ is an $M \times N$
subsampling matrix, i.e., a submatrix of the identity matrix. We dub $\Phi$
a {\em subsampled circulant matrix}.
\subsubsection{Theoretical guarantees}
\label{sec:thtoeplitz}
Even when the sequence defining $\bU$ is drawn at random from the
distributions described in Section~\ref{sec:csbasics}, the particular
structure of the subsampled circulant matrix $\Phi = \bR \bU$ prevents the use
of the proof techniques used in standard CS, which require all entries of the
matrix to be independent. However, it is possible to employ different probabilistic
tools to provide guarantees for subsampled circulant matrices.
The results still require randomness in the selection of the entries of the
circulant matrix.
\begin{theorem}~\cite{RRT10}
Let $\Phi$ be a subsampled circulant matrix whose distinct entries
are independent random variables following a Rademacher distribution,
and $\bR$ is an arbitrary $M \times N$ identity submatrix. Furthermore,
let $\delta$ be the smallest value for which (\ref{eq:rip}) holds for all
$x \in \Sigma_K$. Then for $\delta_0 \in (0,1)$ we have $\E[\delta] \le \delta_0$
provided that
\begin{align}
M \ge \C \max\{\delta_0^{-1}K^{3/2}\log^{3/2}N,\delta_0^{-2}K\log^2N\log^2K\}, \nonumber
\end{align}
where $\C > 0$ is a universal constant. Furthermore, for $0 \le \lambda \le 1$,
\begin{align}
\P(\delta_K \ge \E[\delta] + \lambda) \le e^{\lambda^2/\sigma^2},~\textrm{where}~\sigma^2 = C'\frac{K}{M}\log^2K \log N, \nonumber
\end{align}
for a universal constant $C' > 0$.
\label{th:circulant}
\end{theorem}
In words, if we have $M = \bigo{K^{1.5}\log^{1.5}N}$, then the RIP required
for $\Phi$ by many algorithms for successful recovery is achieved with high
probability.

\subsubsection{Applications}
\label{sec:convimg}

There are several sensing applications where the signal to be acquired
is convolved with the sampling hardware's impulse response before it is
measured. Additionally, because convolution is equivalent to a product
operator in the Fourier domain, it is possible to speed up
the CS recovery process by performing multiplications by the matrices
$\Phi$ and $\Phi^T$ via the FFT. In fact, such
an FFT-based procedure can also be exploited to generate good CS
matrices~\cite{RombergConvolution}. First, we design a matrix $\Phi$ in
the Fourier domain to be diagonal and have entries drawn from a
suitable probability distribution. Then, we obtain the measurement matrix
$\Phibar$ by subsampling the matrix $\Phi$, similarly to the incoherent
basis case. While this formulation assumes that the convolution
during signal acquisition is circulant, this gap between theory and practice
has been studied and shown to be controllable~\cite{MarciaW_ICASSP08}.

\begin{figure}
\begin{center}
\begin{tabular}{cc}
\includegraphics[height=1.5in]{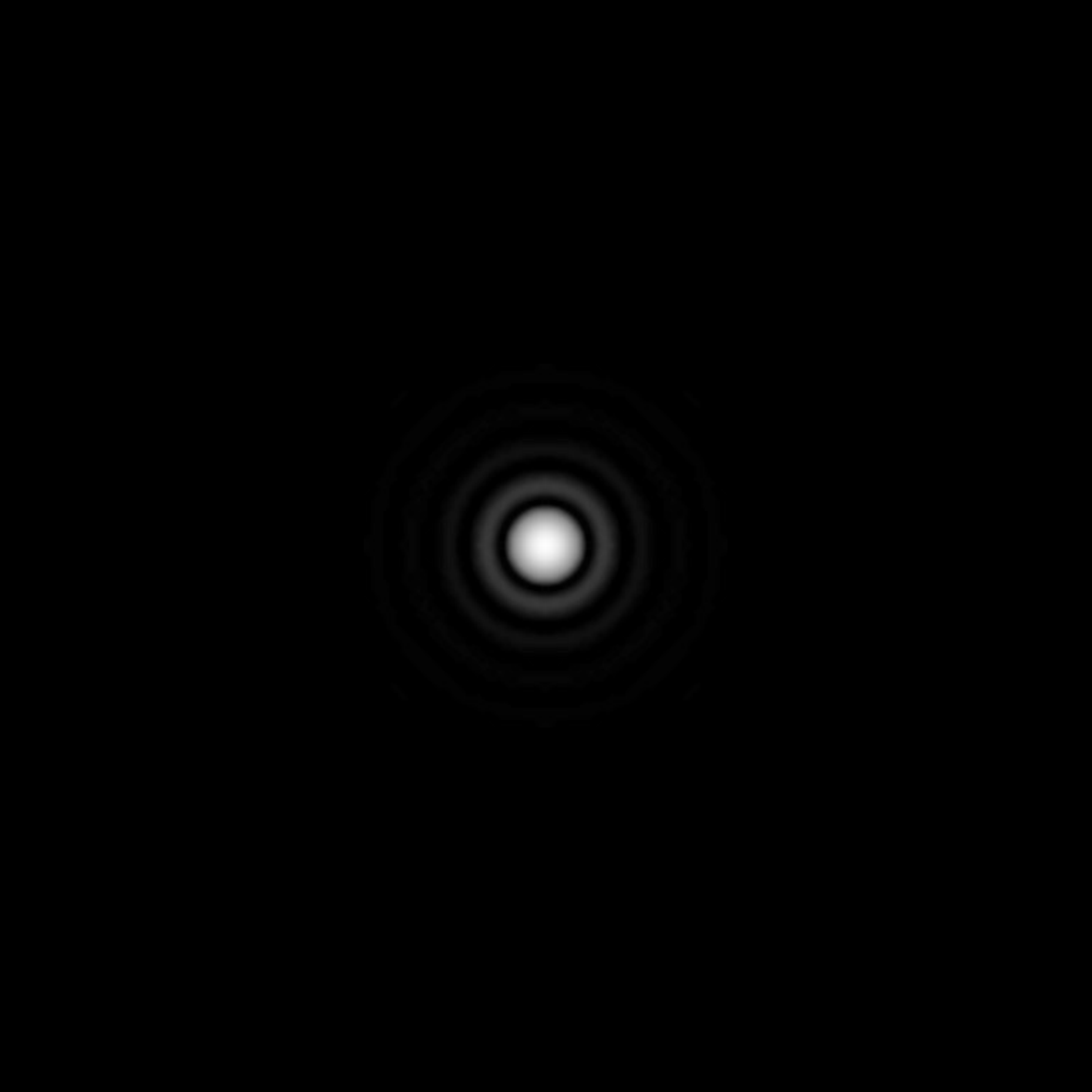} &
\includegraphics[height=1.5in]{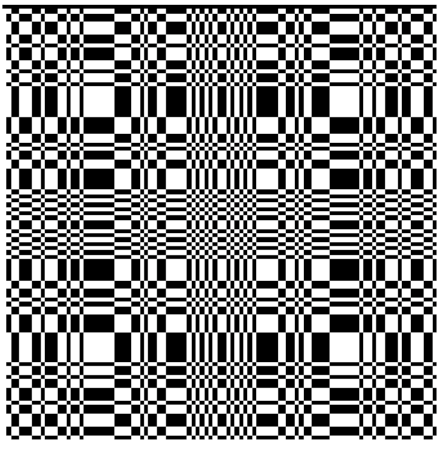} \\
(a) & (b)
\end{tabular}
\end{center}
\caption{{\sl Compressive imaging via coded aperture. (a) Example of an
ideal point spread function for a camera with a pinhole aperture. The size
of the support of the function is dependent on the amount of blur
caused by the imager. (b) Coded aperture used for compressive imaging
(taken from~\cite{MarciaHarmanyWillett}).}}
\label{fig:codedaperture}
\end{figure}

Our first example concerns imaging architectures. The impulse response
of the imaging hardware is known as the point spread function (PSF),
and it represents imaging artifacts such as blurring, distortion, and other
aberrations; an example is shown in Fig.~\ref{fig:codedaperture}(a).
It is possible then to design a compressive imaging architecture by
employing an imaging device that has a dense PSF: an impulse response
having a support that includes all of the imaging field. This dense PSF is
coupled with a downsampling step in the pixel domain to achieve
compressive data acquisition~\cite{MarciaW_ICASSP08,MarciaHarmanyWillett}.
This is in contrast to the random sampling advocated by
Theorem~\ref{th:circulant}. A popular way to achieve such a dense PSF
is by employing so-called coded apertures, which change the pinhole
aperture used for image formation in most cameras to a more complex
design. Figure~\ref{fig:codedaperture}(b) shows an example coded
aperture that has been used successfully in compressive
imaging~\cite{MarciaW_ICASSP08,MarciaHarmanyWillett}.

\begin{figure}
\begin{center}
\includegraphics[height=4in]{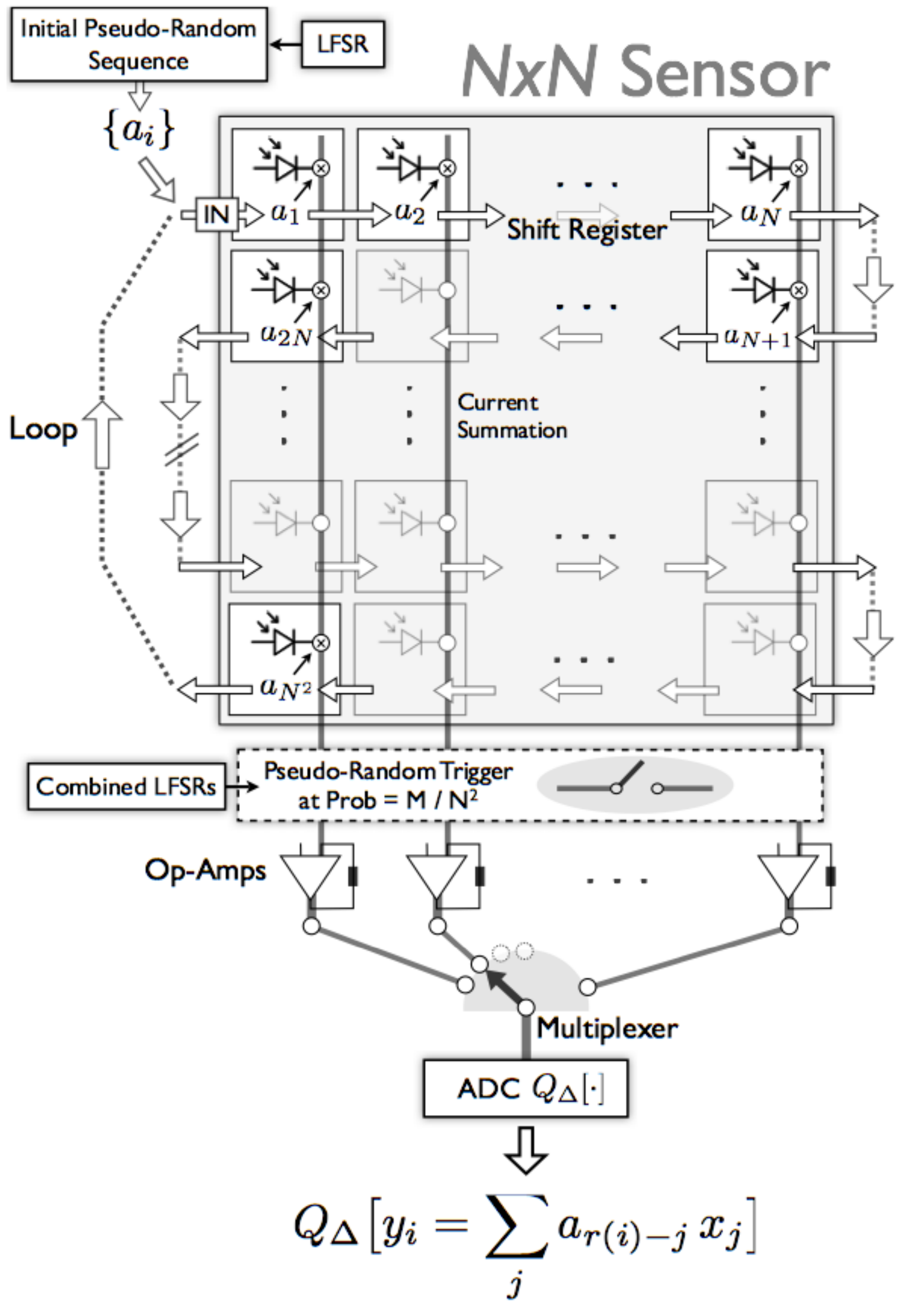}
\end{center}
\caption{{\sl Schematic for the microelectronic architecture of the
convolution-based compressive imager. The acquisition device effectively
implements a subsampled circulant CS matrix. Here, $Q_\Delta$ denotes the
quantization operation with resolution $\Delta$ (taken
from~\cite{JacquesVandergheynstBibetMajidzadehSchmidLeblebici}).}}
\label{fig:convolutionimager}
\end{figure}

Our second example uses special-purpose light sensor arrays that
perform the multiplication with a Toeplitz matrix using a custom
microelectronic architecture~\cite{JacquesVandergheynstBibetMajidzadehSchmidLeblebici}, which is shown in Fig.~\ref{fig:convolutionimager}.
In this architecture, an $N \times N$ pixel light sensor array is coupled
with a linear feedback shift register (LFSR) of length $N^2$ whose input
is connected to a pseudorandom number generator. The bits in the
LFSR control multipliers that are tied to the outputs from the $N^2$
light sensors, thus performing additions and subtractions according to the
pseudorandom pattern generated. The weighted outputs are summed
in two stages: column-wise sums are obtained at an operational
amplifier before quantization, whose outputs are then added together in an
accumulator. In this way, the microelectronic architecture calculates the inner
product of the image's pixel values with the sequence contained in the LFSR.
The output of the LFSR is fed back to its input after the register is full, so that
subsequent measurements correspond to inner products with shifts of previous
sequences. The output of the accumulator is sampled in a pseudorandom
fashion, thus performing the subsampling required for CS.
This results in an effective subsampled circulant CS matrix.

\subsection{Separable Matrices}

Separable matrices~\cite{DuarteBaraniuk,RivensonStern} provide
computationally efficient alternatives to measure very large signals, such
as hypercube samplings from multidimensional data.
These matrices have a succinct mathematical formulation as Kronecker
products, and feature significant structure present as correlations among
matrix subblocks.

\subsubsection{Formulation}
Kronecker product bases are well suited for CS applications
concerning multidimensional signals, that is, signals that capture
information from an event that spans multiple dimensions, such as
spatial, temporal, spectral, etc. These bases can be used both to
obtain sparse or compressible representations or as CS
measurement matrices.

The {\em Kronecker product} of two matrices $A$ and $B$ of sizes
$P\times Q$ and $R \times S$, respectively, is defined as
\begin{align}
A \otimes B := \left[ \begin{array}{cccc}
A(1,1)B & A(1,2)B & \ldots & A(1,Q)B \\
A(2,1)B & A(2,2)B & \ldots & A(2,Q)B \\
\vdots&\vdots&\ddots&\vdots \\
A(P,1)B & A(P,2)B & \ldots & A(P,Q)B
\end{array} \right]. \nonumber
\end{align}
Thus, $A \otimes B$ is a matrix of size $PR \times QS$.
Let $\Psi_1$ and $\Psi_2$ be bases for $\real^{N_1}$ and $\real^{N_2}$,
respectively. Then one can find a basis for
$\real^{N_1} \otimes \real^{N_2} = \real^{N_1N_2}$ as
$\widetilde{\Psi} = \Psi_1 \otimes \Psi_2$. We focus now on CS matrices
that can be expressed as the Kronecker product of $D$ matrices:
\begin{align}
\Phibar = \Phi_1 \otimes \Phi_2 \otimes \ldots \otimes \Phi_D.
\label{eq:phik}
\end{align}
If we denote the size of $\Phi_d$ as $M_d \times N_d$, then the matrix
$\Phi$ has size $M \times N$, with $M = \prod_{d=1}^D M_d$ and
$N = \prod_{d=1}^D N_d$.

Next, we describe the use of Kronecker product matrices in CS of
multidimensional signals. We denote the entries of
a $D$-dimensional signal $x$ by $x(n_1,\ldots,n_d)$. We call the
restriction of a multidimensional signal to fixed indices for all but its
$d^{\textrm{th}}$ dimension a $d$-section of the signal. For example, for
a 3-D signal $\x \in \real^{N_1\times N_2 \times N_3}$, the vector
$\x_{i,j,\cdot} := [\x(i,j,1)$ $\x(i,j,2)~\ldots~\x(i,j,N_3)]$ is a 3-section of $\x$.

Kronecker product sparsity bases make it possible to simultaneously
exploit the sparsity properties of a multidimensional signal along each of
its dimensions. The new basis is simply the Kronecker product of the bases 
used for each of its $d$-sections. More formally, we let 
$\x \in \real^{N_1} \otimes \ldots \otimes \real^{N_D} = \real^{N_1\times \ldots \times N_D} = \real^{\prod_{d=1}^D N_d}$
and assume that each $d$-section is sparse or compressible in a
basis $\Psi_d$. We then define a sparsity/compressibility basis for $x$ as
$\Psibar = \Psi_1 \otimes \ldots \otimes \Psi_D$,
and obtain a coefficient vector $\thetabar$ for the signal ensemble
so that $\xbar = \Psibar~\thetabar$, where $\xbar$ is a column
vector-reshaped representation of $x$. We then have
$y = \Phibar \xbar = \Phibar~\Psibar~\thetabar$.

\subsubsection{Theoretical guarantees}

Due to the similarity between blocks of Kronecker product CS matrices, it is
easy to obtain bounds for their performance metrics. Our first result concerns
the RIP.

\begin{lemma}
~\cite{DuarteBaraniuk,RivensonStern}
Let $\Phi_d$, $1 \le d \le D$, be matrices that have the $(K,\delta_d)$-RIP,
$1 \le d \le D$, respectively. Then $\Phibar$, defined in (\ref{eq:phik}), has
the $(K,\delta)$-RIP, with
\begin{align}
\delta \le \prod_{d=1}^D(1+\delta_d)-1.
\end{align}
\end{lemma}
When $\Phi_d$ is an orthonormal basis, it has the $(K,\delta_d)$-RIP
with $\delta_d = 0$ for all $K \le N$. Therefore the RIP constant
of the Kronecker product of an orthonormal basis and a CS matrix is
equal to the RIP constant of the CS matrix.

It is also possible to obtain results on mutual coherence (described in
Section~\ref{sec:csbasics}) for cases in which the basis used for sparsity
or compressibility can also be expressed as a Kronecker product.
\begin{lemma}~\cite{DuarteBaraniuk,RivensonStern}
Let $\Phi_d$, $\Psi_d$ be bases for $\real^{N_d}$ for $d=1,\ldots,D$. Then
\begin{align}
\mu(\Phibar,\Psibar) = \prod_{d=1}^D\mu(\Phi_d,\Psi_d).
\end{align}
\label{lemma:mukron}
\end{lemma}
Lemma~\ref{lemma:mukron} provides a {\em conservation of mutual coherence}
across Kronecker products.
Since the mutual coherence of each $d$-section's sparsity and
measurement bases is upper bounded by one, the number
of Kronecker product-based measurements necessary for
successful recovery of the multidimensional signal
(following Theorems~\ref{thm:coherence} and~\ref{thm:coherenceRIP})
is always lower than or equal to the corresponding number of
necessary {\em partitioned measurements} that process only a
portion of the multidimensional signal along its $d^{th}$ dimension
at a time, for some $d \in \{1,\ldots,D\}$. This reduction is
maximized when the $d$-section measurement basis $\Phi$ is
maximally incoherent with the $d$-section sparsity basis $\Psi$.

\subsubsection{Applications}
\label{sec:kronimg}

Most applications of separable CS matrices involve multidimensional
signals such as video sequences and hyperspectral datacubes. Our
first example is an extension of the single pixel camera (see
Fig.~\ref{fig:spc}) to hyperspectral imaging~\cite{SunKelly}. The aim
here is to record the reflectivity of a scene at different wavelengths; each
wavelength has a corresponding spectral frame, so that the hyperspectral
datacube can be essentially interpreted as the stacking of a sequence of
images corresponding to the different spectra. The single pixel
hyperspectral camera is obtained simply by replacing the single
photodiode element by a spectrometer, which records the intensity of the
light reaching the sensor for a variety of different wavelenghts.
Because the digital micromirror array reflects all the wavelengths of interest, the
spectrometer records measurements that are each dependent only on a
single spectral band. Since the same patterns are acting as a modulation
sequence to all the spectral bands in the datacube, the resulting
measurement matrix is expressible as $\Phibar = \ident \otimes \Phi$,
where $\Phi$ is the matrix that contains the patterns programmed into the
mirrors. Furthermore, it is possible to compress a hyperspectral datacube
by using a hyperbolic wavelet basis, which itself is obtained as the
Kronecker product of one-dimensional wavelet bases~\cite{DuarteBaraniuk}.

\begin{figure}
\begin{center}
\includegraphics[height=3in]{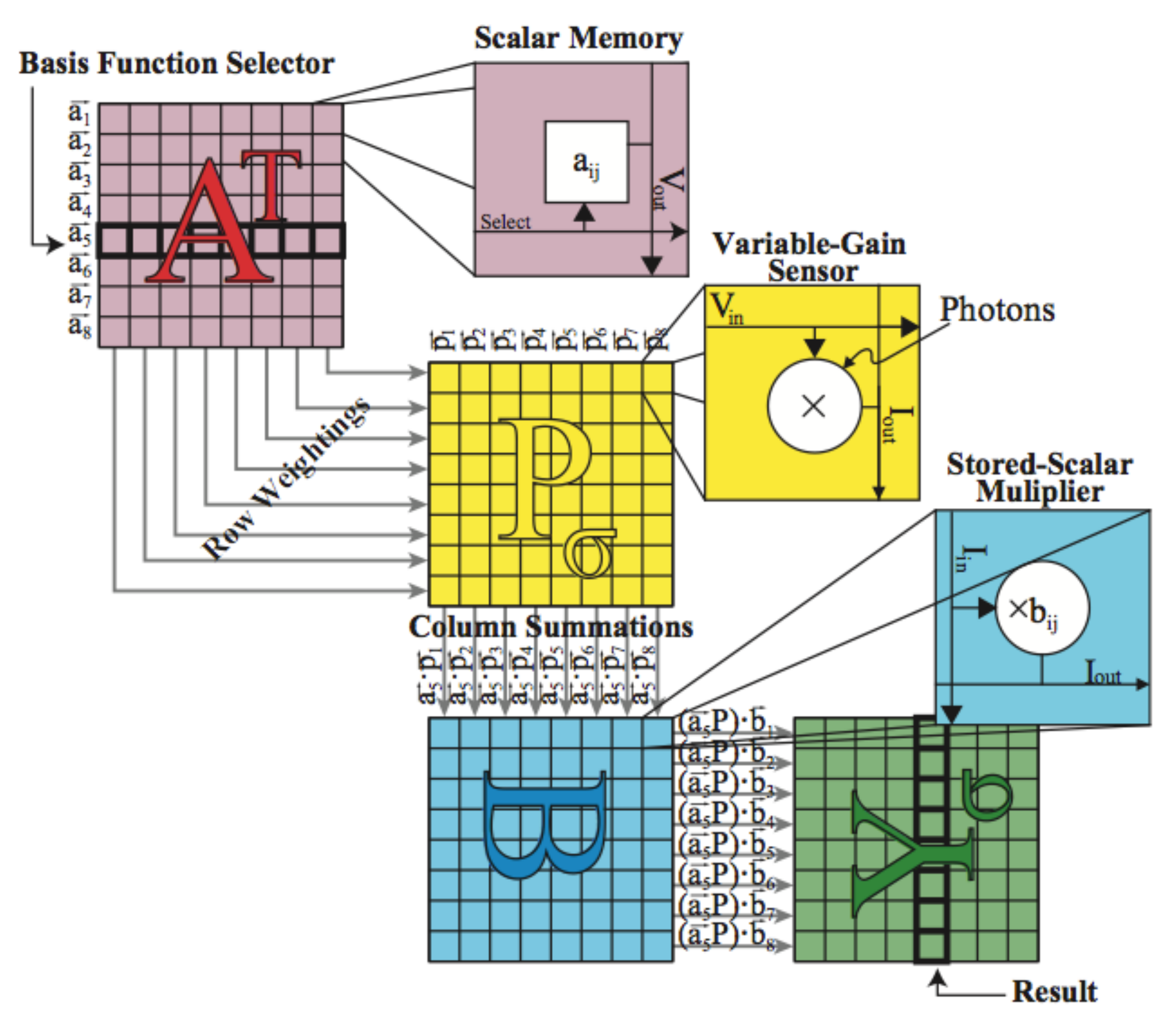}
\end{center}
\caption{{\sl Schematic for the compressive analog CMOS imager.
The acquisition device effectively implements a separable CS matrix (taken
from~\cite{RobucciGrayChiuRombergHasler}).}}
\label{fig:analogcsimaging}
\vspace{-5mm}
\end{figure}

Our second example concerns the transform
imager~\cite{RobucciGrayChiuRombergHasler}, an imaging hardware
architecture that implements a separable CS matrix. A sketch of the
transform imager is shown in Fig.~\ref{fig:analogcsimaging}. The image
$x$ is partitioned into blocks $P_\sigma$ of size $16 \times 16$ to form a
set of tiles; here $\sigma$ indexes the block locations. The transform
imager is designed to perform the multiplication
$y_\sigma = A^T P_\sigma B$, where $A$ and $B$ are fixed matrices.
This product of three matrices can be equivalently rewritten as
$y_\sigma = (B \otimes A)^T \overline{P_\sigma}$, where
$\overline{P_\sigma}$ denotes a column vector reshaping of the block
$P_\sigma$~\cite{HornJohnson}. The CS measurements are then obtained
by randomly subsampling the vector $y_\sigma$. The compressive transform
imager sets $A$ and $B$ to be noiselet bases and uses a 2-D undecimated
wavelet transform to sparsify the image.

\section{Structure in Finite-Dimensional Models}
\label{sec:finitemodels}

Until now we focused on structure in the measurement matrix $\Phi$,
and considered signals $\x$ with finite-dimensional sparse
representations. We now turn to discuss how to exploit structure beyond
sparsity in the input signal in order to reduce the number of measurements
needed to faithfully represent it.
Generalizing the notion of sparsity will allow us to move away from
finite-dimensional models extending the ideas of CS
to reduce sampling rates for infinite-dimensional continuous-time
signals, which is the main goal of sampling theory.

We begin our discussion in this section within the finite-dimensional
context by adding structure to the non-zero values of $\x$. We will
then turn, in Section~\ref{sec:analog}, to more general,
infinite-dimensional notions of structure, that can be applied to a broader class of analog
signals.

\subsection{Multiple Measurement Vectors}
\label{sec:mmv}

Historically, the first class of structure that has been
considered within the CS framework has been that of multiple
measurement vectors (MMVs)~\cite{BWDSB06} which, similarly to sparse
approximation, has been an area of interest in signal processing for
more than a decade~\cite{GR97}. In this setting, rather
than trying to recover a single sparse vector $\x$, the goal is to jointly
recover a set of vectors $\{\x_i\}_{i=1}^L$ that share a common support.
Stacking these vectors into the columns of a matrix $\bX$, there will be
at most $K$ non-zero rows in $\bX$. That is, not only is each vector
$K$-sparse, but the non-zero values occur on a common location set.
We use the notation $\Omega = \supp(\bX)$ to denote the set of indices 
of the non-zero rows.

MMV problems appear quite naturally in many different applications areas.
Early work on MMV algorithms focused on magnetoencephalography,
which is a modality for imaging the brain \cite{GR97,PLM97}.
Similar ideas were also developed in the context of array
processing~\cite{G93,GR97,MCW05}, equalization of sparse
communication channels~\cite{CR02,FGF99}, and more recently cognitive
radio and multiband communications~\cite{YuHoyosSadler,Bazerque,ME09b,ME10,MEDS09}.

\subsubsection{Conditions on measurement matrices}

As in standard CS, we assume that we are given measurements
$\{y_i\}_{i=1}^L$ where each vector is of length $M<N$.
Letting $\bY$ be the $M \times L$ matrix with columns $y_i$, our problem
is to recover $\bX$ assuming a known measurement matrix $\Phi$ so that
$\bY=\Phi \bX$. Clearly, we can apply any CS method to recover $\x_i$ from
$y_i$ as before. However, since the vectors $\{\x_i\}$ all have a common
support, we expect intuitively to improve the recovery ability by exploiting
this joint information. In other words, we should in general be able to
reduce the number of measurements $M L$ needed to represent $\bX$
below $SL$, where $S$ is the number of measurements required to
recover one vector $\x_i$ for a given matrix $\Phi$.

Since $|\Omega|=K$, the rank of $\bX$ satisfies $\rank(\bX) \leq K$.
When $\rank(\bX)=1$, all the sparse vectors $\x_i$ are multiples of each
other, so that there is no advantage to their joint processing. However,
when $\rank(\bX)$ is large, we expect to be able to exploit the diversity in
its columns in order to benefit from joint recovery.
This essential result is captured by the following necessary and
sufficient uniqueness condition:
\begin{theorem}~\cite{DE10}
\label{thm:mmvunique}
A necessary and sufficient condition for the measurements $\bY=\Phi \bX$
to uniquely determine the jointly sparse matrix $\bX$ is that
\begin{equation} \label{eq:nsmmv}
|\supp(\bX)| < \frac{\spark(\Phi)-1+\rank(\bX)}{2}.
\end{equation}
\end{theorem}
The sufficiency result was initially shown for the case $\rank(\bX) = |\supp(\bX)|$~\cite{FengPhD}. 
As shown in \cite{CH06,DE10}, we can replace $\rank(\bX)$ by $\rank(\bY)$
in (\ref{eq:nsmmv}). The sufficient direction of this condition was shown
in~\cite{ME08a} to hold even in the case where there are infinitely many
vectors $\x_i$. A direct consequence of Theorem~\ref{thm:mmvunique}
is that matrices $\bX$ with larger rank can be recovered from fewer
measurements. Alternatively, matrices $\bX$ with larger support can be
recovered from the same number of measurements. When
$\rank(\bX)=K$ and $\spark(\Phi)$ takes on its largest possible value equal
to $M+1$, condition (\ref{eq:nsmmv}) becomes $M \geq K+1$. Therefore,
in this best-case scenario, only $K+1$ measurements per signal are
needed to ensure uniqueness. This is much lower than the value of $2K$
obtained in standard CS via the spark (cf. Theorem~\ref{th:mu}), which we
refer to here as the single measurement vector (SMV) setting. Furthermore,
as we now show, in the noiseless setting $\bX$ can be recovered by a simple algorithm, in
contrast to the combinatorial complexity needed to solve the SMV problem
from $2K$ measurements for general matrices $\Phi$.

\subsubsection{Recovery Algorithms}
\label{sec:imv}

A variety of algorithms have been proposed
that exploit the joint sparsity in different ways. As in the SMV setting,
two main approaches to solving MMV problems are based
on convex optimization and greedy methods.
The analogue of (\ref{eq:L0}) in the MMV case is
\begin{equation}
\bXhat = \arg \min_{\bX \in \real^{N \times L}} \|\bX\|_{0,q}~\textrm{subject to}~\bY = \Phi \bX,
\label{eq:L0mmv}
\end{equation}
where we define the $\ell_{p,q}$ norms for matrices as
\begin{equation}
\| \bX \|_{p,q} = \left( \sum_i \| x^i \|_p^q \right)^{1/q}
\end{equation}
with $x^i$ denoting the $i$th row of $\bX$. With a slight abuse of
notation, we also consider the quasi-norms with $p = 0$ such that
$\|\bX \|_{0,q} = |\supp(\bX)|$ for any $q$. Optimization based
algorithms relax the $\ell_0$ norm in (\ref{eq:L0mmv}) and attempt
to recover $\bX$ by mixed norm
minimization~\cite{TGS06,T06,FR08,CR05,CH06,EM09a}:
\begin{equation}\label{eq:mnmmv}
\bXhat = \arg \min_{\bX \in \real^{N \times L}} \|\bX\|_{p,q}~\textrm{subject to}~\bY = \Phi \bX
\end{equation}
for some $p,q \geq 1$; values of $p,q = 1,2$ and $\infty$ have been
advocated.

The standard greedy approaches in the SMV setting have also been
extended to the MMV case
\cite{BWDSB06,TGS06,GRSV08,ER10,EKB10,DE10}. The basic
idea is to replace the residual vector $r$ by a residual matrix $\bR$, which
contains the residuals with respect to each of the measurements, and
to replace the surrogate vector $\Phi^Tr$ by the $q$-norms of the
rows of $\Phi^T \bR$. For example, making these changes to OMP
(Algorithm~\ref{alg:OMP}) yields a variant known as simultaneous
orthogonal matching pursuit, shown as
Algorithm~\ref{alg:SOMP}, where $\bX|_\Omega$ denotes the
restriction of $\bX$ to the rows indexed by $\Omega$.
\begin{algorithm*}[!t]
\caption{Simultaneous Orthogonal Matching Pursuit \label{alg:SOMP}}
\begin{tabbing}
Input: CS matrix $\Phi$, MMV matrix $\bY$ \\
Output: Row-sparse matrix $\bXhat$ \\
Initialize: $\bXhat_0=0$, $\bR = \bY$, $\Omega = \emptyset$, $i = 0$. \\
{\bf while} \= halting criterion false {\bf do} \hspace{10mm}\= \\
\> $i \leftarrow i+1$ \\
\> $\b(n) \leftarrow \|\phi_n\trans \bR\|_q$, $1\le n \le N$ \> \{form residual matrix $\ell_q$-norm vector\} \\
\> $\Omega \leftarrow \Omega \cup \supp(\thresh(\b,1))$ \> \{update row support with index of residual's row with largest magnitude\} \\
\> $\bXhat_i\vert_\Omega \leftarrow \Phi_\Omega\pinv \bY$, $\bXhat_i\vert_{\Omega^C} \leftarrow 0$ \> \{update signal estimate\} \\
\> $\bR \leftarrow \bY - \Phi \bXhat_i$ \> \{update measurement residual\} \\
{\bf end while} \\
return $\bXhat \leftarrow \bXhat_i$
\end{tabbing}
\end{algorithm*}

An alternative MMV strategy is the ReMBo (reduce MMV and boost)
algorithm~\cite{ME08a}. ReMBo first reduces the problem to an SMV
that preserves the sparsity pattern, and then recovers the signal support
set; given the support, the measurements can be inverted to recover the
input. The reduction is performed by merging the measurement columns
with random coefficients.  The details of the approach together with a recovery guarantee are given in the following theorem.
\begin{theorem}
\label{thm:rembo}
Let $\overline{\bbx}$ be the unique $K$-sparse solution matrix of $\bby=\Phi \bbx$ and let $\Phi$ have spark greater than $2K$.
Let $\ba$ be a random vector with an absolutely continuous distribution and define
the random vectors $\by=\bby \ba$ and $\overline{\bx}=\overline{\bbx} \ba$.
Consider the random
SMV system $\by=\Phi \overline{\bx}$. Then:
\begin{enumerate}
\item for every realization of $\ba$, the vector $\overline{\bx}$ is the unique
$K$-sparse solution of the SMV;
\item $\Omega(\overline{\bx})=\Omega(\overline{\bbx})$ with probability one.
\end{enumerate}
\end{theorem}
According to Theorem~\ref{thm:rembo}, the MMV problem is first reduced
to an SMV counterpart, for which the optimal solution $\overline{\bx}$ is
found. We then choose the support of $\overline{\bbx}$ to be equal to that
of $\overline{\bx}$, and invert the measurement vectors $\bby$ over this
support. In practice, computationally efficient methods are used to solve
the SMV counterpart, which can lead to recovery errors in the presence of
noise, or when insufficient measurements are taken. By repeating the
procedure with different choices of $\ba$, the empirical recovery rate can be
boosted significantly~\cite{ME08a}, and lead to superior performance over
alternative MMV methods.

The MMV techniques discussed so far are {\em rank blind}, namely, they
do not explicitly take the rank of $\bX$, or that of $\bY$, into account.
Theorem~\ref{thm:mmvunique} highlights the role of the rank of
$\bX$ in the recovery guarantees. If $\rank(\bX)=K$ and
(\ref{eq:nsmmv}) is satisfied, then every $K$ columns of $\Phi$ are
linearly independent. This in turn means that
$\range(\bY)=\range(\Phi_\Omega)$, where $\range(\bY)$ denotes the
column range of the matrix $\bY$. We can therefore identify the
support of $\bX$ by determining the columns $\phi_n$ that lie in
$\range(\bY)$. One way to accomplish this is by minimizing the norm
of the projections onto the orthogonal complement of $\R(\bY)$:
\begin{equation}\label{eq:musicmmv}
\min_n \|(\bI-P_\R(\bY))\phi_n \|_2,
\end{equation}
where $P_\R(\bY)$ is the orthogonal projection onto the range of $\bY$.
The objective in (\ref{eq:musicmmv}) is equal to zero if and only if
$n \in \Omega$. Since, by assumption, the columns of $\Phi_\Omega$ are
linearly independent, once we find the support we can determine $\bX$ as
$\bX_\Omega=\Phi_\Omega^\dagger \bY$.
We can therefore formalize the following guarantee.
\begin{theorem}~\cite{DE10}
If $\rank(\bX)=K$ and (\ref{eq:nsmmv}) holds, then the algorithm
($\ref{eq:musicmmv}$) is guaranteed to recover $\bX$ from $\bY$ exactly.
\end{theorem}
In the presence of noise, we choose the $K$ values
of $n$ for which the expression ($\ref{eq:musicmmv}$) is minimized.
Since ($\ref{eq:musicmmv}$) leverages the rank to achieve recovery, we say
that this method is {\em rank aware}. More generally, any method whose
performance improves with increasing rank will be termed rank aware.
It turns out that it is surprisingly simple to modify existing greedy
methods, such as OMP and thresholding, to be rank aware: instead of
taking inner products with respect to $\bY$ or the residual $\bR$, at each
stage the inner products are computed with respect to an orthonormal
basis $\bU$ for the range of $\bY$ or $\bR$~\cite{DE10}.

The criterion in (\ref{eq:musicmmv}) is similar in spirit to the MUSIC
algorithm~\cite{S79}, popular in array signal processing, which also
exploits the signal subspace properties. As we will see below in
Section~\ref{sec:analog}, array processing algorithms can be used
to treat a variety of other structured analog sampling problems.

The MMV model can be further extended to include the case in which
there are possibly infinitely many measurement vectors
\begin{equation}
\label{eq:imv}
\by(\lambda)=\Phi \bx(\lambda), \quad \lambda \in \Lambda,
\end{equation}
where $\Lambda$ indicates an appropriate index set (which can be
countable or uncountable). Here again the assumption is that the set of
vectors $\{\bx(\lambda),\lambda \in \Lambda\}$ all have common support
$\Omega$ of size $K$. Such an infinite-set of equations is referred to as
an infinite measurement vector (IMV) system.
The common, finite, support set can be exploited in order to recover
$\bx(\lambda)$ efficiently by solving an MMV problem \cite{ME08a}
Reduction to a finite MMV counterpart is performed via the continuous to
finite (CTF) block, which aims at robust detection of $\Omega$. The CTF
builds a frame (or a basis) from the measurements using
\begin{align}
  \by(\lambda) \quad \xrightarrow{\mbox{\footnotesize Frame construct}} \quad &\bbq=\sum_{\lambda \in \Lambda} \by(\lambda)\by^H(\lambda)\nonumber \\
  \xrightarrow{\mbox{\footnotesize Decompose}}  \quad  &\bbq=\bbv\bbv^H.
\label{eq:qv}
\end{align}
Typically, $\bbq$ is constructed from roughly $2K$ snapshots
$\by(\lambda)$, and the (optional) decomposition allows removal of the
noise space \cite{ME10}. Once the basis $\bbv$ is determined, the CTF
solves an MMV system  $\bbv=\Phi\bbu$ with $\supp(\bbu)\leq K$. An 
alternative approach based on the MUSIC algorithm was suggested in~\cite{FB961,FengPhD}.

The crux of the CTF is that the indices of the nonidentically-zero rows
of the matrix $\bbu$ that solves the finite underdetermined system
$\bbv=\Phi\bbu$ coincide with the index set $\Omega$ that is associated
with the infinite signal-set $\bx(\lambda)$~\cite{ME08a}, as incoporated in
the follwoing theorem.
\begin{theorem}
\label{thm:imv}~\cite{ME10}
Suppose that the system of equations (\ref{eq:imv}) has a unique $K$-sparse
solution set with support $\Omega$, and that the matrix $\Phi$ satisfies
(\ref{eq:nsmmv}). Let $\bbv$ be a matrix whose column span is equal to the
span of $\{\by(\lambda),\lambda \in \Lambda\}$. Then, the linear system
$\bbv=\Phi \bbu$ has a unique $K$-sparse solution with row support equal
$\Omega$.
\end{theorem}
Once $\Omega$ is found, the IMV system reduces to
$\by(\lambda)=\Phi_\Omega\bx_\Omega(\lambda)$, which can be solved 
simply by computing the pseudo-inverse $\bx_\Omega(\lambda) = \Phi_\Omega^\dag \by(\lambda)$.

\subsubsection{Performance guarantees}

In terms of theoretical guarantees, it can be shown that MMV extensions
of SMV algorithms will recover $\bX$ under similar conditions to the SMV
setting in the worst-case scenario~\cite{CH06,EM09a,ER10} so that
theoretical equivalence results for arbitrary values of $\bX$ do not predict
any performance gain with joint sparsity. In practice, however, multichannel
reconstruction techniques perform much better than recovering each
channel individually. The reason for this discrepancy is that these results
apply to all possible input signals, and are therefore worst-case
measures. Clearly, if we input the same signal to each channel,
namely when $\rank(\bX)=1$, no additional information on the joint
support is provided from multiple measurements. However, as we have
seen in Theorem~\ref{thm:mmvunique}, higher ranks of the input $\bX$
improve the recovery ability. In particular, when $\rank(\bX)=K$, rank-aware
algorithms such as (\ref{eq:musicmmv}) recover the true value of $\bX$ from the
minimal number of measurements given in Theorem~\ref{thm:mmvunique}.
This property is not shared by the other MMV methods.

Another way to improve performance guarantees is by posing a probability
distribution on $\bX$ and developing conditions under which $\bX$
is recovered with high probability \cite{BWDSB06,ER10,GRSV08,SV07}.
Average case analysis can be used to show that fewer measurements
are needed in order to recover $\bX$ exactly~\cite{ER10}.
\begin{theorem}~\cite{ER10}
Let $\bX \in \real^{N\times L}$ be drawn from a probabilistic model in which
the indices for its $K$ nonzero rows are drawn uniformly at random from the
${N \choose K}$ possibilities and its nonzero rows (when concatenated)
are given by $\Sigma \Delta$, where $\Sigma$ is an arbitrary diagonal
matrix and each entry of $\Delta$ is an i.i.d.\ standard Gaussian random
variable. If $K \le \min(C_1/\mu^2(\Phi),C_2N/\|\Phi\|^2)$ then $\bX$ can be
recovered exactly from $\bY = \Phi\bX$ via (\ref{eq:mnmmv}), with $p=2$ and $q=1$, with high probability.
\end{theorem}
In a similar fashion to the SMV case (cf. Theorem~\ref{th:randomsubdic}),
while worst-case results limit the sparsity level to $K = \mathcal{O}(\sqrt{M})$,
average-case analysis shows that sparsity up to order $K = \mathcal{O}(M)$ may
enable recovery with high probability.
Moreover, under a mild condition on the sparsity and on the matrix $\Phi$,
the failure probability decays exponentially in the number of channels
$L$~\cite{ER10}.

\subsubsection{Applications}
\begin{figure}[t]\centering
\includegraphics[scale=0.6]{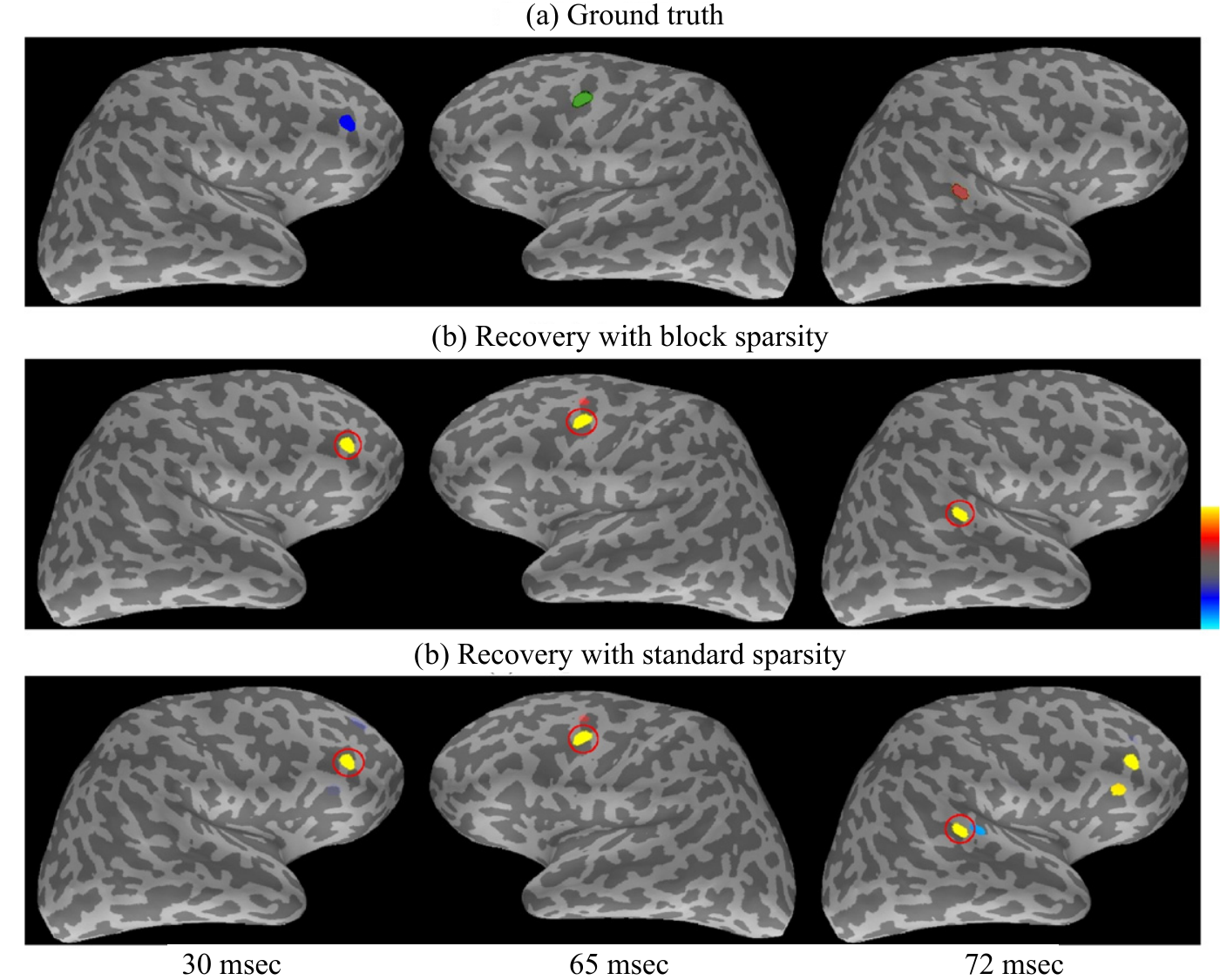}
\caption{\sl Example performance of the MMV model for EEG
inversion with temporal regularization. The setup consists of a configuration
of 306 sensors recording simulated brain activity consisting of three vertices
peaking at distinct times over a 120 ms period. The figures show (a) the
ground truth EEG activity, (b) the inversion
obtained by applying MMV recovery on all the data recorded by the sensors,
and (c) the inversion obtained by independently applying sparse recovery
on the data measured at each particular time instance.  In each case, we
show three representative time instances (30 ms, 65 ms, and 72 ms,
respectively). The results show that MMV is able to exclude spurious activity
detections successfully through implicit temporal regularization (taken
from~\cite{OuEEG}).}
\label{fig:Ou}
\vspace{-5mm}
\end{figure}

The MMV model has found several applications in the applied CS
literature. One example is in the context of electroencephalography and magnetoencephalography (EEG/MEG)~\cite{GR97,PLM97}.
As mentioned earlier, sparsity-promoting inversion algorithms have been
popular in EEG/MEG literature due to their ability to accurately
localize electromagnetic source signals. It is also possible to further improve
estimation performance by introducing temporal regularization when a
sequence of EEG/MEG is available. For example, one may apply the MMV
model on the measurements obtained over a coherent time period, effectively
enforcing temporal regularity on the brain activity~\cite{OuEEG}.
Such temporal regularization can correct estimation errors that appear as
temporal ``spikes'' in EEG/MEG activity.
The example in Fig.~\ref{fig:Ou} shows a test MEG activation signal with
three active vertices peaking at separate time instances. A 306-sensor
acquisition configuration was simulated with $\mathrm{SNR} = 3dB$.
The performance of MEG inversion with independent recovery of each
time instance exhibits spurious activity detections that are removed
by the temporal regularization enforced by the MMV model. Additionally,
the accuracy of the temporal behavior for each vertex is improved;
see~\cite{OuEEG} for details.

MMV models are also used during CS recovery for certain
infinite-dimensional signal models~\cite{E09}. We will discuss this
application in more detail in Section~\ref{sec:finiteinfinite}.

\subsection{Unions of Subspaces}

To introduce more general models of structure on the input signal,
we turn now to extend the notion of sparsity to a much broader
class of signals which can incorporate both finite-dimensional and
infinite-dimensional signal representations.
The enabling property that allows recovery of a sparse vector
$\x \in \Sigma_K$ from $M<N$ measurements is the fact that the
set $\Sigma_K$ corresponds to a union of
$K$-dimensional subspaces within $\real^N$. More specifically, if
we know the $K$ nonzero locations of $\x$, then we can represent
$\x$ by a $K$-dimensional vector of coefficients and therefore only
$K$ measurements are needed in order to recover it.
Therefore, for each {\em fixed} location set, $\x$ is restricted to
a $K$-dimensional subspace, which we denote by $\U_i$. Since the
location set is unknown, there are $N \choose K$ possible subspaces
$\U_i$ in which $\x$ may reside. Mathematically, this means that
sparse signals can be represented by a {\em union} of subspaces \cite{LD08}:
\begin{equation}
\label{eq:union} \x \in \U=\bigcup_{i=1}^m \U_i,
\end{equation}
where each subspace $\U_i$, $1\le i \le m$, is a $K$-dimensional
subspace associated with a specific set of $K$ nonzero values.

For canonically sparse signals, the union $\Sigma_K$ is composed
of canonical subspaces $\U_i$ that are aligned with $K$ out of the
$N$ coordinate axes of $\real^N$.
Allowing for more general choices of $\U_i$ leads to powerful
representations that accommodate many interesting signal priors. 
It is important to note that union models are not closed under linear operations:
The sum of two signals from a union $\U$ is generally no longer in $\U$.
This nonlinear behavior of the signal set renders sampling and recovery
more intricate. To date, there is no general methodology to treat all unions in a unified manner. Therefore, we focus our attention on some specific classes of union models, in order
of complexity.

The simplest class of unions result when the number of subspaces
comprising the union is finite, and each subspace has finite
dimensions. We call this setup a finite union of subspaces (FUS) model.
Within this class we consider below two special cases:
\begin{itemize}
\item {\em Structured sparse supports}: This class consists of
sparse vectors that meet additional restrictions on the support (i.e.,
the set of indices for the vector's nonzero entries). This corresponds
to only certain subspaces $\U_i$ out of the $N \choose K$ subspaces
present in $\Sigma_K$ being allowed~\cite{BCDH10}.\footnote{While the 
description of structured sparsity given via unions of subspaces is deterministic, 
there exists many different CS recovery approaches that leverage probabilistic 
models designed to promote structured sparsity, such as Markov random fields, 
hidden Markov Trees, etc.~\cite{csmrf,Schniter,BSB10,WaveletBCS,JDC09,HCC10,FEE10}, 
as well as deterministic approaches that rely on structured sparsity-inducing norm 
minimization~\cite{JAB09,EM09a}.}
\item {\em Sparse sums of subspaces} where each subspace $\U_i$
comprising the union is a direct sum of $K$ low-dimensional
subspaces~\cite{EM09a}
\begin{equation}
\label{eq:unionv} \U_i=\bigoplus_{|j|=K} \A_j.
\end{equation}
Here $\{\A_j,1 \leq j \leq m\}$ are a given set of subspaces with
dimensions $\dim(\A_j)=d_j$, and $|j|=K$ denotes a sum over $K$
indices. Thus, each subspace $\U_i$ corresponds to a different choice
of $K$ subspaces $\A_j$ that comprise the sum. The dimensionality
of the signal representation in this case will be $N = \sum_{j=1}^m d_j$;
for simplicity, we will often let $d_j = d$ for all $j$ so that $N = dm$.
As we show, this model leads to block sparsity in which certain blocks in
a vector are zero, and others are not~\cite{PV07,EKB10,YL06}.
This framework can model standard sparsity by letting $\A_j$, $j=1,\ldots,N$
be the one-dimensional subspace spanned by the $j^{th}$ canonical vector.
\end{itemize}
The two FUS cases above can be combined to allow only certain
sums of $K$ subspaces to be part of the union $\U$.

More complicated is the setting in which the number of possibilities is
still finite while each underlying subspace has infinite dimensions.
Finally, there may be infinitely many possibilities of finite or infinite
subspaces. These last two classes allow us to treat different families of
analog signals, and will be considered in more detail in
Section~\ref{sec:analog}.

\subsubsection{Conditions on measurement matrices}

Guarantees for signal recovery using a FUS model can be developed
by extending tools used in the standard sparse setting. For example,
the $(\U,\delta)$-RIP for FUS models~\cite{LD08,BD09a,EM09a,BCDH10}
is similar to the standard RIP where instead of the inequalities in
(\ref{eq:rip}) having to be satisfied for all sparse vectors $\x \in \Sigma_K$,
they have to hold only for vectors $\x \in \U$. If the constant $\delta$ is
small enough, then it can be shown that recovery algorithms tailored to
the FUS model will recover the true underlying vector $\x$.

An important question is how many samples are needed roughly in
order to guarantee stable recovery. This question is addressed in
the following proposition \cite{BDDW08,BD09a,EM09a,BCDH10}.
\begin{proposition}[{\cite[Theorem 3.3]{BDDW08}}]\label{Corr:Thomas}
Consider a matrix $\Phi$ of size $M\times N$ with entries independently
drawn from a subgaussian distribution, and let $\U$ be composed of $L$
subspaces of dimension $D$. Let $t>0$ and $0 < \delta < 1$ be constant
numbers. If
\begin{equation}\label{eq:nummeasrandom}
M \geq
\frac{36}{7\delta}\left(\ln(L)+D\ln\bl\frac{12}{\delta}\br+t\right),
\end{equation}
then $\Phi$ satisfies the $(\U,\delta)$-RIP with probability at least $1-e^{-t}$.
\end{proposition}
As observed in \cite{BDDW08}, the first term in
(\ref{eq:nummeasrandom}) has the dominant impact on the required
number of measurements for sparse signals in an asymptotic sense.
This term quantifies the amount of measurements that are needed to
code the exact subspace where the sparse signal resides.
We now specialize this result to the two FUS classes given earlier.
\begin{itemize}
\item In the structured sparse supports case, $L$ corresponds to the
number of distinct $K$-sparse supports allowed by the constraints,
with $L \le \binom{N}{K}$; this implies a reduction in the number of
measurements needed as compared to the traditional sparsity model.
Additionally, $D = K$, since each subspace has dimension $K$.
\item For the sparse sum of subspaces setting, we focus on the
case where each $\A_j$ is of dimension $d$; we then have
$L = \binom{N/d}{K}$. Using the approximation
\begin{equation}
(N/dK)^{K} \leq L={N/d \choose K} \leq (e\,N/dK)^{K},
\end{equation}
we conclude that for a given fraction of nonzeros $r=dK/N$, roughly $M\approx
K\log(N/dK)=-K\log(r)$ measurements are needed. For comparison, to
satisfy the standard RIP a larger number $n\approx -Kd\log(r)$ is
required. The FUS model reduces the total number of subspaces and
therefore requires $d$ times less measurements to code the signal
subspace. The subspace dimension $D = Kd$ equals the number of
degrees of freedom in $\x$.
\end{itemize}
Since the number of nonzeros is the same regardless of the sparsity
structure, the term $D=\bigo{K}$ is not reduced in either setting.

\begin{figure}[t]
\centering
\vspace*{10pt}
\includegraphics[scale=1]{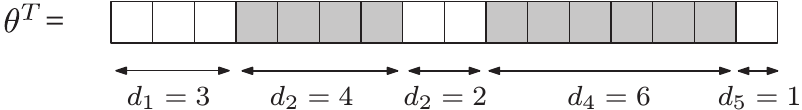}
\caption{\sl A block-sparse vector $\theta$ over $\I=\{d_1,\dots,d_5\}$.
The gray areas represent 10 non-zero entries which occupy two
blocks.}\label{FigBlockSparsity}
\vspace{-5mm}
\end{figure}

There also exists an extension of the coherence property to the sparse
sum of subspaces case~\cite{EKB10}. We must first elaborate on the
notation for this setup.
Given a signal $x \in \U$ our goal is to recover it from measurements
$y= Ax$ where $A$ is an appropriate measurement matrix of size
$M \times N$.
To recover $x$ from $y$ we exploit the fact that any $x \in \U$ can be
represented in an appropriate basis as a block-sparse vector~\cite{EM09}.
This follows by first choosing a basis $\Psi_j$ for each of the subspaces
$\A_j$. Then, any $x \in \U$ can be expressed as
\begin{equation}
\label{eq:models} x=\sum_{|j|=K} \Psi_j \theta[j],
\end{equation}
following the notation of (\ref{eq:unionv}), where $\theta[j] \in \real^{d_j}$
are the representation coefficients in $\Psi_j$. Let $\Psi$ be the column 
concatenation of $\Psi_j$, and denote by $\theta[j]$ the $j$th sub-block of a
length-$N$ vector $\theta$ over $\I=\{d_1,\ldots,d_m\}$. The $j$th
sub-block is of length $d_j$, and the blocks are formed sequentially so
that
\begin{equation}
\label{eq:xblock} \theta^T=[\underbrace{\theta_1 \,\, \ldots \,\,
\theta_{d_1}}_{\theta[1]} \,\, \ldots\,\,
\underbrace{\theta_{N-d_m+1}\,\,\ldots \,\,\theta_{N}}_{\theta[m]}]^T.
\end{equation}
If for a given $x$ the $j$th subspace $\A_j$ does not appear in the
sum (\ref{eq:unionv}), then $\theta[j]=0$. Finally, we can write
$x=\Psi\theta$, where there are at most $K$ non-zero blocks
$\theta[i]$. Consequently, our union model is equivalent to the
model in which $x$ is represented by a block-sparse vector $\theta$.
The CS measurements can then be written as
$y = Ax = A\Psi \theta = \Phi \theta$, where we denoted $\Phi = A\Psi$.
An example of a block-sparse vector with $k=2$ is depicted in
Fig.~\ref{FigBlockSparsity}.
When $d_i=1$ for all $i$, block sparsity reduces to conventional
sparsity in the dictionary $\Psi$.
We will say that a vector $\theta \in \real^N$ is block $K$-sparse if
$\theta[i]$ is nonzero for at most $K$ indices $i$.

The block-sparse model we present here has also been studied in
the statistical literature, where the objective is often quite
different. Examples include group selection
consistency~\cite{B08,NR08}, asymptotic prediction
properties~\cite{NR08,SPH09}, and block sparsity for logistic
regression~\cite{MGB08}. Block sparsity models are also
interesting in their own right (namely, not only as an equivalence
with an underlying union), and appear naturally in several problems.
Examples include DNA microarray analysis~\cite{ES05,PVMH08},
equalization of sparse communication channels \cite{CR02}, and
source localization \cite{MCW05}.

We now introduce the adaptation of coherence to sparse sums of
subspaces: the \textit{block-coherence} of a matrix $\Phi$ is defined
as
\begin{equation}
\label{eq:bc} \mub(\Phi)=\max_{\ell, r \neq \ell}
\frac{1}{d}\rho(\Phi^H[\ell]\Phi[r])
\end{equation}
with $\rho(A)$ denoting the spectral norm of the matrix $A$.
Here $\Phi$ is represented as a concatenation of
column-blocks $\Phi[\ell]$ of size $M \times d$:
\begin{equation}
\label{eq:dblock} \Phi=[\underbrace{\phi_1 \,\, \ldots \,\,
\phi_d}_{\Phi[1]} \,\,\underbrace{\phi_{d+1}\,\,\ldots \,\,
\phi_{2d}}_{\Phi[2]}\,\, \ldots\,\,
\underbrace{\phi_{N-d+1}\,\,\ldots \,\,\phi_{N}}_{\Phi[m]}].
\end{equation}
When $d=1$, as expected, $\mub(\Phi)=\mu(\Phi)$. More generally,
$\mub(\Phi) \le \mu(\Phi)$.

While $\mub(\Phi)$ quantifies global properties of the matrix $\Phi$,
local properties are characterized by the \textit{sub-coherence} of
$\Phi$, defined as
\begin{align}\label{eq:subcoherence}
 \nu(\Phi) = \max_\ell \max_{i,j\neq i}|\phi_{i}^H\phi_j|,\quad \phi_i,\phi_j\in\Phi[\ell].
\end{align}
We define $\nu(\Phi)=0$ for $d=1$. In addition, if the columns of
$\Phi[\ell]$ are orthogonal for each $\ell$, then $\nu(\Phi)=0$.

\subsubsection{Recovery algorithms}

Like in the MMV setting, it is possible to extend standard sparsity-based
signal recovery algorithms to the FUS model. For example, greedy
algorithms may be modified easily by changing the thresholding
$\thresh(x,K)$ (which finds the best approximation of $x$ in the union of
subspaces $\Sigma_K$) to a structured sparse approximation step:
\begin{equation}
\aalg_\U(x) = \arg \min_{x' \in \U} \|x-x'\|_2.
\label{eq:modelapprox}
\end{equation}
For example, the CoSaMP algorithm (see Algorithm~\ref{alg:cosamp})
is modified according to the FUS model~\cite{BCDH10} by changing the
following two steps:
\begin{itemize}
\item Prune residual: $\Omega \leftarrow \mathrm{supp}(\aalg_{\U_2}(\e))$.
\item Prune signal: $\xhat_i \leftarrow \aalg_\U(\b)$.
\end{itemize}
\noindent A similar change can be made to the IHT
algorithm~(\ref{eq:iht}) to obtain a model-based IHT variant:
\begin{equation*}
\xhat_i = \aalg_\U(\xhat_{i-1}+\Phi^T(\y-\Phi\xhat_{i-1})).
\label{eq:miht}
\end{equation*}
Structured sparse approximation algorithms of the form~(\ref{eq:modelapprox})
are feasible and computationally efficient for a variety of structured sparse
support models~\cite{BCDH10,HDC09,CIHB09,DCB09}. For example, the
approximation algorithm $\aalg_\U$ under block sparsity is equivalent
to block thresholding, with the $K$ blocks with the largest energies (or
$\ell_2$ norms) being preserved; we will show another example in
greater detail later in this section.

For the sparse sum of subspaces setting, it is possible to formulate
optimization-based algorithms for signal recovery.
A convex recovery method can be obtained by minimizing the sum of
the energy of the blocks $\theta[i]$. To write down the problem explicitly,
we define the mixed $\ell_2/\ell_1$ norm as
\begin{equation}
\|\theta\|_\mixed = \sum_{i=1}^m \|\theta[i] \|_2.
\end{equation}
We may then recover $\theta$ by solving \cite{YL06,B08,NR08,EM09a}
\begin{equation}
\label{mainalg}
\thetahat = \arg \min_{\theta \in \real^{N }} \|\theta\|_\mixed~\textrm{subject to}~y = \Phi \theta.
\end{equation}
The optimization constraints can be relaxed to address the case
of noisy measurements, in a way similar to the standard BPIC algorithm.
Generalizations of greedy algorithms to the block sparse setting have been developed in \cite{BWDSB06,EKB10}.

\subsubsection{Recovery guarantees}
Many recovery methods for the FUS model inherit the guarantees of
their standard counterparts. Our first example deals with the model-based
CoSaMP algorithm. Since CoSaMP requires RIP of order $4K$, here we
must rely on enlarged unions of subspaces.
\begin{definition}
For an integer $J$ and a FUS model $\U$, denote the $J$-sum of $\U$ as
the sum of subspaces
\begin{equation}
\label{eq:unionv2} S_J(\U)=\bigoplus_{|j|=J} \U_j,
\end{equation}
following the notation of (\ref{eq:unionv}), which contains all sums of
$J$ signals belonging in $U$.
\end{definition}
\noindent We can then pose the following guarantee.
\begin{theorem}~\cite{BCDH10}
Let $\x \in \U$ and let $\y = \Phi \x + \n$ be a set of noisy CS measurements.
If $\Phi$ has the $(S_4(\U),\delta)$-RIP with $\delta \le 0.1$,
then the signal estimate $\xhat_i$ obtained from iteration $i$ of
the model-based CoSaMP algorithm satisfies
\begin{equation}
\|\x - \xhat_i\|_2 \le 2^{-i} \|\x\|_2 + 15\|\n\|_2. \label{eq:rcosamp}
\end{equation}
\label{theo:rcosamp}
\end{theorem}
\noindent One can also show that under an additional
condition on $\Phi$ the algorithm is stable to signal mismodeling~\cite{BCDH10}.
Similar guarantees exist for the model-based IHT algorithm~\cite{BCDH10}.

Guarantees are also available for the optimization-based approach used in
the sparse sum of subspaces setting.
\begin{theorem}~\cite{EM09a}
Let $\x \in \real^N$ and let $\y = \Phi \x + \n$ be a set of noisy CS
measurements, with $\|\n\|_2 \le \epsilon$.
If $\Phi$ has the $(S_2(\U),\delta)$-RIP with $\delta \le \sqrt{2}-1$,
then the signal estimate $\xhat$ obtained from (\ref{mainalg}) with relaxed
constraints $\|\y-\Phi \x\|_2 \le \epsilon$ satisfies
\begin{equation}
\|\x - \xhat\|_2 \le C_1K^{-1/2}\|\x-\aalg_\U(\x)\|_2 + C_2\epsilon, \label{eq:mainalg}
\end{equation}
where $\aalg_\U(x)$ is defined in (\ref{eq:modelapprox}), and $\U$ is a sparse sum of subspaces.
\end{theorem}
Finally, we point out that recovery guarantees can also be obtained based
on block coherence.
\begin{theorem}~\cite{EKB10}
Both the greedy methods and the
optimization-based approach of (\ref{mainalg}) recover a block-sparse
vector $\theta$ from measurements $y=\Phi \theta$ if the block-coherence
satisfies
\begin{equation}
\label{eq:muc1} Kd<\frac{1}{2}
\left(\frac{1}{\mub(\Phi)}+d-(d-1)\frac{\nu(\Phi)}{\mub(\Phi)}\right).
\end{equation}
\end{theorem}
In the special case in which the columns of $\Phi[\ell]$ are orthonormal
for each $\ell$, we have $\nu(\Phi)=0$ and therefore the recovery condition
becomes $Kd<(\mub^{-1}(\Phi)+d)/2$. Comparing to Theorem~\ref{th:mu} for
recovery of a conventional $Kd$-sparse signal we can see that, thanks to
$\mub(\Phi)\,\le\,\mu(\Phi)$, making explicit use of block-sparsity leads to
guaranteed recovery for a potentially higher sparsity level.
These results can also be extended to both adversarial and random additive
noise in the measurements~\cite{BenHaimEldar10}.

\subsubsection{Applications}

\begin{figure*}[t]
\centering
\begin{tabular}{ccccc}
{\includegraphics[width=0.34\hsize]{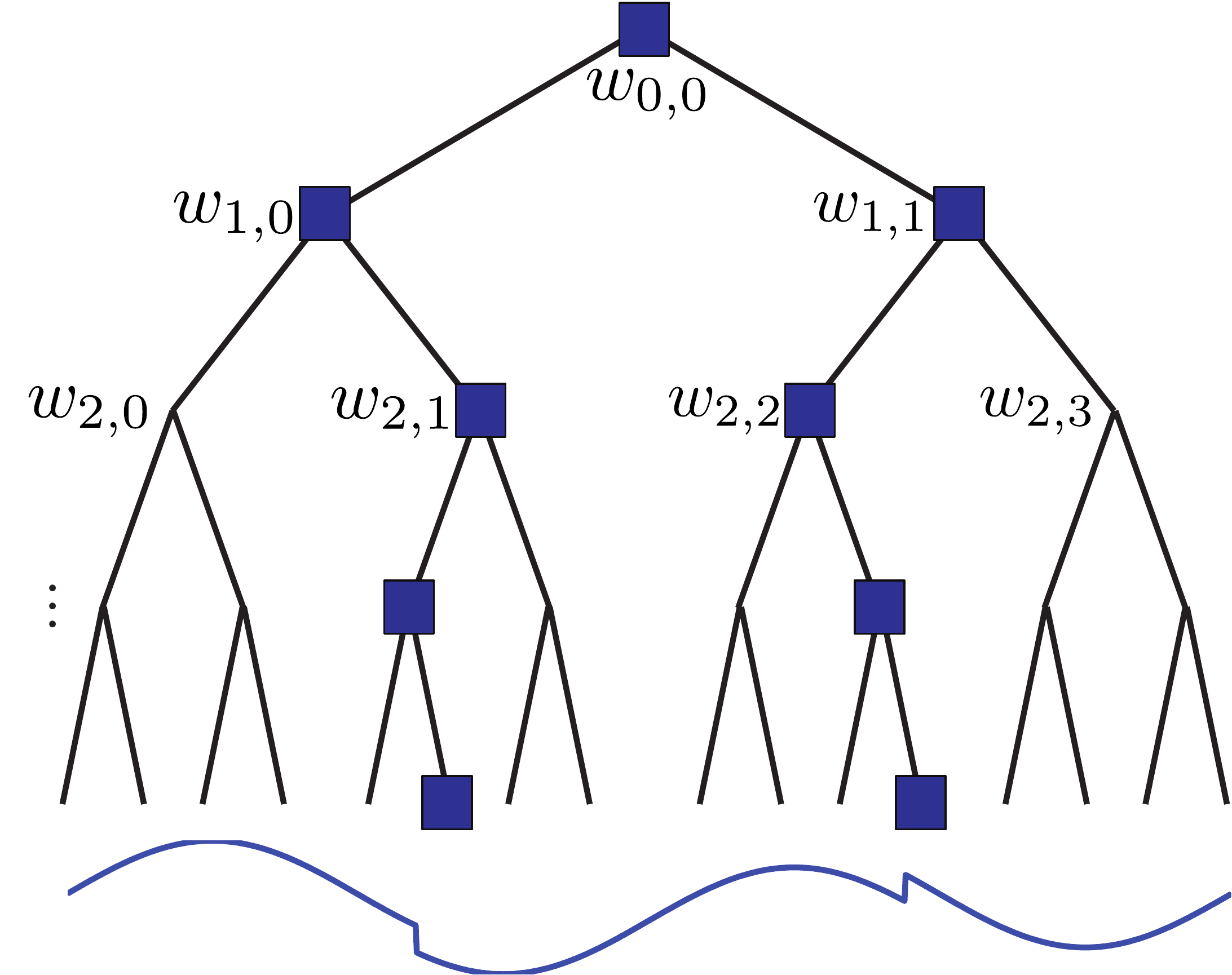}}&
{\includegraphics[width=0.3\hsize]{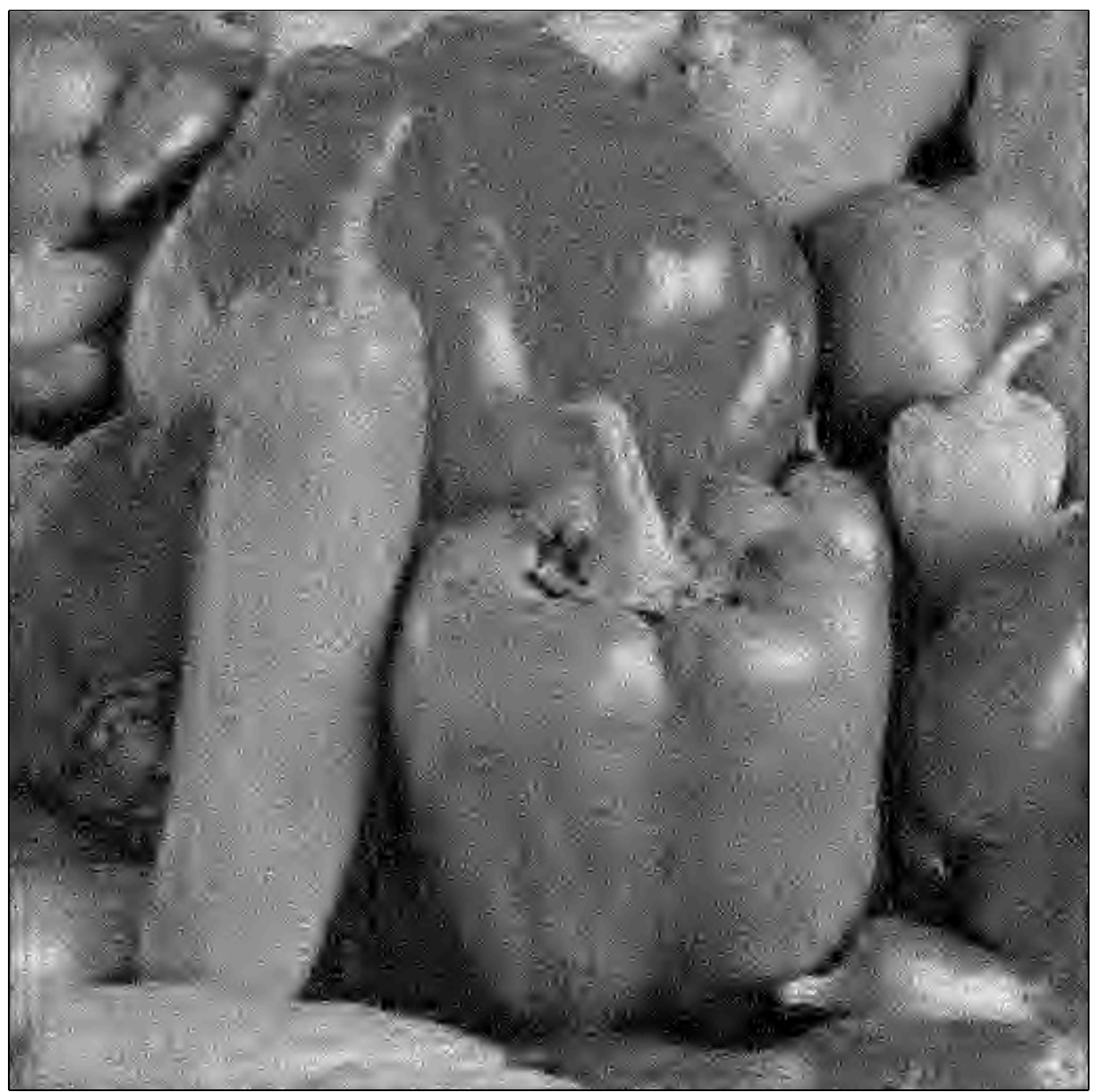}} &
{\includegraphics[width=0.3\hsize]{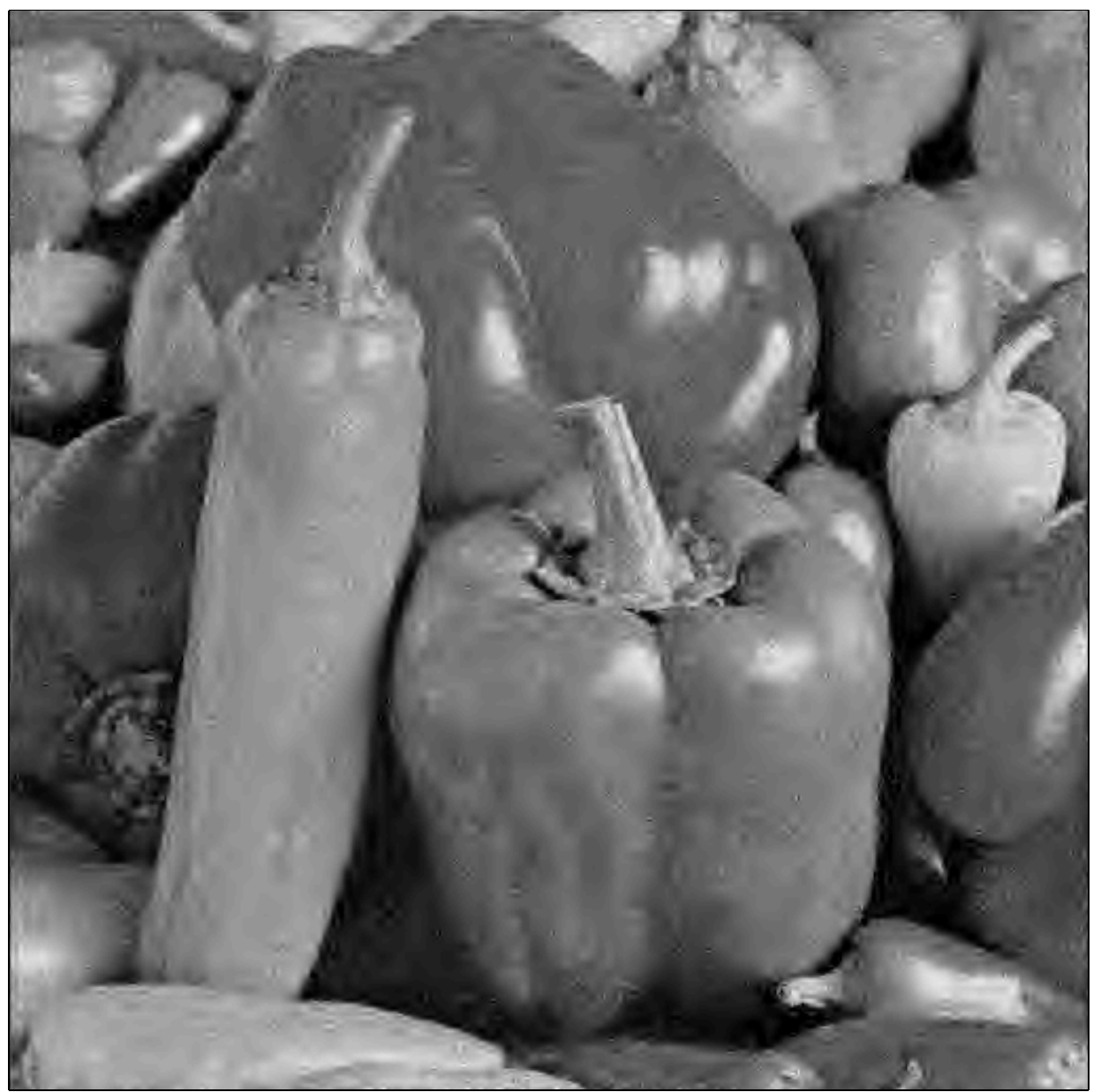}} \\
(a) & (b) & (c)
\end{tabular}
\caption{\sl 
The class of tree-sparse signals is an example of a FUS
that enforces structure in the sparse signal support. (a) Example of a
tree-structured sparse support for a 1-D piecewise smooth signal's
wavelet coefficients (taken from~\cite{BCDH10}). (b) Example recovery of the
$N=512 \times 512=262144$-pixel {\em Peppers} test image from
$M = 40000$ random measurements (15\%) using standard CoSaMP
(SNR =17.45dB). (c) Example recovery of the same image from the
same measurements using model-based CoSaMP for the tree-structured
FUS model (SNR = 22.6dB). In both cases, the Daubechies-8 wavelet basis
was used as the sparsity/compressibility transform.
\label{fig:treesparse}}
\end{figure*}

A particularly interesting example of structured sparse supports
corresponds to tree-structured supports~\cite{BCDH10}. For
signals that are smooth or piecewise smooth, including natural
images, sufficiently smooth wavelet bases provide sparse or
compressible representations $\theta$. Additionally, because each
wavelet basis element acts as a discontinuity detector in a local
region of the signal at a certain scale, it is possible to link together
wavelet basis elements corresponding to a given location at
neighboring scales, forming what are known as branches that
connect at the coarsest wavelet scale. The corresponding full graph
is known as a wavelet tree. For locations where there exist discontinuities,
the corresponding branches of wavelet coefficients tend to be large,
forming a connected subtree inside the wavelet tree. Thus, the restriction
of $\U \subseteq \Sigma_K$ includes only the subspaces
corresponding to this type of structure in the signal's support.
Figure~\ref{fig:treesparse}(a) shows an example of
a tree-sparse 1-D signal support for a wavelet coefficient vector.
Since the number of possible supports containing this structure
is limited to $L \le C^K/K$ for a constant $C$, we obtain that
$M = \bigo{K}$ random measurements are needed to recover
these signals using model-based recovery algorithms.
Additionally, there exist approximation algorithms to implement
$\aalg$ for this FUS model based on both greedy and
optimization-based approaches; see~\cite{BCDH10} for more
details.  Figure~\ref{fig:treesparse}(b) shows an example of
improved signal recovery from random measurements leveraging
the FUS model.

\begin{figure*}[t]
\begin{center}
\includegraphics[height=0.13\textheight]{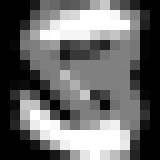} 
\includegraphics[height=0.13\textheight]{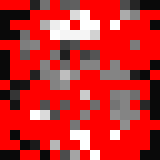} 
\includegraphics[height=0.13\textheight]{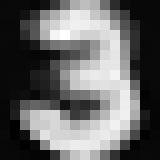} 
\includegraphics[width=0.13\textheight]{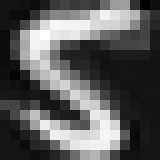} 
\includegraphics[height=0.13\textheight]{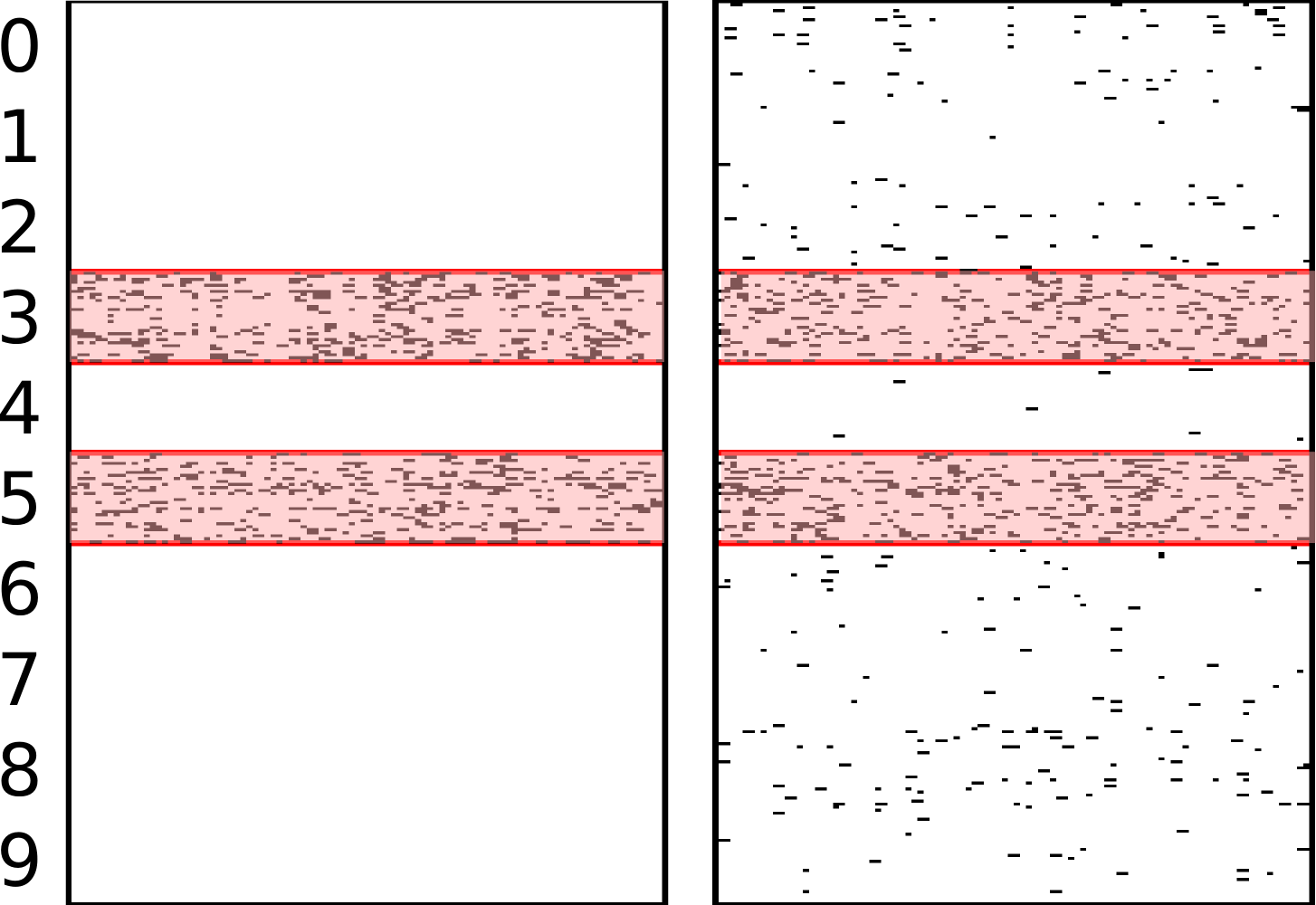} 
\end{center}
\vspace{-1ex}\caption{
\label{fig:missing}
\sl
Example of recovered digits (3 and 5) from a mixture with 60\% of missing
components. From left to right: noiseless mixture, observed mixture with
missing pixels highlighted in red, recovered digits 3 and 5, and active
set recovered for 180 different mixtures using the C-HiLasso
and~(\ref{eq:BPDN}) respectively. The last two figures show the active sets
of the recovered coefficients vectors $\theta$ as a matrix (one column per
mixture), where black dots indicate nonzero coefficients. The coefficients
corresponding to the subspace bases for digits 3 and 5 are marked as pink
bands. Notice that C-HiLasso efficiently exploits the FUS model,
succeeding in recovering the correct active groups in all the samples. The
standard approach (\ref{eq:BPDN}), which lacks this stronger signal model,
clearly is not capable of doing so, and active sets spread all over the
groups~(taken from~\cite{SRSE10}).}
\vspace{-5mm}
\end{figure*}

In our development so far, we did not consider any structure within the
subspaces comprising the unions. In certain applications it may be beneficial
to add internal structure. For example, the coefficient vectors $\theta[j]$
may themselves be sparse. Such scenarios can be accounted for by adding
an $\ell_1$ penalty in (\ref{mainalg}) on the individual blocks~\cite{FHT10},
an approach that is dubbed C-HiLasso in~\cite{SRSE10}. This allows
to combine the sparsity-inducing property of $\ell_1$ optimization at the
individual feature level, with the block-sparsity property of (\ref{mainalg})
on the group level, obtaining a hierarchically structured sparsity pattern.
An example FUS model featuring sparse sums of subspaces that exploits
the additional structure described above is shown
in Fig.~\ref{fig:missing}. This example is based on identification of digits; a
separate subspace is trained for each digit $0,\ldots,9$, and the task
addressed is separation of a mixture of digits from subsampled information.
We collect bases that span each of the 10 subspaces $\A_0,\ldots,\A_9$ into
a dictionary $\Phi$, and apply the FUS model where the subspaces $\U_i$
considered correspond to mixtures of pairs of digits. The FUS model allows
for successful source identification and separation.

\section{Structure in Infinite-Dimensional Models}
\label{sec:analog}

One of the prime goals of CS is to allow for reduced-rate sampling
of analog signals. In fact, many of the original papers in the
field state this as a motivating drive for the theory of CS.
In this section we focus on union models that include a degree of
infiniteness: This can come into play either by allowing for infinitely
many subspaces, by letting the subspaces have infinite dimension,
or both. As we will show, the resulting models may be used to describe
a broad class of structured continuous-time signals. We will then
demonstrate how such priors can be translated into concrete hardware
solutions that allow sampling and recovery of analog signals at rates
far below that dictated by Nyquist.

The approach we present here to reduced-rate sampling is based
on viewing analog signals in unions of subspaces, and is therefore
fundamentally different than previous attempts to treat similar
problems,~\cite{TroppLaskaDuarteRombergBaraniuk,MCW05,kay:icassp97,bajwa:allerton08,herman:tsp09}. 
The latter typically rely on discretization 
of the analog input, and are often not concerned with the actual hardware.
Furthermore, in some cases the reduced rate analog stage results in
high rate DSP. In contrast, in all of the examples below we treat the
analog formulation directly, the DSP
can be performed in real time at the low rate, and the analog front end
can be implemented in hardware using standard analog design
components such as modulators, low rate analog-to-digital converters
(ADCs) and low-pass filters (LPFs). A more detailed discussion on the
comparison to discretized approaches can be found in~\cite{MEE10,BGE10}.

In our review, we consider three
main cases:
\begin{itemize}
\item finite unions of infinite dimensional spaces;
\item infinite unions of finite dimensional spaces;
\item infinite unions of infinite dimensional spaces.
\end{itemize}
In each one of the three settings above there is an element that can
take on infinite values. We present general theory and results behind each of these cases, and focus in additional detail on a representative example application
for each class.

Before describing the three cases, we first briefly introduce the notion of sampling in shift-invariant (SI) subspaces, which plays a key role in the development of standard (subspace) sampling theory \cite{U00,EM09}. We then discuss how to incorporate structure into SI settings, leading to the union classes outlined above.

\subsection{Shift-invariant spaces for analog signals}
\label{sec:si}
A signal class that plays an important role in sampling theory are
signals in SI spaces~\cite{DDR94,RS95b,U00,AG01}.
Such signals are characterized by a set of generators
$\{h_{\ell}(t),1 \leq \ell \leq N\}$ where in principle $N$ can be finite or
infinite (as is the case in Gabor or wavelet expansions of $L_2$). Here
we focus on the case in which $N$ is finite. Any signal in such a SI space
can be written as
\begin{equation}
\label{eq:si} x(t)=\sum_{\ell=1}^N \sum_{n \in \ZZ}
d_\ell[n]h_\ell(t-nT),
\end{equation}
for some set of sequences $\{d_\ell[n] \in \ell_2,1 \leq \ell \leq N\}$ and period $T$.
This model encompasses many signals used in communication and
signal processing including bandlimited functions, splines~\cite{U00},
multiband signals~\cite{LV98,HW99,ME10,ME09b} and pulse amplitude
modulation signals.

The subspace of signals described by (\ref{eq:si}) has infinite
dimensions, since every signal is associated with infinitely many
coefficients $\{d_{\ell}[n],1 \leq \ell \leq N\}$. Any such signal can be
recovered from samples at a rate of $N/T$; one possible sampling
paradigm at the minimal rate is given in Fig.~\ref{fig:fbs}.
 \setlength{\unitlength}{.1in}
\begin{figure*}[t]
\begin{center}
\begin{picture}(77,31)(0,0)

\put(-6,0){

 \put(5,0){
\put(5,6.5){{\ar{\filt{$s_N(-t)$}{\ar}}}} \put(21,6.5){
   \put(0,0){\line(3,2){3}}
   \qbezier(0,2)(2,2)(3,-1)
   \put(2.95,-1.3){\vector(0,-1){0.05}}}
\put(24,3.5){\makebox(0,0){$t=nT$}} \put(25,6.5){\ar}
\put(40,6.5){\ar{\rmult{\ar{\filt{$h_N(t)$}{\ar}}}}}
\put(27,8){\makebox(0,0){$c_N[n]$}}
\put(46,1.5){\makebox(0,0){$\sum_{n \in \ZZ} \delta(t-nT)$}}
\put(45,3){\vector(0,1){2.5}} \put(42,8){\makebox(0,0){$d_N[n]$}}
}

 \put(20,15){{\Large \vdots}} \put(50,15){{\Large \vdots}}

 \put(5,20){

 \put(5,6.5){{\ar{\filt{$s_1(-t)$}{\ar}}}} \put(21,6.5){
   \put(0,0){\line(3,2){3}}
   \qbezier(0,2)(2,2)(3,-1)
   \put(2.95,-1.3){\vector(0,-1){0.05}}}
\put(24,3.5){\makebox(0,0){$t=nT$}} \put(25,6.5){\ar}
\put(40,6.5){\ar{\rmult{\ar{\filt{$h_1(t)$}{\ar}}}}}
\put(27,8){\makebox(0,0){$c_1[n]$}}
\put(46,1.5){\makebox(0,0){$\sum_{n \in \ZZ} \delta(t-nT)$}}
\put(45,3){\vector(0,1){2.5}} \put(42,8){\makebox(0,0){$d_1[n]$}}

}

\put(5,16){\st{$x(t)$}{\ar}} \put(10,6.5){\line(0,1){20}}
\put(34,4){\line(0,1){25}} \put(45,4){\line(0,1){25}}
\put(34,4){\line(1,0){11}} \put(34,29){\line(1,0){11}}
\put(40,17){\makebox(0,0){$\bbg(e^{j\omega})$}}
\put(66,16){\radder{\ar{\et{$x(t)$}}}}
\put(67,6.5){\vector(0,1){8.5}} \put(67,26.5){\vector(0,-1){9.3}}

}
\end{picture}
\end{center}
\caption{\sl Sampling and reconstruction in shift-invariant spaces.
A compressive signal acquisition scheme can be obtained by
reducing the number of sampling filters from $N$ to $p < N$ and replacing
the filter bank $\bbg(e^{j\omega})$ with a CTF block with real-time
recovery (taken from~\cite{E09}).}
\label{fig:fbs}
\vspace{-5mm}
\end{figure*}
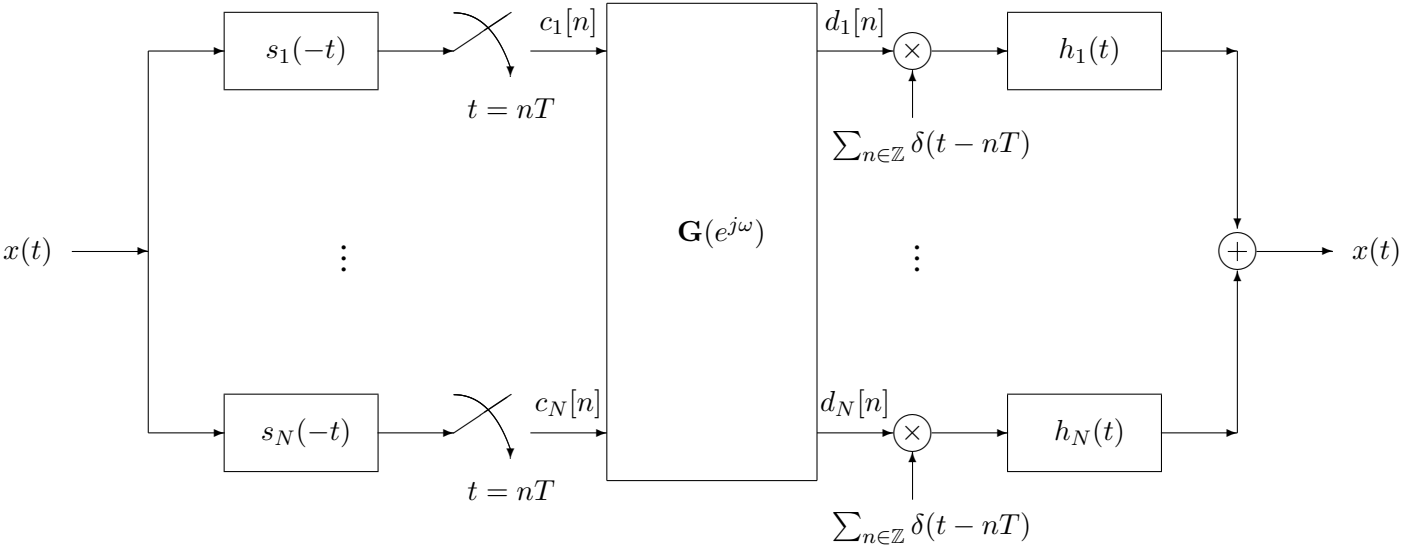
Here $x(t)$ is filtered with a bank of $N$ filters, each with impulse
response $s_\ell(t)$ which can be almost arbitrary, and the outputs are
uniformly sampled with period $T$. Denote by $\bc(\omega)$ a vector
collecting the frequency responses of $c_\ell[n]$,
$1 \le \ell \le N$. The signal is then recovered by first processing the
samples with a filter bank with frequency response $\bbg(e^{j\omega})$,
which depends on the sampling filters and the generators $h_\ell(t)$
(see~\cite{E09} for details). In this way we obtain the vectors
\begin{equation}
\bd(\omega) = \bbg(e^{j\omega})\bc(\omega)
\label{eq:sirecon}
\end{equation}
containing the frequency responses of the sequences $d_\ell[n]$,
$1 \le \ell \le N$. Each output sequence is then modulated by a
periodic impulse train $\sum_{n \in \ZZ}\delta(t-nT)$ with period $T$,
followed by filtering with the corresponding analog filter $h_\ell(t)$.

In the ensuing subsections we consider settings in which further
structure is incorporated into the generic SI model (\ref{eq:si}). In
Sections~\ref{sec:finiteinfinite} and \ref{sec:infiniteinfinite} we treat
signals of the form (\ref{eq:si}) involving a small number $K$ of
generators, chosen from a finite or infinite set, respectively, while in
Section~\ref{sec:infinitefinite} we consider a finite-dimensional
counterpart of (\ref{eq:si}) in which the generators are chosen
from an infinite set. All of these examples lead to union of subspaces
models for analog signals. Our goal is to exploit the available structure
in order to reduce the sampling rate.

Before presenting the more detailed applications we point out that
in all the examples below the philosophy is similar: we develop an
analog sensing stage that consists of simple hardware devices designed
to spread out (alias) the signal prior to sampling, in
such a way that the samples contain energy from all subspace
components. The first step in the digital reconstruction stage
identifies the underlying subspace structure. Recovery is then
performed in a subspace once the parameters defining the subspace
are determined. The difference between the examples is in how the
aliasing is obtained and in the digital recovery step which identifies
the subspace. This framework has been dubbed Xampling in~\cite{MEDS09,MEE10},
which combines CS and sampling, emphasizing that this is an analog
counterpart to discrete CS theory and methods.

\subsection{Finite union of infinite-dimensional subspaces}
\label{sec:finiteinfinite}

In this first case, we follow the model~(\ref{eq:union}-\ref{eq:unionv})
where $\U$ is composed of a finite number $m$ of subspaces, and
each subspace has infinite dimension
(i.e., $d_j = \infty$ for $1 \le j \le m$).

\subsubsection{Analog signal model}

To incorporate structure into (\ref{eq:si}) we proceed in two ways
(see Section~\ref{sec:infiniteinfinite} for the complement). In the
first, we assume that only $K$ of the $N$ generators are active,
leading to the model
\begin{equation}
\label{eq:modelfsi} x(t)=\sum_{|\ell|=K} \sum_{n \in \ZZ}
d_{\ell}[n]h_\ell(t-nT),
\end{equation}
where the notation $|\ell|=K$ means a union (or sum) over at most $K$
elements. If the $K$ active generators are known, then it suffices to
sample at a rate of $K/T$ corresponding to uniform samples with
period $T$ at the output of $K$ appropriate filters. However, a more
difficult question is whether the rate can be reduced if we know that
only $K$ of the generators are active, but do not know in advance which
ones. This is a special case of the model (\ref{eq:unionv}) where now
each of the subspaces $\A_\ell$ is a single-generator SI subspace
spanned by $h_\ell(t)$. For this model, it is possible to reduce the
sampling rate to as low as $2K/T$~\cite{E09}. Such rate reduction is
achieved by using a bank of appropriate sampling filters that replaces
the filters $s_\ell(t)$ in Fig.~\ref{fig:fbs}, followed by postprocessing via
subspace reduction and solving an MMV problem (cf.
Section~\ref{sec:mmv}). We now describe the main ideas underlying this
approach.

\subsubsection{Compressive signal acquisition scheme}
A block diagram of the basic architecture is very similar to the one given in
Fig.~\ref{fig:fbs}. We simply change the $N$ sampling filters
$s_1(-t),\ldots,s_N(-t)$ to $p < N$ sampling filters $w_1(t),\ldots,w_p(t)$ and
replace the filter bank $\bbg(e^{j\omega})$ with the CTF block with real-time
recovery (cf. Section~\ref{sec:imv}).
The design of the filters $w_\ell(t),\,1\leq \ell \leq p$ relies on two
main ingredients:
\begin{enumerate}
\item a $p \times N$ matrix $\bPhi$ chosen such that it solves a
discrete MMV problem with sparsity $K$;
\item a set of functions $\{s_\ell(t),1 \leq \ell \leq N\}$ which can be
used to sample and reconstruct the entire set of generators
$\{h_\ell(t),1 \leq \ell \leq N\}$ according to the Nyquist-rate scheme of
Fig.~\ref{fig:fbs}.
\end{enumerate}
Based on these ingredients, the compressive sampling filters $w_\ell(t)$
consist of linear combinations of $s_\ell(t)$, with coefficients that depend
on the matrix $\bPhi$ through \cite{E09}
\begin{equation}
\label{eq:wsn}
  \bw(\omega) = \bbm^*(e^{j\omega T})\bPhi^*\bbg^*(e^{j\omega T})\bh(\omega),
\end{equation}
where $(\cdot)^*$ denotes the conjugate,
$\bw(\omega),\bh(\omega)$ concatenate the frequency responses of
$w_\ell(t)$ ($1 \le \ell \le p$) and $h_\ell(t)$ ($1 \le \ell \le N$),
respectively, and $\bbm(e^{j\omega T})$ is a $p \times p$ arbitrary
invertible matrix representing the discrete-time Fourier transform (DTFT)
of a bank of filters. Since this matrix can be chosen arbitrarily, it allows
for freedom in selecting the sampling filters.

\subsubsection{Reconstruction algorithm}

Directly manipulating the expression for the sampled sequences leads
to a simple relationship between the measurements $y_\ell[n]$ and the
unknown sequences $d_\ell[n]$ \cite{E09}:
\begin{equation}\label{eq:yphid}
  \by[n]=\bPhi\bd[n],\quad \|\bd[n]\|_0\leq K,
\end{equation}
where the vector $\by[n]=[y_1[n],\cdots,y_p[n]]^T$ collects the
measurements at $t=nT$ and the vector $\bd[n]=[d_1[n],\cdots,d_N[n]]^T$
collects the unknown generator coefficients for time period $n$.
Since only $K$ out of the $N$ sequences $d_\ell[n]$ are
identically non-zero by assumption,
 the vectors $\{\bd[n]\}$ are jointly $K$-sparse.
  Equation (\ref{eq:yphid}) is valid when
$\bbm(e^{j\omega T})=\bbi$ for all $\omega$;
otherwise, the samples must first be pre-filtered with the inverse filter bank
to obtain (\ref{eq:yphid}). Therefore, by properly choosing the sampling
filters, we have reduced the recovery problem to an IMV problem, as studied
in Section~\ref{sec:imv}. To recover $\bd[n]$ we therefore rely on the CTF and
then reconstruct $x(t)$ by interpolating $d_\ell[n]$ with their generators
$h_\ell(t)$.

\subsubsection{Recovery guarantees}

From Theorem~\ref{thm:imv}, it follows that $p=2K$ filters are needed in
order to ensure recovery for all possible input signals. In practice, since
polynomial-time algorithms will be used to solve the equivalent MMV, we
will need to increase $p$ slightly beyond $2K$.
\begin{theorem}~\cite{E09}
Consider a signal of the form (\ref{eq:modelfsi}). Let $w_\ell(t),1 \leq \ell \leq p$
be a set of sampling filters defined by (\ref{eq:wsn}) where $\bPhi$ is a size
$p \times N$ matrix. Then, the signal $x(t)$ can be recovered exactly using
the scheme of Fig.~\ref{fig:fbs} as long as $\bPhi$ allows the solution
of an MMV system of size $p \times N$ with sparsity $K$.
\end{theorem}

Since our approach to recovering $x(t)$ relies on solving an MMV system,
the quality of the reconstruction depends directly on the properties of the
underlying MMV problem. Therefore, results regarding noise, mismodeling,
and suboptimal recovery using polynomial techniques developed in the
MMV context can immediately be adapted to this setting.

\subsubsection{Example application}
\label{sec:mb}

We now describe an application of the general analog CS
architecture of Fig.~\ref{fig:fbs}: sampling of multiband signals
at sub-Nyquist rates. We also expand on practical alternatives to this
system, which reduce the hardware complexity.

\begin{figure}
\centering \mbox {
\includegraphics[width=8cm]{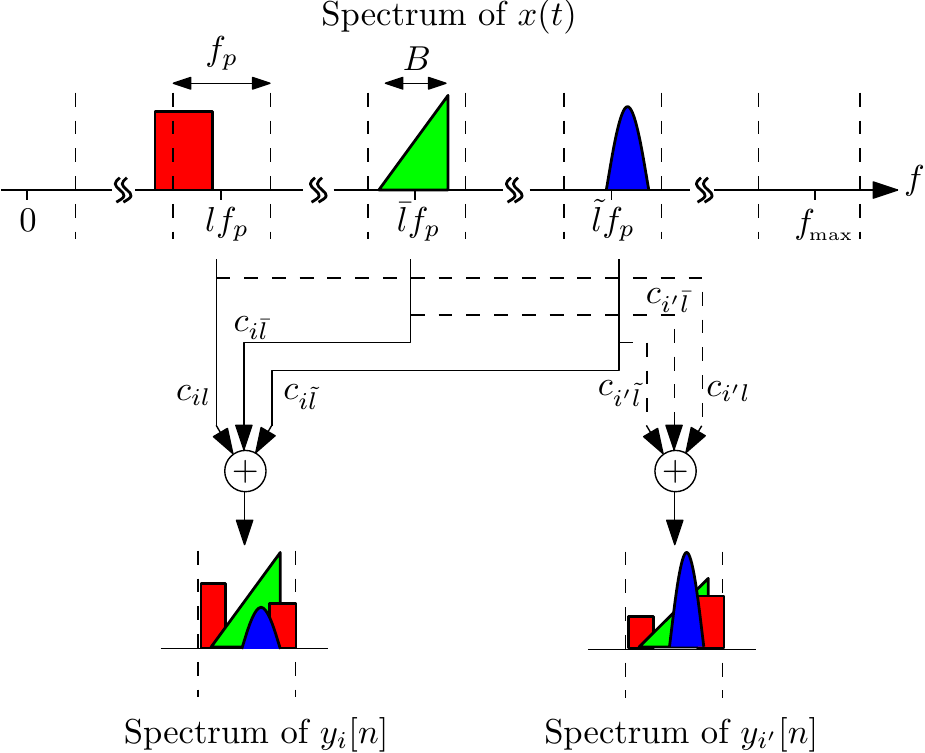}}
\caption{\sl Spectrum slices of $x(t)$ are overlayed in the spectrum of the
output sequences $y_i[n]$. In the example, channels $i$ and $i'$ realize
different linear combinations of the spectrum slices centered around
$lf_p,\bar{l}f_p,\tilde{l}f_p$. For simplicity, the aliasing of the negative
frequencies is not drawn (taken from~\cite{MEDS09}).}
\label{fig:mixtures}
\vspace{-5mm}
\end{figure}

The class of multiband signals models a scenario in which $x(t)$ consists
of several concurrent radio-frequency (RF) transmissions. A receiver that
intercepts a multiband $x(t)$ sees the typical spectral support that is
depicted in Fig.~\ref{fig:mixtures}. We assume that the signal contains at
most $N$ (symmetric) frequency bands with carriers $f_i$, each of maximal
width $B$. The carriers are limited to a maximal frequency of $f_{\max}$.
When the carrier frequencies $f_i$ are fixed, the resulting signal model
can be described as a subspace, and standard demodulation techniques
may be used to sample each of the bands at low rate. A more challenging
scenario is when the carriers $f_i$ are unknown. This situation arises, for
example, in spectrum sensing for mobile cognitive radio (CR)
receivers~\cite{M01}, which aim at utilizing unused frequency regions on
an opportunistic basis.

The Nyquist rate associated with $x(t)$ is $\fnyq = 2\fmax$, which can be
quite high in modern applications -- on the order of several GHz.
On the other hand, by exploiting the multiband structure, it can be shown
that a lower bound on the sampling rate with unknown carriers is $2NB$,
as incorporated in the following theorem. We refer
to a sampling scheme that does not exploit the carrier frequencies as blind.
\begin{theorem}
\label{thm:mb}~\cite{ME09b}
Consider a multiband model with maximal frequency $f_{\max}$ and total
support $\Omega$. Let $D(R)$ denote the sampling density of the blind
sampling set\footnote{For a formal definition of these parameters,
see~\cite{ME09b}. Intuitively, $D(R)$ describes the average sampling rate
where $R=\{r_n\}$ are the sampling points.} $R$. Then,
$D(R) \geq \min\{\Omega,2f_{\max}\}$.
\end{theorem}
In our case the support $\Omega$ is equal to $2NB$. Therefore, as long as
$NB < f_{\max}$ we can potentially reduce the sampling rate below Nyquist.

Our goal now is to use the union of subspaces framework in order to
develop a sampling scheme which achieves the lower bound of
Theorem~\ref{thm:mb}. To describe a multiband signal as a union of
subspaces, we divide the Nyquist range $[-f_{\max},f_{\max}]$ into
$M=2L+1$ consecutive, non-overlapping, slices of individual widths $f_p$
as depicted in Fig.~\ref{fig:mixtures}, such that $L/T\geq \fmax$. Each
spectrum slice represents a single bandpass subspace $\U_i$. By
choosing $f_p\geq B$, we ensure that no more than $2N$ spectrum slices
are active, i.e., contain signal energy~\cite{E09}. The conceptual division
to spectrum slices does not restrict the band positions; a single band can
split between adjacent slices.

One way to realize the sampling scheme of Fig.~\ref{fig:fbs} is
through periodic nonuniform sampling (PNS)~\cite{ME09b}. This
strategy corresponds to choosing
\begin{equation}
  w_i(t)=\delta(t-c_i\Tnyq),\quad 1 \leq i \leq p,
\end{equation}
where $\Tnyq=1/\fnyq$ is the Nyquist period, and using a sampling period of
$T=M\Tnyq$ with $M>p$.
Here  $c_i$ are integers which select part of the uniform sampling grid, resulting in $p$ uniform sequences
\begin{equation}
  y_i[n] = x((nM+c_i)\Tnyq).
\end{equation}
The IMV model (\ref{eq:yphid}) that results from PNS has sequences 
$d_\ell[n]$ representing the contents of the $\ell$th bandpass subspace 
of the relevant spectrum slice \cite{ME09b}. The sensing matrix $\Phi$ is 
a partial discrete Fourier transform (DFT), obtained by taking only the
row indices $c_i$ from the full $M \times M$ DFT matrix.

We note that PNS was utilized for multiband sampling already in classic 
studies, though the traditional goal was to approach a rate of $NB$ 
samples/sec. This rate is optimal according to the Landau 
theorem~\cite{Landau}, though achieving it for all input signals is possible 
only when the spectral support is known and fixed. When the carrier 
frequencies are unknown, the optimal rate is $2NB$ \cite{ME09b}. Indeed, 
\cite{LV98,K53} utilized knowledge of the band positions to design a PNS 
grid and the required interpolation filters for reconstruction. The approaches 
in~\cite{HW99,VB00} were semi-blind: a sampler design independent of 
band positions combined with the reconstruction algorithm of~\cite{LV98} 
which requires exact support knowledge. Other techniques targeted the 
rate $NB$ by imposing alternative constraints on the input 
spectrum~\cite{FB961,FengPhD}. Here we demonstrate how analog CS 
tools~\cite{E09,ME08a} can lead to a fully-blind sampling system of 
multiband inputs with unknown spectra at the appropriate optimal 
rate~\cite{ME09b}. A more thorough discussion in~\cite{ME09b} studies 
the differences between the analog CS method presented here based 
on~\cite{ME09b,E09,ME08a} and earlier approaches.

\subsubsection{Example hardware}
\label{sec:mwc}

As shown in~\cite{ME09b}, the PNS strategy can approach the minimal rate
of Theorem~\ref{thm:mb}. However, it requires pointwise samplers that are
capable of accommodating the wide bandwidth of the input. This necessitates
a high bandwidth track and hold device, which is difficult to build at high
frequencies. An alternative architecture referred to as the modulated wideband
converter (MWC) was developed in~\cite{ME10}. The MWC replaces the need
for a high bandwidth track and hold device by using high bandwidth
modulators, which are easier to manufacture. Indeed, off-the-shelf modulators
at rates of tens of GHz are readily available.

\begin{figure}
\centering
\includegraphics[width=7cm]{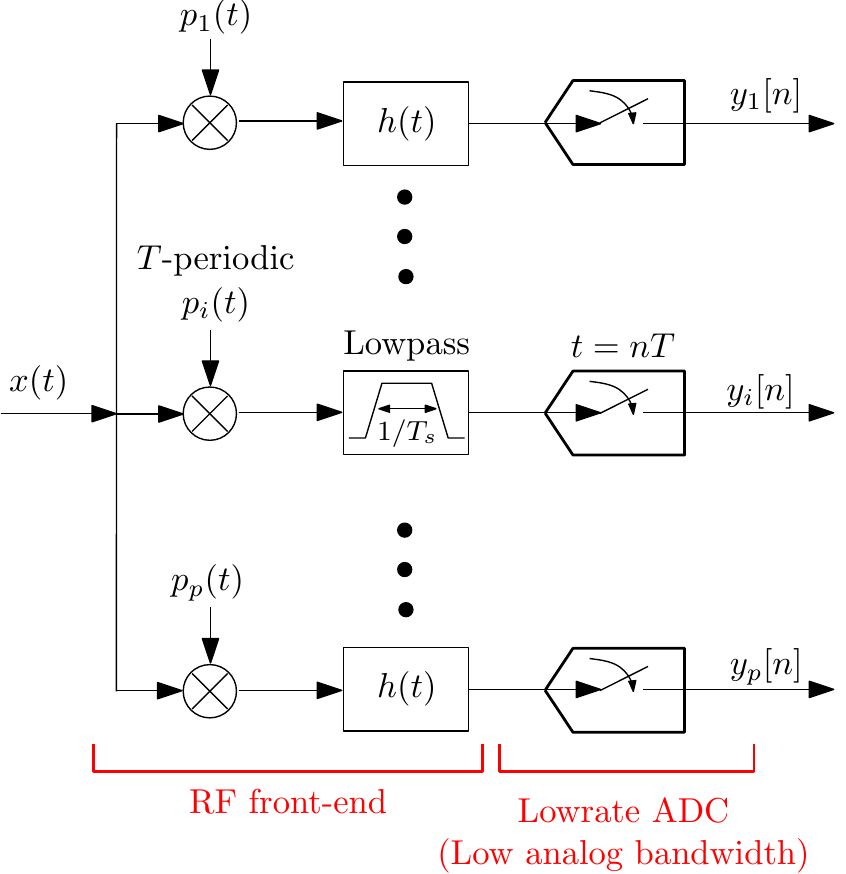}
\caption{\sl Block diagram of the modulated wideband
converter. The input passes through $p$ parallel branches, where it is
mixed with a set of periodic functions $p_i(t)$, lowpass filtered and
sampled at a low rate (taken from~\cite{MEDS09}).} \label{fig:mwc}
\vspace{-5mm}
\end{figure}

The MWC combines the spectrum slices $d_\ell[n]$ according to the scheme
depicted in Fig.~\ref{fig:mwc}. 
Its design is modular so that when the carrier frequencies are known the
same receiver can be used with fewer channels or lower sampling rate.
Furthermore, by increasing the number of channels or the rate on each
channel the same realization can be used for sampling full band signals at
the Nyquist rate.

The MWC consists of an analog front-end with $p$ channels. In the
$i$th channel, the input signal $x(t)$ is multiplied by a periodic waveform
$p_i(t)$ with period $T$, lowpass filtered by a generic filter with impulse
response $h(t)$ with cutoff $1/2T$, and then sampled at rate $f_s=1/T$.
The mixing operation scrambles the spectrum of $x(t)$, so that as before,
the spectrum is conceptually divided into slices of width $1/T$, and a
weighted-sum of these slices is shifted to the origin~\cite{ME10}. The
lowpass filter $h(t)$ transfers only the narrowband frequencies up to
$f_s/2$ from that mixture to the output sequence $y_i[n]$. The output has
the aliasing pattern illustrated in Fig.~\ref{fig:mixtures}. Sensing with the
MWC leads to a matrix $\bPhi$ whose entries are the Fourier expansion
coefficients $c_{il}$ of the periodic sequences $p_i(t)$.

The MWC can operate with as few as $p=2N$ channels and with a
sampling rate $f_s=\frac{1}{T}\geq B$ on each channel, so that it
approaches the minimal rate of $2NB$. Advanced configurations enable
additional hardware savings by collapsing the number of branches $p$ by
a factor of $q$ at the expense of increasing the sampling rate of each channel
by the same factor~\cite{ME10}. The choice of periodic functions $p_i(t)$ is
flexible: The highest Dirac frequency needs to exceed $\fmax$. In principle,
any periodic function with high-speed transitions within the period $T$ can
satisfy this requirement. One possible choice for $p_i(t)$ is a sign-alternating
function, with $M=2L+1$ sign intervals within the period
$T$~\cite{ME10,ME09EXRIP}. Imperfect sign alternations are allowed as long
as periodicity is maintained~\cite{MEDS09}. This property is crucial since
precise sign alternations at high speeds are extremely difficult to maintain,
whereas simple hardware wirings ensure that $p_i(t)=p_i(t+T_p)$ for every
$t\in\RR$. The waveforms $p_i(t)$ need low mutual correlation in order to
capture different mixtures of the spectrum. Popular binary patterns, \eg the
Gold or Kasami sequences, are especially suitable for the
MWC~\cite{ME09EXRIP}. Another important practical design aspect is that
the lowpass filter $h(t)$ does not have to be ideal. A nonflat frequency
response can be compensated for in the digital domain, using the algorithm
developed in~\cite{CMEH10}.

The MWC has been implemented as a board-level hardware
prototype~\cite{MEDS09}.\footnote{A video of experiments and additional
documentation for the MWC hardware are available at
\phantom{a}\url{http://webee.technion.ac.il/Sites/People/YoninaEldar/hardware.html}.
A graphical package demonstrating the MWC numerically is available at
\url{http://webee.technion.ac.il/Sites/People/YoninaEldar/software_det3.html}.}
The hardware specifications cover inputs with a
2 GHz Nyquist rate with spectrum occupation $NB=120$ MHz. The total
sampling rate is $280$ MHz, far below the 2 GHz Nyquist rate.
In order to save analog components, the hardware realization incorporates
the advanced configuration of the MWC~\cite{ME10} with a collapsing
factor $q=3$. In addition, a single shift-register provides a basic periodic
pattern, from which $p$ periodic waveforms are derived using delays, that
is, by tapping $p$ different locations of the register.
A nice feature of the recovery stage is that it interfaces seamlessly with
standard DSPs by providing (samples of) the narrowband
information signals. This capability is
provided by a digital algorithm that is developed in~\cite{MEE10}.\footnote{The algorithm is  available
online at~\url{http://webee.technion.ac.il/Sites/People/YoninaEldar/software_det4.html}.}

The MWC board is a first hardware example of the use of ideas borrowed
from CS for sub-Nyquist sampling and low-rate recovery
of wideband signals where the sampling rate is directly proportional to the
actual bandwidth occupation and not the highest frequency. Existing
implementations of the random demodulator (RD) (cf. Section~\ref{sec:rd})
recover signals at effective sampling rates below 1 MHz, falling outside of
the class of wideband samplers. Additionally, the signal representations
used by the RD have size proportional to the Nyquist frequency, leading to
recovery problems that are much larger than those posed by the MWC.
See~\cite{MEDS09,MEE10} for additional information on the similarities and
differences between the MWC, the RD, and other comparable architectures.

\subsection{Infinite union of finite-dimensional subspaces}
\label{sec:infinitefinite}

The second union class we consider is when $\U$ is composed of an infinite number
$m$ of subspaces, and each subspace has finite dimension.

\subsubsection{Analog signal model}
\label{sec:fri}
As we have seen in Section~\ref{sec:si}, the SI model (\ref{eq:si}) is a
convenient way to describe analog signals in infinite-dimensional spaces.
We can use a similar approach to describe analog signals that lie within
finite-dimensional spaces by restricting the number of unknown gains
$a_\ell[n]$ to be finite. In order to incorporate structure into this model, we
assume that each generator $h_{\ell}(t)$ has an unknown parameter
$\theta_\ell$ associated with it, which can take on values in a continuous
interval, resulting in the model
\begin{equation}
\label{eq:pfri}
x(t)=\sum_{\ell=1}^L a_\ell h_\ell(t,\theta_\ell).
\end{equation}
 Each possible choice of the set $\{\theta_\ell\}$
leads to a different $L$-dimensional subspace of signals $\U_i$,
spanned by the functions $\{h(t,\theta_\ell)\}$. Since
$\theta_\ell$ can take on any value in a given interval, the model
(\ref{eq:pfri}) corresponds to an infinite union
of finite dimensional subspaces (i.e., $m = \infty$).

An important example of (\ref{eq:pfri}) is when $h_{\ell}(t,\theta_\ell)=h(t-t_\ell)$
for some unknown time delay $\theta_l = t_\ell$, leading to a stream of pulses
\begin{equation}\label{SingleCh:eq_sig_model_periodict}
  x(t) = \sum_{\ell=1}^L a_\ell h(t - t_\ell).
\end{equation}
Here $h(t)$ is a known pulse shape
and $\{t_\ell,a_\ell\}_{\ell=1}^L,\, t_\ell \in [0,\tau)$, $a_\ell \in
\mathbb{C}$ are unknown delays and amplitudes. This model was first 
introduced and studied by Vetterli et al.~\cite{VMB02,DVB07,BDVMC08} as a
special case of signals having a finite number of degrees of freedom
per unit time, termed finite rate of innovation (FRI) signals.

Our goal is to sample $x(t)$ and reconstruct it from a minimal
number of samples. The primary interest is in pulses which have small
time-support, and therefore the required Nyquist rate would be very high.
However, since the pulse shape is known, the signal has only $2L$
degrees of freedom, and therefore, we expect the minimal number of
samples to be $2L$, much lower than the number of samples resulting
from Nyquist rate sampling.

A simpler version of the problem is when the signal $x(t)$ of
(\ref{SingleCh:eq_sig_model_periodict})  is repeated periodically leading
to the model
\begin{equation}\label{SingleCh:eq_sig_model_periodic}
  x(t) = \sum_{m\in \mathbb{Z}} \sum_{\ell=1}^L a_\ell h(t - t_\ell - m\tau),
\end{equation}
where $\tau$ is the known period. This periodic setup is easier to treat
because we can exploit the properties of the Fourier series representation
of $x(t)$ due to the periodicity. The dimensionality and number of
subspaces included in the model (\ref{eq:union}) remain unchanged.

\subsubsection{Compressive signal acquisition}
To date, there are no general acquisition methods for signals of the
form (\ref{eq:pfri}). Instead, we focus on the special case of
(\ref{SingleCh:eq_sig_model_periodic}).

Our sampling scheme follows the ideas of~\cite{VMB02,DVB07,BDVMC08} 
and consists of a filter $s(t)$ followed by uniform sampling of the output with 
period $T=\tau/N$ where $N$ is the number of samples in one period, and 
$N \geq 2L$. The resulting samples can be written as inner products 
$c[n] = \inner{s(t-nT)}{x(t)}$. The following theorem establishes
properties of the filter $s(t)$ that allow recovery from $2L$ samples.
\begin{theorem}
\label{thm:fri}~\cite{RonenSingleChannel}
Consider the $\tau$-periodic stream of pulses of order $L$ given
by~(\ref{SingleCh:eq_sig_model_periodic}). Choose a set $\mathcal{K}$
of consecutive indices for which $H(2\pi k/\tau) \neq 0$ where $H(\omega)$
is the Fourier transform of the pulse $h(t)$. Then the samples
$c[n]$ for $n = 0,\ldots,N-1$, uniquely determine the
signal $x(t)$ as long as $N \geq |\mathcal{K}| \geq 2L$ for any $s(t)$
satisfying
\begin{equation}\label{eq_S_omega}
S(\omega) = \left\{
\begin{array}{ll}
0 & \omega = 2\pi k/\tau, k \notin \mathcal{K} \\
\mbox{nonzero} & \omega = 2\pi k/\tau, k \in \mathcal{K} \\
\textrm{arbitrary} & \quad \mbox{otherwise}.\\
\end{array} \right.
\end{equation}
\end{theorem}
\noindent Theorem~\ref{thm:fri} was initially focused on low pass 
filters in~\cite{VMB02}, and was later extended to arbitrary filters 
in~\cite{RonenSingleChannel}.

To see how we can recover $x(t)$ from the samples of  Theorem~\ref{thm:fri}, we note that the Fourier series coefficients $X[k]$ of the periodic pulse
stream $x(t)$ are given by \cite{VMB02}:
\begin{align}
X[k]&=\frac{1}{\tau}H(2\pi k/\tau)\sum_{\ell=1}^L a_\ell e^{-j2\pi kt_\ell/\tau} \nonumber \\
&= \frac{1}{\tau}H(2\pi k/\tau)\sum_{\ell=1}^L a_\ell u_\ell^k,
\label{eq:exp}
\end{align}
where $u_\ell = e^{-j2\pi t_\ell/\tau}$. Given $X[k]$, the problem of retrieving
$a_\ell$ and $t_\ell$ in (\ref{eq:exp})
is a standard problem in array processing \cite{Stoica1997,VMB02}, and can
be solved using methods developed in that context such as the matrix
pencil~\cite{hua-sarkar90-1}, subspace-based
estimators~\cite{kung-etal83,kao-arun93-1}, and the annihilating
filter~\cite{BDVMC08}. These methods require $2L$ Fourier coefficients to
determine $a_\ell$ and $u_\ell$. In the next subsection, we show that the vector
$\mathbf{X}$ of Fourier coefficients $X[k]$ can be computed from the $N$ samples $c[n]$
of Theorem~\ref{thm:fri}.

Since the LPF has infinite time support, the approach of (\ref{eq_S_omega}) cannot work with
time-limited signals, such as those of the form
(\ref{SingleCh:eq_sig_model_periodict}). A class of filters satisfying
(\ref{eq_S_omega}) that have finite time support are Sum of Sincs
(SoS) \cite{RonenSingleChannel}, which are given in the Fourier domain by
\begin{equation}\label{SingleCh:eq_G_omega}
  G(\omega) = \frac{\tau}{\sqrt{2\pi}} \sum_{k \in \mathcal{K}} b_k \sinc\left(\frac{\omega}{2\pi/\tau} - k\right),
\end{equation}
where $b_k \neq 0,\, k \in \mathcal{K}$. Switching to the time domain
\begin{equation}\label{SingleCh:eq_g_t}
  g(t) = \mbox{rect}\left(\frac{t}{\tau}\right) \sum_{k \in \mathcal{K}} b_k e^{j2\pi kt/\tau},
\end{equation}
which is clearly a time compact filter with support $\tau$.
For the special case in which $\mathcal{K} =
\{-p,\ldots,p\}$ and $b_k=1$,
\begin{equation}\label{SingleCh:eq_Dirichlet_filter}
  g(t) = \mbox{rect}\left(\frac{t}{\tau}\right) \sum_{k=-p}^p e^{j2\pi kt/\tau} = \mbox{rect}\left(\frac{t}{\tau}\right) D_p(2\pi
  t/\tau), \nonumber
\end{equation}
where $D_p(t)$ denotes the Dirichlet kernel.

Alternative approaches to sample finite pulse streams of
the form \eqref{SingleCh:eq_sig_model_periodic} rely on the use of
splines~\cite{DVB07}; this enables obtaining moments of the signal rather
than its Fourier coefficients. The moments are then processed in a similar
fashion (see the next subsection for details).
However, this approach is unstable for high values of $L$~\cite{DVB07}.
In contrast, the SoS class can be used for stable reconstruction
even for very high values of $L$, e.g., $L=100$.

\begin{figure}[t]
\centering
\includegraphics[scale=0.65]{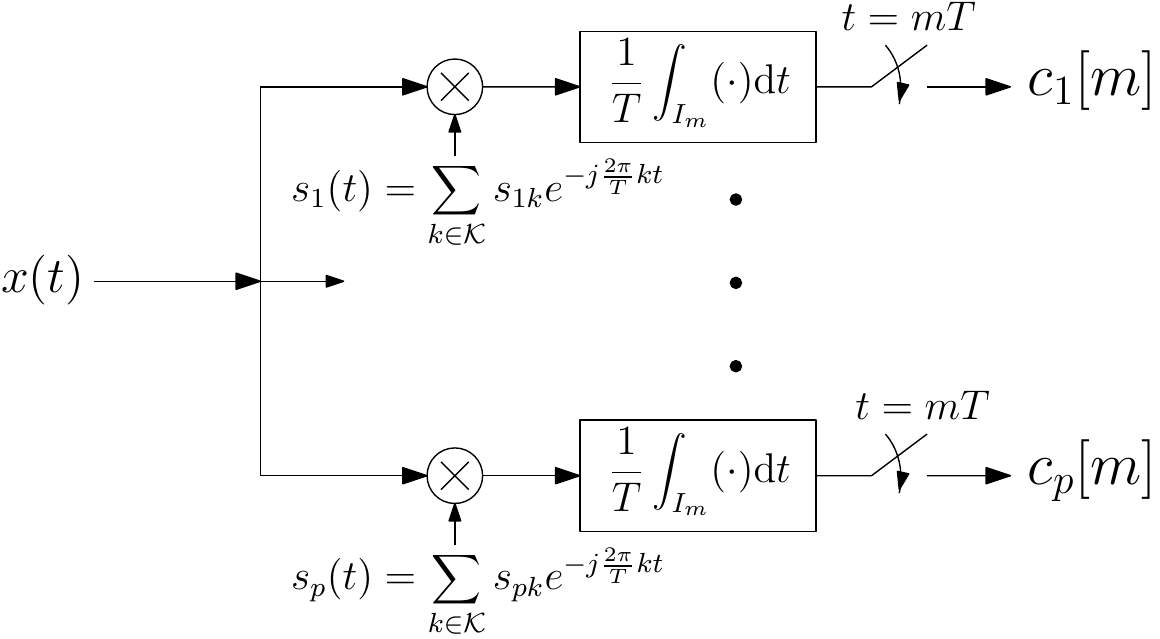}
\caption{\sl Extended sampling scheme using modulating waveforms (taken from~\cite{ronen_kfir}).}
\label{fig_infinite_mixing_waveforms_sampling}
\vspace{-5mm}
\end{figure}

Multichannel schemes can also be used to sample pulse streams. This
approach was first considered for Dirac streams, where a successive
chain of integrators allows obtaining moments of the
signal~\cite{kusuma2006multichannel}. Unfortunately, the method is
highly sensitive to noise. A simple sampling and reconstruction scheme
consisting of two channels, each with an RC circuit, was presented
in~\cite{UnserFRI2008} for the special case where there is no more than
one Dirac per sampling period. A more general multichannel architecture
that can treat a broader class of pulses, while being much more stable, is
depicted in Fig.~\ref{fig_infinite_mixing_waveforms_sampling}~\cite{ronen_kfir}.
The system is very similar to the MWC presented in the previous section.
By correct choice of the mixing coefficients, the Fourier
coefficients $X[k]$ may be extracted from the samples by a simple matrix
inversion.

\subsubsection{Recovery algorithms}
\label{sec:iirec}

In both the single-channel and multichannel approaches, recovery of the
unknown delays and amplitudes proceeds in two steps. First, the vector
of samples $\mathbf{c}$ is related to the Fourier coefficient vector
$\mathbf{x}$ through a $p \times |\mathcal{K}|$ mixing matrix
$\mathbf{Q}$, as $\mathbf{c}=\mathbf{Q} \mathbf{x}$.
Here $p \geq 2L$ represents the number of samples. When using
the SoS approach with a filter $S(\omega)$, $\mathbf{Q}=\bbv\bbs$
where $\mathbf{S} $ is a $p \times p$ diagonal matrix with diagonal
elements $S^*(-2\pi \ell/\tau)$, $1 \le \ell \le p$, and $\bbv$ is a
$p \times |\mathcal{K}|$ Vandermonde matrix with $\ell^{th}$ element
given by $e^{j2 \pi \ell T/\tau}$, $1 \le \ell \le p$, where $T$ denotes
the sampling period. For the multichannel architecture of
Fig.~\ref{fig_infinite_mixing_waveforms_sampling},
$\mathbf{Q}$ consists of the modulation coefficients $s_{\ell k}$. The
Fourier coefficient vector $\mathbf{x}$ can be obtained from the
samples as $\mathbf{x}=\mathbf{Q}^{\dagger} \mathbf{c}$.

The unknown parameters $\{t_l,a_l\}_{l=1}^L$ are recovered from
$\bx$ using, e.g., the annihilating filter method.
The annihilating filter $\{r[k]\}_{k=0}^L$ is defined by
its $z$-transform
\begin{equation}
\label{eq:Rz}
R(z) = \sum_{k=0}^L r[k] z^{-k} = \prod_{\ell=1}^{L}(1-u_\ell z^{-1}).
\end{equation}
That is, the roots of $R(z)$ equal the values $u_\ell$ through which
the delays $t_\ell$ can be found. It then follows that
\begin{align}
r[k] * X[k] &= \sum_{l=0}^{L} r[l] X[k-l] = \sum_{l=0}^L\sum_{\ell=1}^L r[l]a_\ell u_\ell^{k-l} \nonumber \\
&= \sum_{\ell=1}^L a_\ell u_\ell^k \sum_{l=0}^L r[l] u_\ell^{-l} = 0,
\label{eq:ann}
\end{align}
where the last equality is due to $R(u_\ell) = 0$.
Assuming without loss of generality that $R[0] = 1$, the identity in
(\ref{eq:ann}) can be written in matrix/vector form as
{\small
\begin{align}
\left(\begin{array}{ccc}
X[-1] &  \ldots & X[-L] \\
X[0] & \ldots & X[-L+1] \\
\vdots & \ddots & \vdots \\
X[L-2] & \ldots & X[-1]
\end{array}\right)
\left(\begin{array}{c}
r[1] \\
r[2] \\
\vdots\\
r[L]
\end{array}\right)=
\left(\begin{array}{c}
X[0] \\
X[1] \\
\vdots\\
X[L-1]
\end{array}\right). \nonumber
\end{align}}
Thus, we only need $2L$ consecutive values of $X[k]$
to determine the annihilating filter. Once the filter is found, the values $t_\ell$ are retrieved from the zeros $u_\ell$ of the $z$-transform in (\ref{eq:Rz}).
Finally, the Fourier coefficients $a_\ell$ are computed using~(\ref{eq:exp}).
For example, if we have the coefficients $X[k]$, $0 \le k
\le 2L-1$, then~(\ref{eq:exp}) may be written as
{\small \begin{align}
\frac{1}{\tau}
\left(\begin{array}{cccc}
1 & 1 & \ldots & 1 \\
u_0 & u_1 & \ldots & u_L \\
\vdots & \vdots & \ddots & \vdots \\
u_0^{2L} & u_1^{2L} & \ldots & u_L^{2L}
\end{array}\right)
\left(\begin{array}{c}
a_1 \\
a_2 \\
\vdots\\
a_L
\end{array}\right)=
\left(\begin{array}{c}
\frac{X[0]}{H(0)} \\
\frac{X[1]}{H(2\pi/\tau)} \\
\vdots\\
\frac{X[2L-1]}{H(2\pi(2L-1)/\tau)}
\end{array}\right).
\nonumber
\end{align}}
\noindent Since this Vandermonde matrix is left-invertible,
the values $a_\ell$ can be computed by matrix inversion.

\begin{figure*}
\centering \mbox {
\subfigure[]{\includegraphics[scale=0.65]{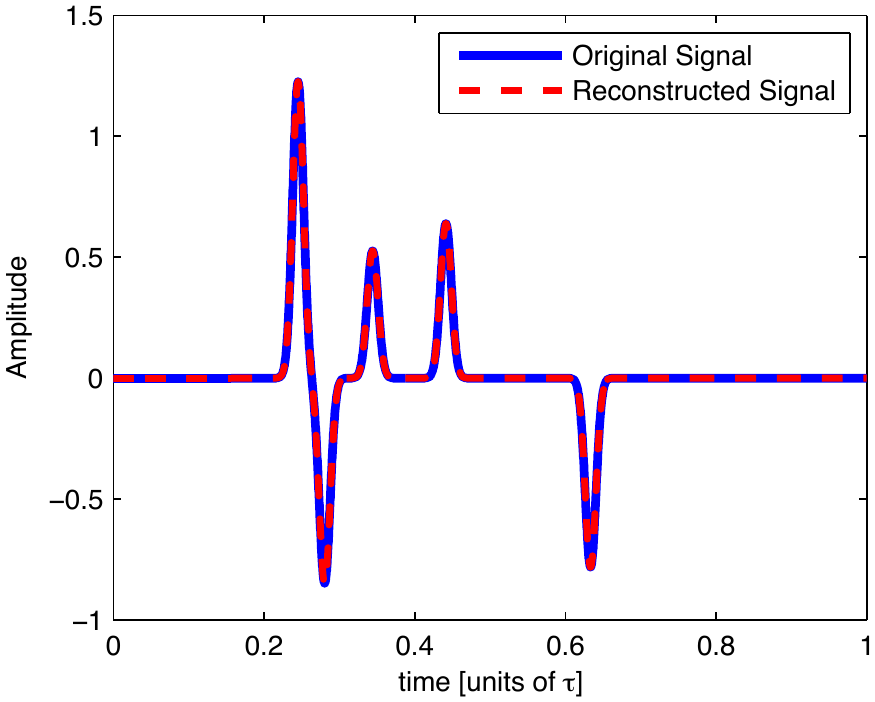}}}
\subfigure[$L=3$]{\includegraphics[scale=0.7]{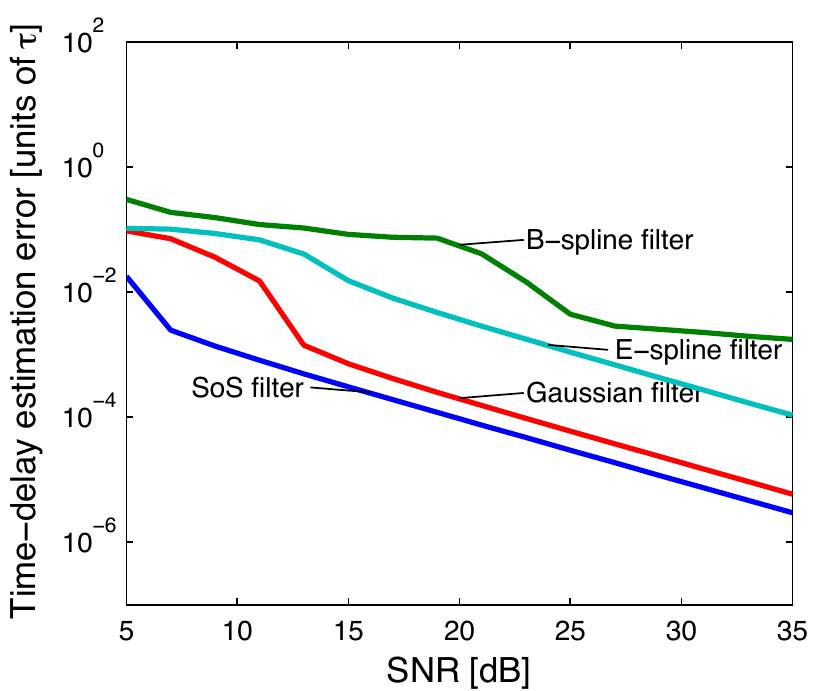}}
\subfigure[$L=5$]{\includegraphics[scale=0.7]{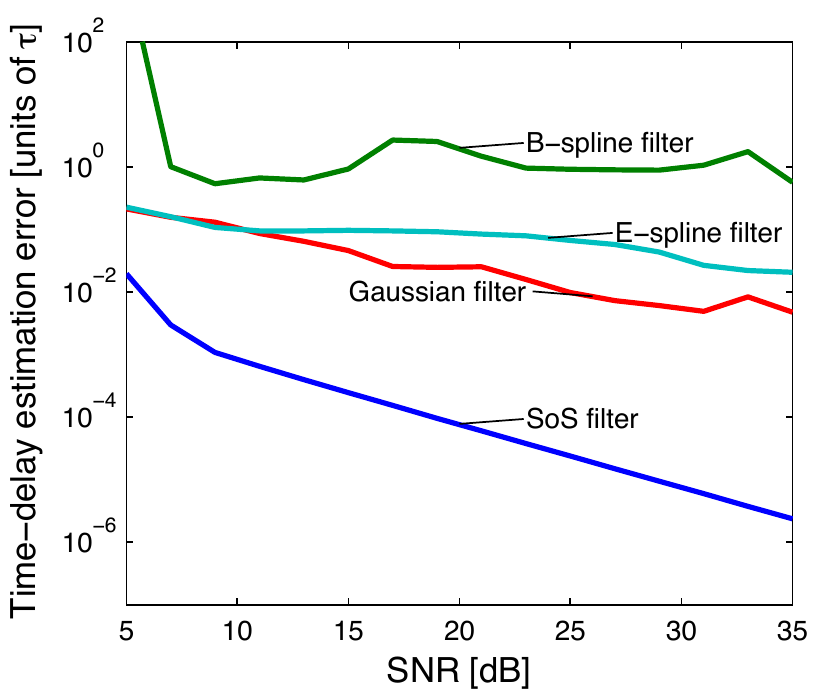}}
\caption{\sl Performance comparison of finite pulse stream recovery using
Gaussian \cite{VMB02}, B-spline, E-spline \cite{DVB07}, and SoS sampling kernels. 
(a) Reconstructed
signal using SoS filters vs. original one. The reconstruction is exact to
numerical precision. (b) $L=3$ dirac pulses are present, (c) $L=5$ pulses
(taken from~\cite{RonenSingleChannel}).}
\label{SingleCh:fig_periodic_demonstration_reconstruction}
\vspace{-5mm}
\end{figure*}

Reconstruction results for the sampling scheme based on the SoS
filter with $b_k=1$ are depicted in
Fig.~\ref{SingleCh:fig_periodic_demonstration_reconstruction}. The
original signal consists of $L=5$ Gaussian pulses, and $N=11$ samples
were used for reconstruction. The reconstruction is exact to numerical
precision. A comparison of the performance of various methods in the
presence of noise is depicted in
Fig.~\ref{SingleCh:fig_periodic_demonstration_reconstruction} for a
finite stream consisting of 3 and 5 pulses. The pulse-shape is a Dirac
delta, and white gaussian noise is added to the samples with a proper
level in order to reach the desired SNR for all methods. All approaches
operate using $2L+1$ samples.

\subsubsection{Applications}

As an example application of FRI, we consider multiple image
registration for superresolution imaging. A superresolution
algorithm aims at creating a single detailed image, called a
super-resolved image (SR), from a set of low-resolution input images of
the same scene. If different images from the same scene are
taken such that their relative shifts are not integer multiples of the pixel
size, then sub-pixel information exists among the set. This allows to obtain
a higher resolution accuracy of the scene once the images are
properly registered.

\begin{figure*}[t]
\centering \mbox { \subfigure[]{\includegraphics[scale=0.39]{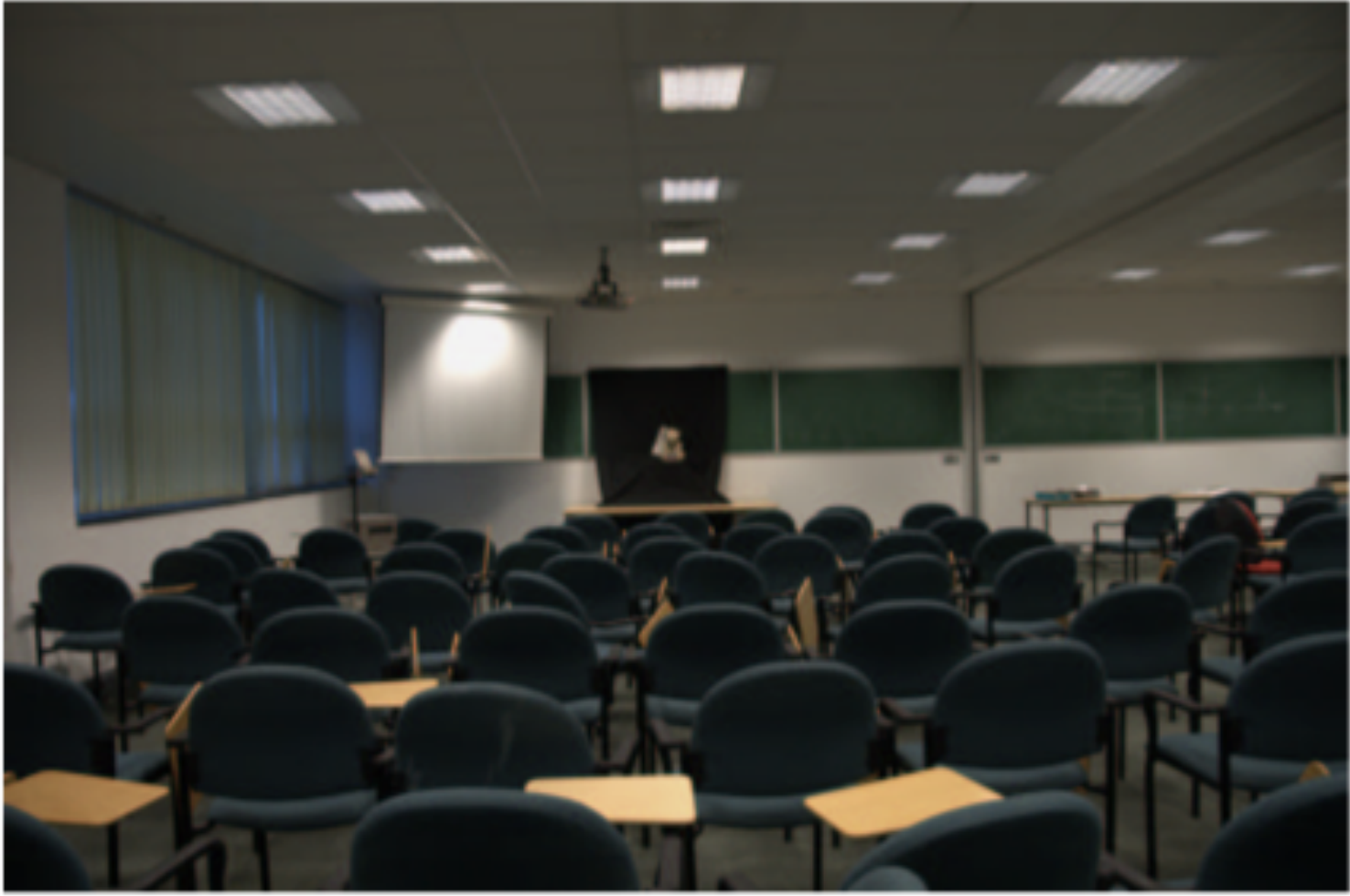}}
\subfigure[]{\includegraphics[scale=0.39]{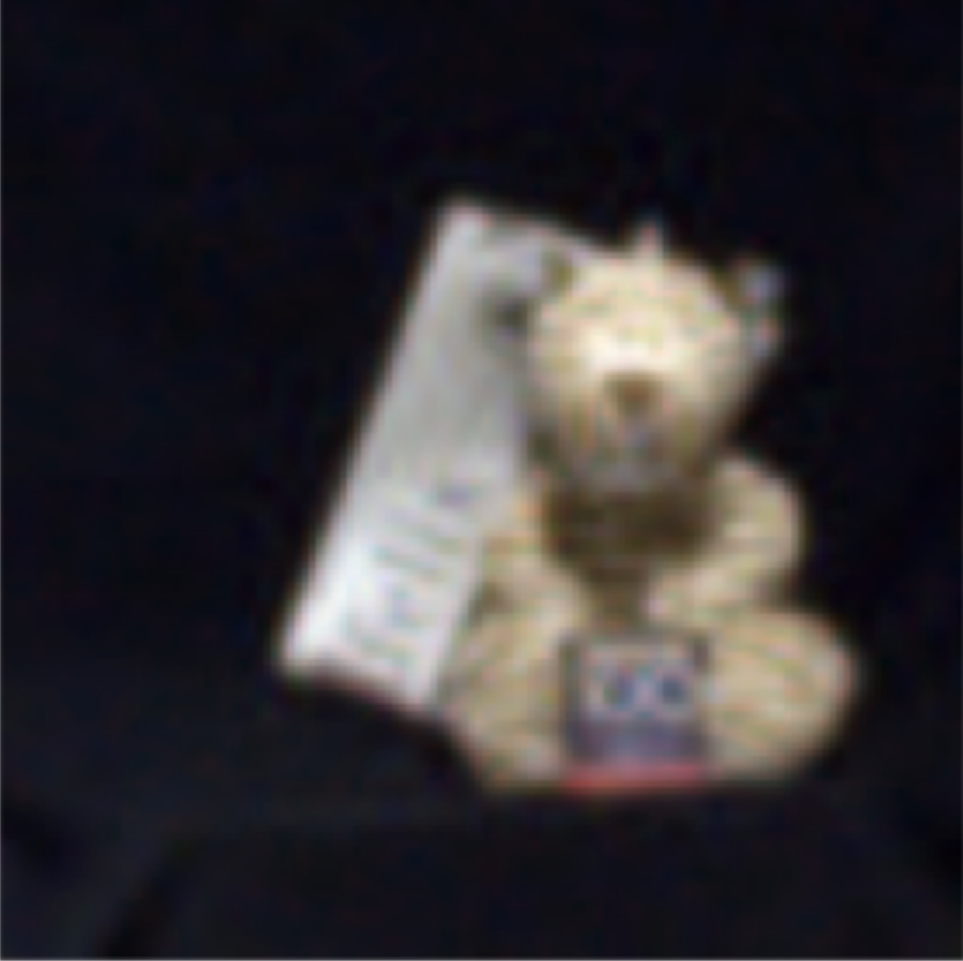}}
\subfigure[]{\includegraphics[scale=0.39]{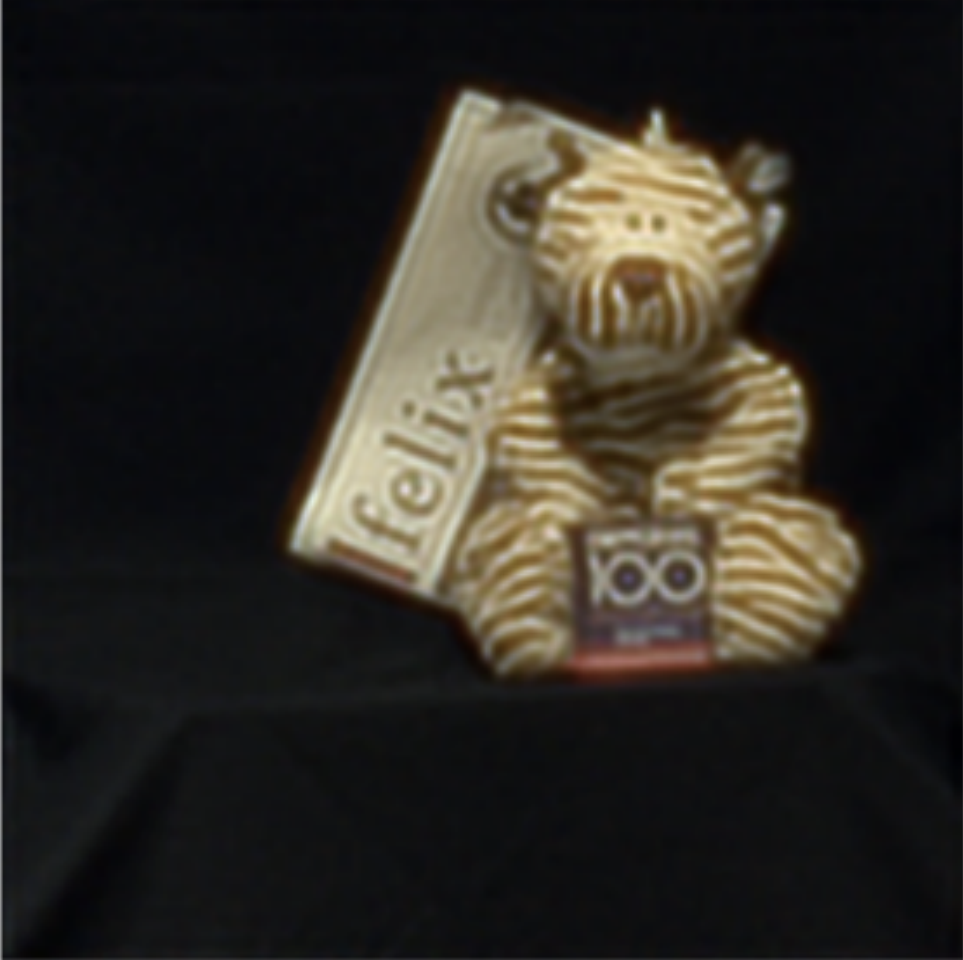}}
\subfigure[]{\includegraphics[scale=0.39]{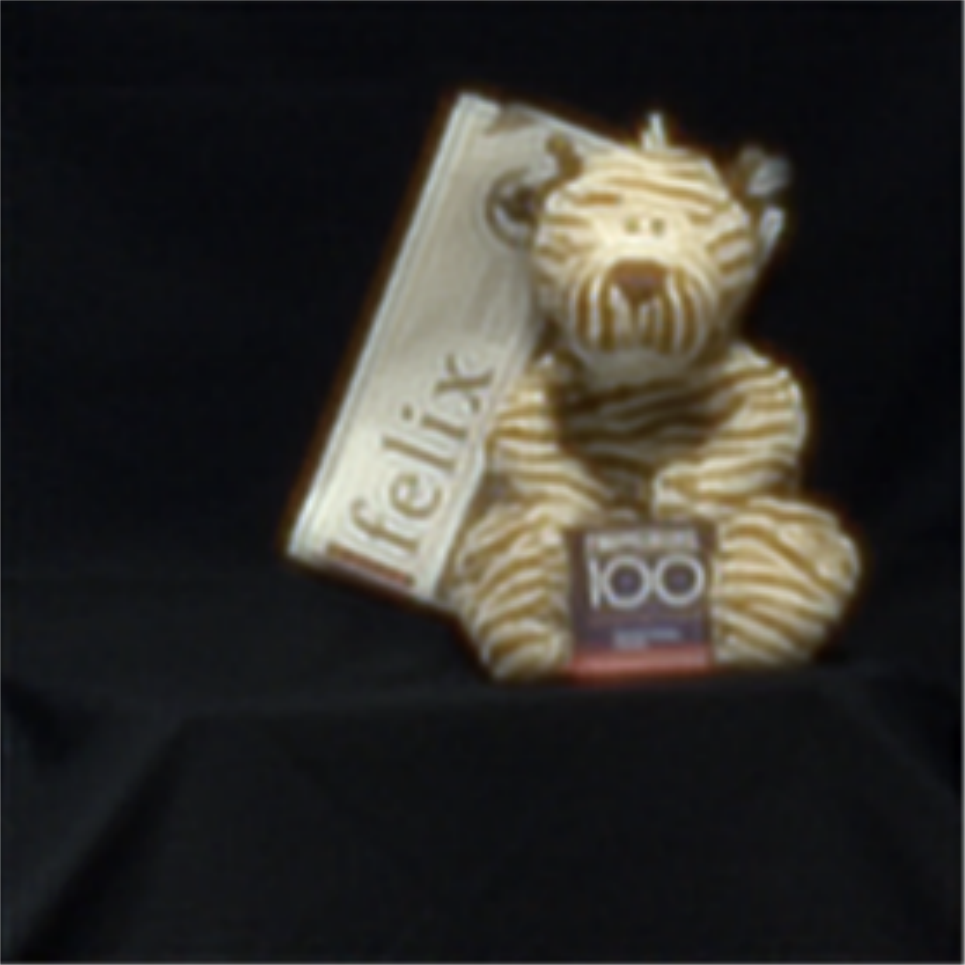}} }
\caption{\sl Example of FRI-based image superresolution. 40 images of a
target scene were acquired with a digital camera. (a) Example acquired image.
(b) Region of interest (128x128 pixels) used for superresolution.
(c) Superresolved image of size $1024\times 1024$ pixels (SR factor = 8).
The PSF in this case is modeled by a B-spline of order 7 (scale 1).
(d) Superresolved image of size $1024\times 1024$ pixels (SR factor = 8).
The PSF in this case is modeled by a B-spline of order 3 (scale 2)
(taken from~\cite{BD10}).}
\label{fig_superres}
\vspace{-2mm}
\end{figure*}

Image registration involves any group of transformations that removes
the disparity between two low resolution (LR) images. This is followed
by image fusion, which blends the properly aligned LR images into a higher
resolution output, possibly removing blur and noise introduced by the
system~\cite{BD10}. The registration step is crucial is order to obtain
a good quality SR image. The theory of FRI can be extended to provide
superresolution imaging.
The key idea of this approach is that, using a proper model for the
point spread function (PSF) of the scene acquisition system, it is possible to
retrieve the underlying continuous geometric moments of the irradiance
light-field. From this information, and assuming the disparity between any
two images can be characterized by a global affine transformation, the set
of images may be registered. The parameters obtained via FRI
correspond to the shift vectors that register the different images before
image fusion.
Figure~\ref{fig_superres} shows a real-world
superresolution example in which 40 low-resolution images allow an
improvement in image resolution by a factor of 8.

A second example application of the FRI model is ultrasonic imaging.
In this imaging modality, an ultrasonic pulse is transmitted into a tissue,
e.g., the heart, and a map of the underlying tissues is created by locating
the echoes of the pulse. Correct location of the tissues and their
edges is crucial for medical diagnosis. Current technology uses
high rate sampling and processing in order to construct the image,
demanding high computational complexity.
Noting that the received signal is a stream of delayed and weighted
versions of the known transmitted pulse shape, FRI sampling schemes
can be exploited in order to reduce the sampling rate and the subsequent
processing rates by several orders of magnitude while still locating the echoes. The received ultrasonic
signal is modeled as a finite FRI problem, with a Gaussian pulse-shape.

\begin{figure*}[t]
\centering \mbox { \subfigure[]{\includegraphics[scale=0.6]{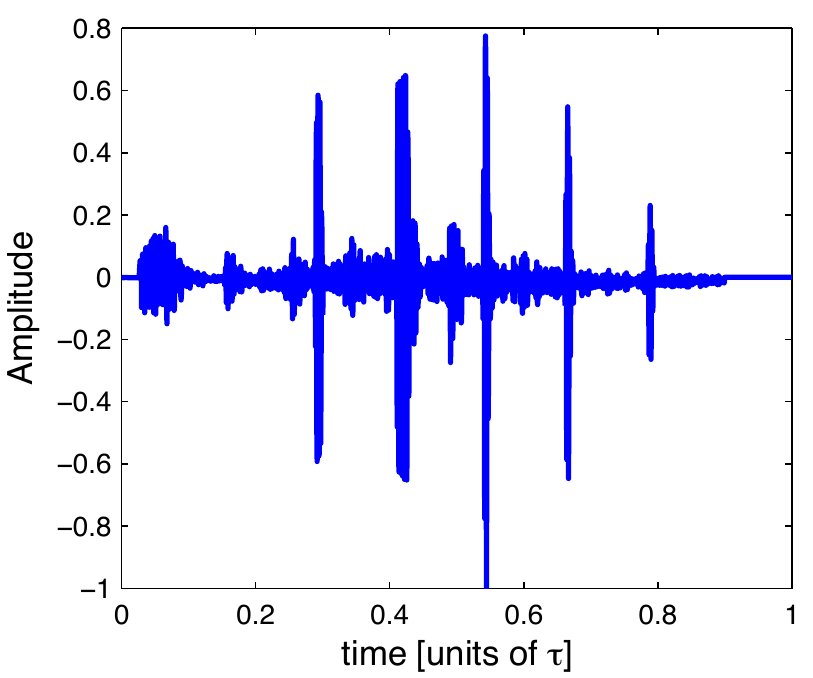}}
\subfigure[]{\includegraphics[scale=0.6]{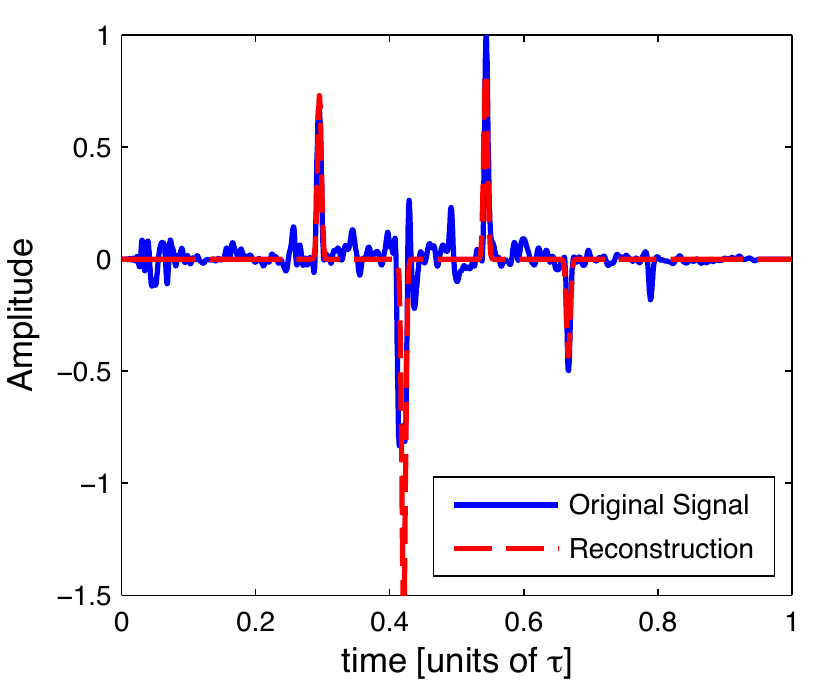}}
\subfigure[]{\includegraphics[scale=0.6]{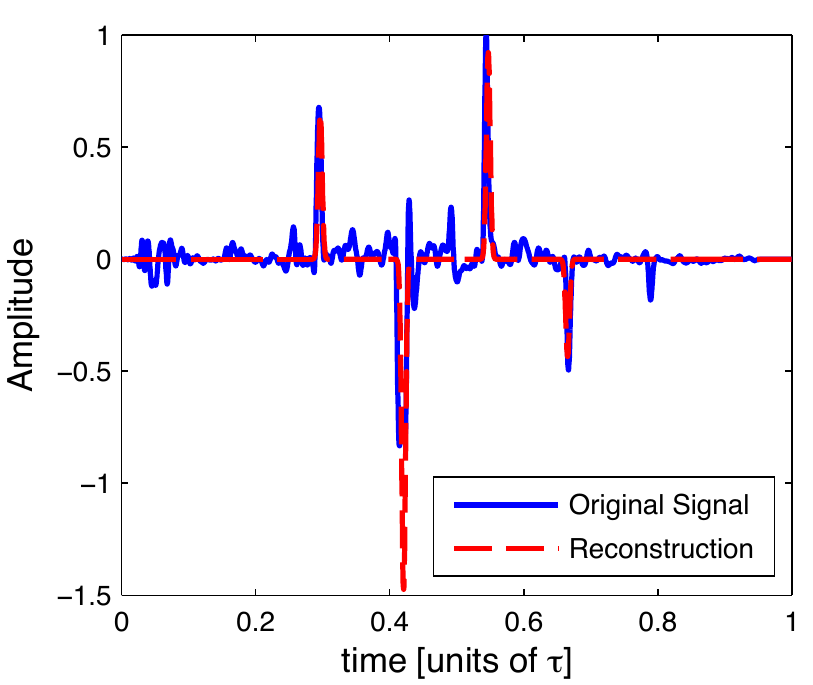}} }
\caption{\sl Applying the SoS sampling scheme with $b_k=1$ on real
ultrasound imaging data. Results are shown vs. original signal
which uses $4160$ samples. (a) Recorded ultrasound imaging signal.
The data was acquired by GE healthcare's Vivid-i ultrasound imaging
system. Reconstructed signal (b) using $N=17$
samples (c) using $N=33$ samples (taken from~\cite{RonenSingleChannel}).}
\label{SingleCh:fig_RealDataReconstruction}
\vspace{-5mm}
\end{figure*}

In Fig.~\ref{SingleCh:fig_RealDataReconstruction} we consider a signal 
obtained using a phantom consisting of uniformly spaced pins, mimicking 
point scatterers, which is scanned by GE Healthcare's Vivid-i portable 
ultrasound imaging system. The data recorded by a single element in the 
probe is modeled as a 1D stream of pulses. The
recorded signal is depicted in Fig.~\ref{SingleCh:fig_RealDataReconstruction}(a).
The reconstruction results using the SoS filter with $b_k=1$ are depicted
in Fig.~\ref{SingleCh:fig_RealDataReconstruction}(b--c). The algorithm looked
for the $L=4$ strongest echoes, using $N=17$ and $N=33$ samples.
In both simulations, the estimation error in the location of the pulses is
around $0.1\textrm{ mm}$. These ideas have also been extended recently
to allow for 2D imaging with multiple received elements~\cite{wagner2011xampling}.

\subsection{Infinite Union of Infinite-Dimensional Subspaces}
\label{sec:infiniteinfinite}

Extending the finite-dimensional model of Section~\ref{sec:infinitefinite} to
the SI setting (\ref{eq:si}), we now incorporate structure by assuming
that each generator $h_{\ell}(t)$ has an unknown parameter $\theta_\ell$
associated with it, leading to an infinite union of infinite-dimensional
spaces. As with its finite counterpart, there is currently no general
sampling framework available to treat such signals. Instead, we focus
on the case in which $h_{\ell}(t)=h(t-\tau_\ell)$.

\subsubsection{Analog signal model}

Suppose that
\begin{equation}\label{SingleCh:eq_sig_model_infinite}
  x(t) = \sum_{l\in\mathbb{Z}} a_\ell h(t - t_\ell),\quad t_\ell, a_\ell \in \mathbb{R},
\end{equation}
where there are no more than $L$ pulses in an interval of length $T$, and that the pulses do not overlap interval boundaries. In this case,
the signal parameters in each interval can be treated separately, using the schemes of the previous section. In particular, we can use the sampling system of Fig.~\ref{fig_infinite_mixing_waveforms_sampling} where now the integral is obtained over intervals of length $T$ \cite{ronen_kfir}. This requires obtaining a sample from each of the channels once every $T$ seconds, and using $p \geq 2K$ channels, resulting in samples taken at the minimal possible rate.

We next turn to treat the more complicated scenario in which $h(t)$ may have support larger than $T$. This setting can no longer be treated by considering independent problems over periods of $T$. To simplify, we consider the special case in which
the time delays in  (\ref{SingleCh:eq_sig_model_infinite}) repeat periodically (but not the amplitudes) \cite{GE10,BGE10}. As we will
show in this special case, efficient sampling and recovery is
possible even using a single filter, and without requiring the pulse $h(t)$ to
be time limited. Under our assumptions, the input signal can be written as
\begin{align}\label{eq_sig_model_kfir}
x(t) = \sum_{n\in\mathbb{Z}}\sum_{\ell=1}^{L} a_{\ell}[n] h(t - t_{\ell}-nT),
\end{align}
where $\tau=\{t_{\ell}\}_{\ell=1}^{L}$ is a set of unknown time delays
contained in the time interval $[0,T)$,  $\{a_{\ell}[n]\}$ are arbitrary bounded
energy sequences, and $h(t)$ is a known pulse shape.

\subsubsection{Compressive signal acquisition}

We follow a similar
approach to that in Section~\ref{sec:finiteinfinite}, which treats a
structured SI setting where there are $N$ possible generators. The
difference is that in this current case there are infinitely many
possibilities. Therefore, we replace the CTF in Fig.~\ref{fig:fbs}
with a block that supports this continuity: we will see that the ESPRIT
method essentially replaces the CTF block~\cite{ESPRIT_Kailath}.

\begin{figure}[t]
\centering
\includegraphics[scale=0.4]{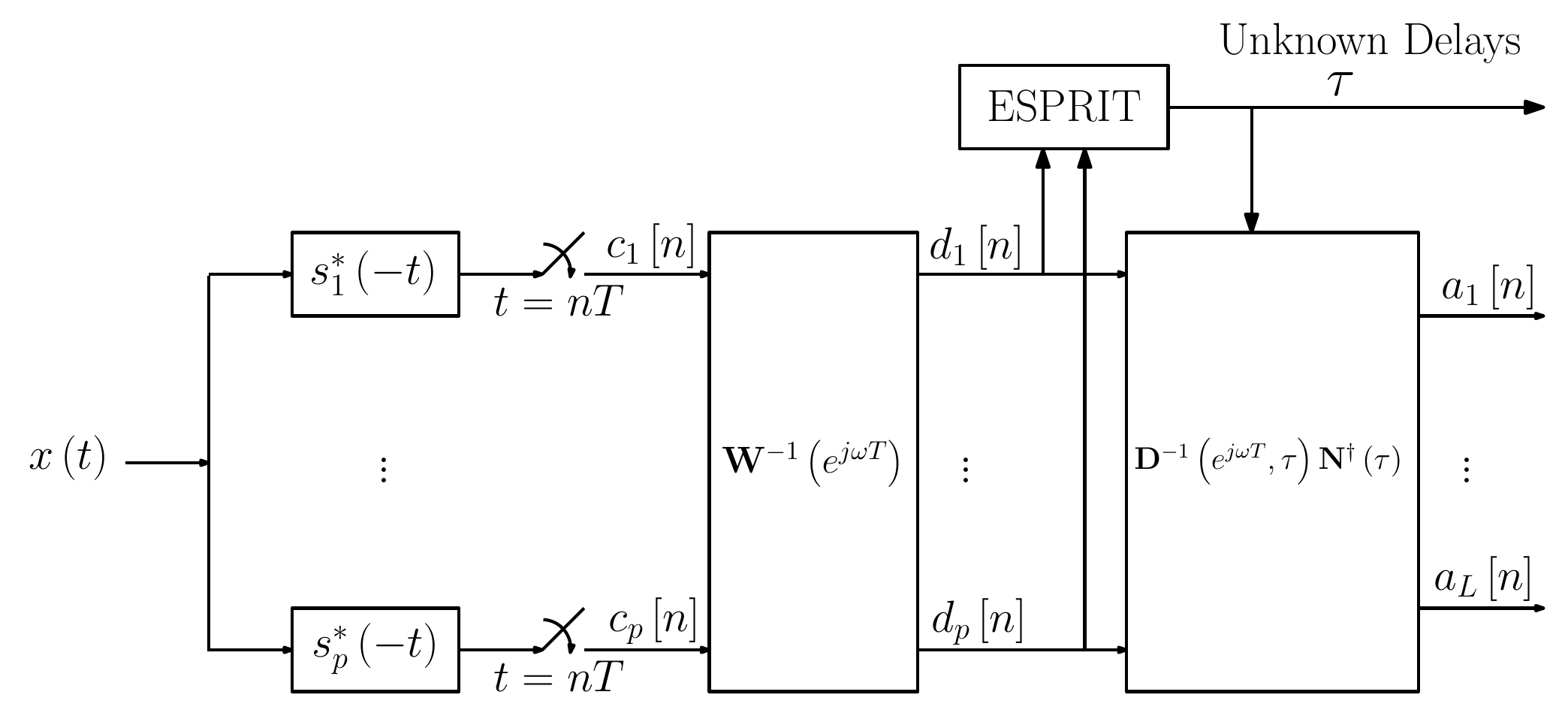}
\caption{\sl Sampling and reconstruction scheme for signals of the
form~\eqref{eq_sig_model_kfir}.}
\label{fig_tde_scheme}
\vspace{-10mm}
\end{figure}

A sampling and reconstruction scheme for signals of the form
(\ref{eq_sig_model_kfir}) is depicted in Fig.~\ref{fig_tde_scheme}~\cite{GE10}.
The analog sampling stage is comprised of $p$ parallel sampling channels.
In each channel, the input signal $x(t)$ is filtered by a band-limited sampling
kernel $s_{\ell}^*(-t)$ with frequency support contained in an interval
of width $2\pi p/T$, followed by a uniform sampler operating at a rate of $1/T$,
thus providing the sampling sequence $c_{\ell}[n]$. Note that just as in the
MWC (Section~\ref{sec:mwc}), the sampling filters can be collapsed to a single
filter whose output is sampled at $p$ times the rate of a single channel.
The role of the sampling kernels is to spread out the energy of the signal in
time, prior to low rate sampling.

\subsubsection{Recovery algorithms}

To recover the signal from the samples, a properly designed digital filter
correction bank, whose frequency response in the DTFT domain is given
by $\mathbf{W}^{-1}(e^{j\omega T})$, is applied to the sampling
sequences in a manner similar to~(\ref{eq:sirecon}). The matrix
$\bbw(e^{j\omega T})$ depends on the choice of the sampling kernels
$s_{\ell}^*(-t)$ and the pulse shape $h(t)$. Its entries are defined for
$1 \le \ell,m \le p$ as
\begin{equation}
\label{eq:wt}
\bbw\left(e^{j\omega T}\right)_{\ell,m} = \frac{1}{T} S_\ell^*(\omega+2 \pi m/T) H(\omega+2 \pi m/T).
\end{equation}
After the digital correction stage, it can be shown that the corrected sample
vector $\bd[n]$ is related to the unknown amplitude vector $\ba[n]$ by a
Vandermonde matrix which depends on the unknown delays~\cite{GE10}.
Therefore, we can now exploit  known tools taken from the direction
of arrival~\cite{krim1996tda} and spectral estimation~\cite{Stoica1997}
literature to recover the delays $\tau = \{t_1,\ldots, t_L\}$, such as the
well-known ESPRIT algorithm~\cite{ESPRIT_Kailath}. Once the delays are
determined, additional filtering operations are applied on the samples to
recover the sequences ${a_{\ell}[n]}$.  In particular, referring to
Fig.~\ref{fig_tde_scheme}, the matrix $\bbd$ is a diagonal matrix with
diagonal elements equal to $e^{-j\omega t_k}$, and $\bbn(\tau)$ is a
Vandermonde matrix with elements $e^{-j2\pi m t_k/T}$.
The ESPRIT algorithm is summarized for our setting as Algorithm~\ref{alg:ESPRIT}, 
where $\mathbf{E}_{s\downarrow}$ and $\mathbf{E}_{s\uparrow}$
denote the sub matrices extracted from $\mathbf{E}_{s}$
by deleting its last/first row, respectively.

\subsubsection{Recovery guarantees}

The proposed algorithm is guaranteed to recover the unknown delays and amplitudes as long as $p$ is large enough \cite{GE10}.

\begin{algorithm}[t]
\caption{ESPRIT Algorithm} \label{alg:ESPRIT}
\begin{algorithmic}
\STATE Input: Signal $\d$, number of parameters $L$.
\STATE Output: Delays $\tau = \{t_1,\ldots, t_L\}$.
\STATE $\mathbf{R}_{dd}=\sum_{n\in\mathbb{Z}}\mathbf{d}\left[n\right]\mathbf{d}^{H}\left[n\right]$
\COMMENT{construct correlation matrix}
\STATE $(\mathbf{E},\sigma) \leftarrow \mathrm{SVD}(\mathbf{R}_{dd})$
\COMMENT{calculate SVD}
\STATE $\mathbf{E}_{s} \leftarrow $ $L$ singular vectors of $\mathbf{R}_{dd}$ associated with non-zero singular values.
\STATE $\Phi=\mathbf{E}_{s\downarrow}^{\dagger}\mathbf{E}_{s\uparrow}$
\COMMENT{compute ESPRIT matrix}
\STATE $\{\lambda_1,\ldots,\lambda_L\} \leftarrow \mathrm{eig}(\Phi)$
\COMMENT{obtain eigenvalues}
\STATE $t_i \leftarrow -\frac{T}{2\pi}\textrm{arg}\left(\lambda_{i}\right)$, $1 \le i \le L$
\COMMENT{map eigenvalues to delays}
\STATE return $\tau = \{t_1,\ldots, t_L\}$
\end{algorithmic}
\end{algorithm}

\begin{theorem}
\label{thm:td}\cite{GE10}
The sampling scheme of Fig.~\ref{fig_tde_scheme} is guaranteed to recover 
any signal of the form \eqref{eq_sig_model_kfir} as long as 
$p \geq 2L-\eta+1$, where $\eta$ is the dimension of the minimal 
subspace containing the vector set $\{{\bf{d}}[n],n \in \mathbb{Z}\}$. 
In addition, the filters $s_\ell(t)$ are supported on $2\pi p/T$ and must be 
chosen such that $\bbw(e^{j\omega T})$ of (\ref{eq:wt}) is invertible.
\end{theorem}
The sampling rate resulting from Theorem~\ref{thm:td} is no larger than $2L/T$ since $0 \leq \eta \leq L$. For certain signals, the
rate can be reduced to $(L+1)/T$. This is due to the fact that the outputs are processed jointly, similar to the MMV model.
Evidently, the minimal sampling rate is not related to the Nyquist rate
of the pulse $h(t)$. Therefore, for wideband pulse shapes, the reduction
in rate can be quite substantial. As an example, consider the setup
in~\cite{uwb_char}, used for characterization of ultra-wide band wireless
indoor channels. Under this setup, pulses with bandwidth of $W=1$GHz
are transmitted at a rate of $1/T=2$MHz. Assuming  that there are $10$
significant multipath components, we can reduce the sampling rate
down to $40$MHz compared with the $2$GHz Nyquist rate.

\subsubsection{Applications}

Problems of the form (\ref{eq_sig_model_kfir}) appear in a variety of different
settings. For example, the model \eqref{eq_sig_model_kfir} can describe
multipath medium identification problems, which arise in applications
such as radar~\cite{quazi1981otd}, underwater acoustics~\cite{urick1983pus},
wireless communications, and more.
In this context, pulses with known shape are transmitted through a multipath
medium, which consists of several propagation paths, at a constant rate.
As a result the received signal is composed of delayed and weighted replicas
of the transmitted pulses. The delays $t_{\ell}$ represent the propagation
delays of each path, while the sequences $a_{\ell}[n]$ describe the
time-varying gain coefficient of each multipath component.

Another important application of (\ref{eq_sig_model_kfir}) is in the context of
radar. In this example, we translate the rate reduction to increased resolution
with a fixed time-bandwidth product (TBP), thus enabling super-resolution radar
from low rate samples. In radar, the goal is to identify the range and velocity
of $K$ targets. The delay in this case captures the range while the time
varying coefficients are a result of the Doppler delay related to the target
velocity~\cite{BGE10}.
More specifically, we assume that several targets can have the same delays
but possibly different Doppler shifts so that  $\{t_{\ell}\}_{\ell=1}^{L}$ denote
the set of distinct delays. For each delay value $t_{\ell}$ there are $K_{\ell}$
values of associated Doppler shifts $\nu_{\ell k}$ and reflection coefficients
$\alpha_{\ell k}$. We also assume that the system is highly underspread,
namely $\nu_{\textrm{max}} T \ll 1$, where $\nu_{\textrm{max}}$ and $T$ 
denote the maximal Doppler shift and delay. To identify the targets we 
transmit the signal
\begin{align}
x_{T}=\sum_{n=0}^{N-1} x_n h(t-nT),
\end{align}
where $x_n$ is a known $N$-length probing sequence, and $h(t)$ is a
known pulse shape. The received signal can then be described in the
form \eqref{eq_sig_model_kfir}, where the sequences $a_{\ell}[n]$ satisfy
\begin{align}
a_{\ell}[n]=x_n \sum_{k=1}^{K_{\ell}} \alpha_{\ell k} e^{j 2 \pi \nu_{\ell k} n T}.
\end{align}

The delays and the sequences  $a_{\ell}[n]$ can be recovered using the
general scheme for time delay recovery. The Doppler
shifts and reflection coefficients are then determined from the sequences
$a_{\ell}[n]$ using standard spectral estimation tools~\cite{Stoica1997}.
The targets can be exactly identified as long as the bandwidth $\mathcal{W}$
of the transmitted pulse satisfies $\mathcal{W} \geq \frac{4\pi L}{T}$, and the
length of the probing sequence satisfies
$N\geq 2 \max K_{\ell}$~\cite{BGE10}. This leads to a minimal
TBP of the input signal of
$\mathcal{WT} \geq {8\pi L \max K_{\ell}}$, which is much lower than that
obtained using standard radar processing techniques, such as
matched-filtering (MF).

\begin{figure*}[t]
\begin{center}
\subfigure[]{\includegraphics[scale=0.22]{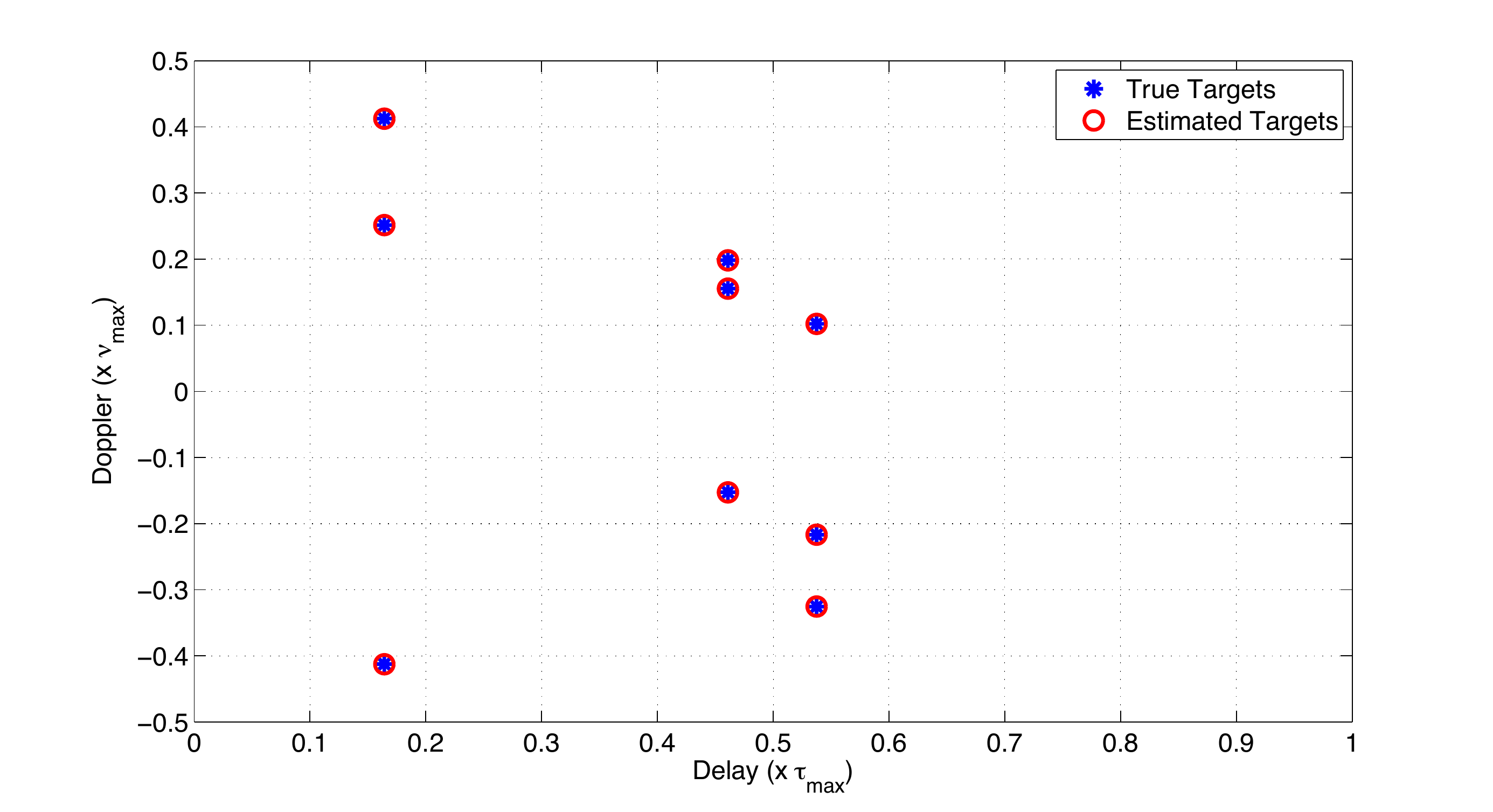}}\subfigure[]{\includegraphics[scale=0.22]{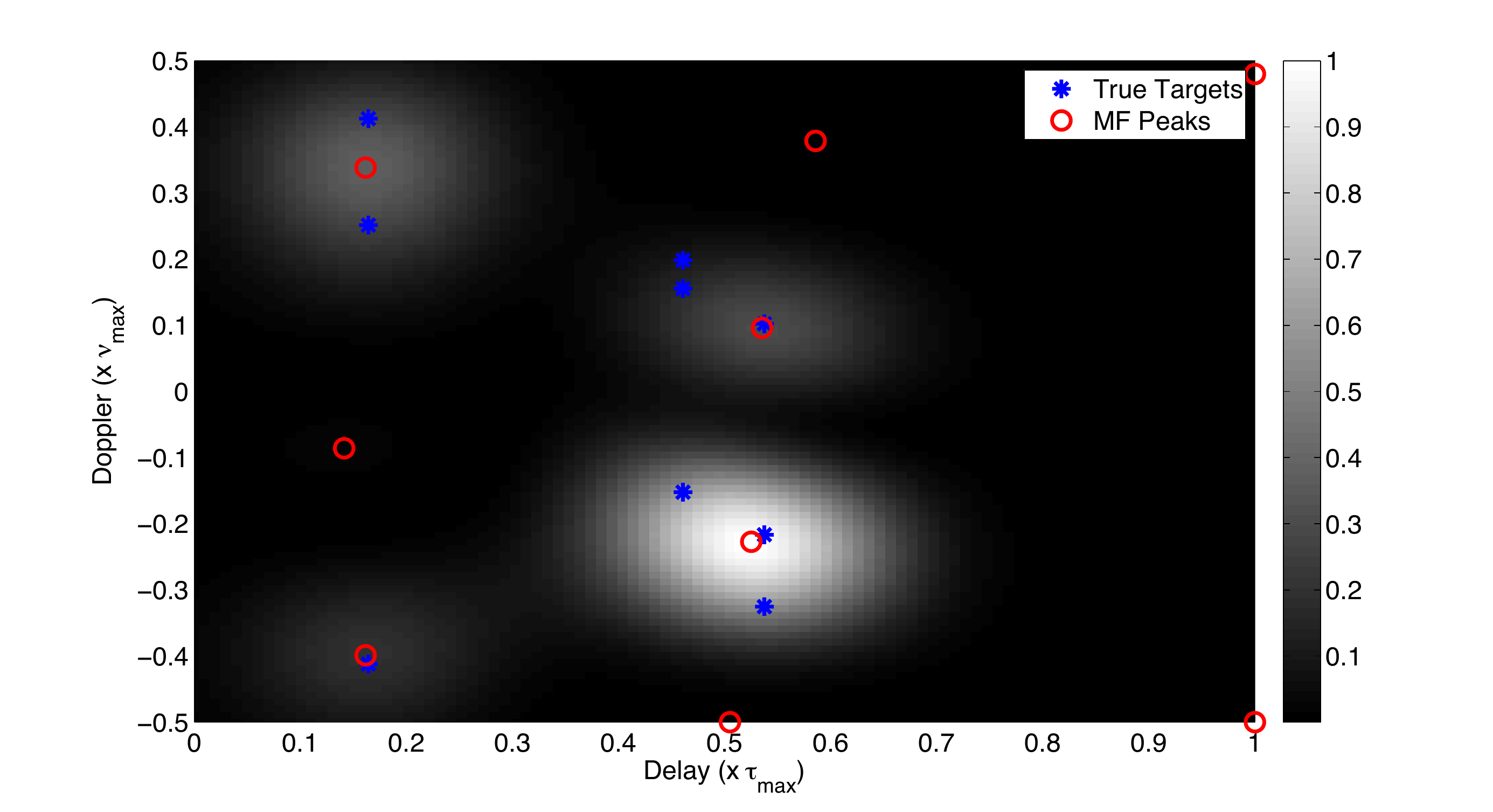}}\subfigure[]{\includegraphics[scale=0.22]{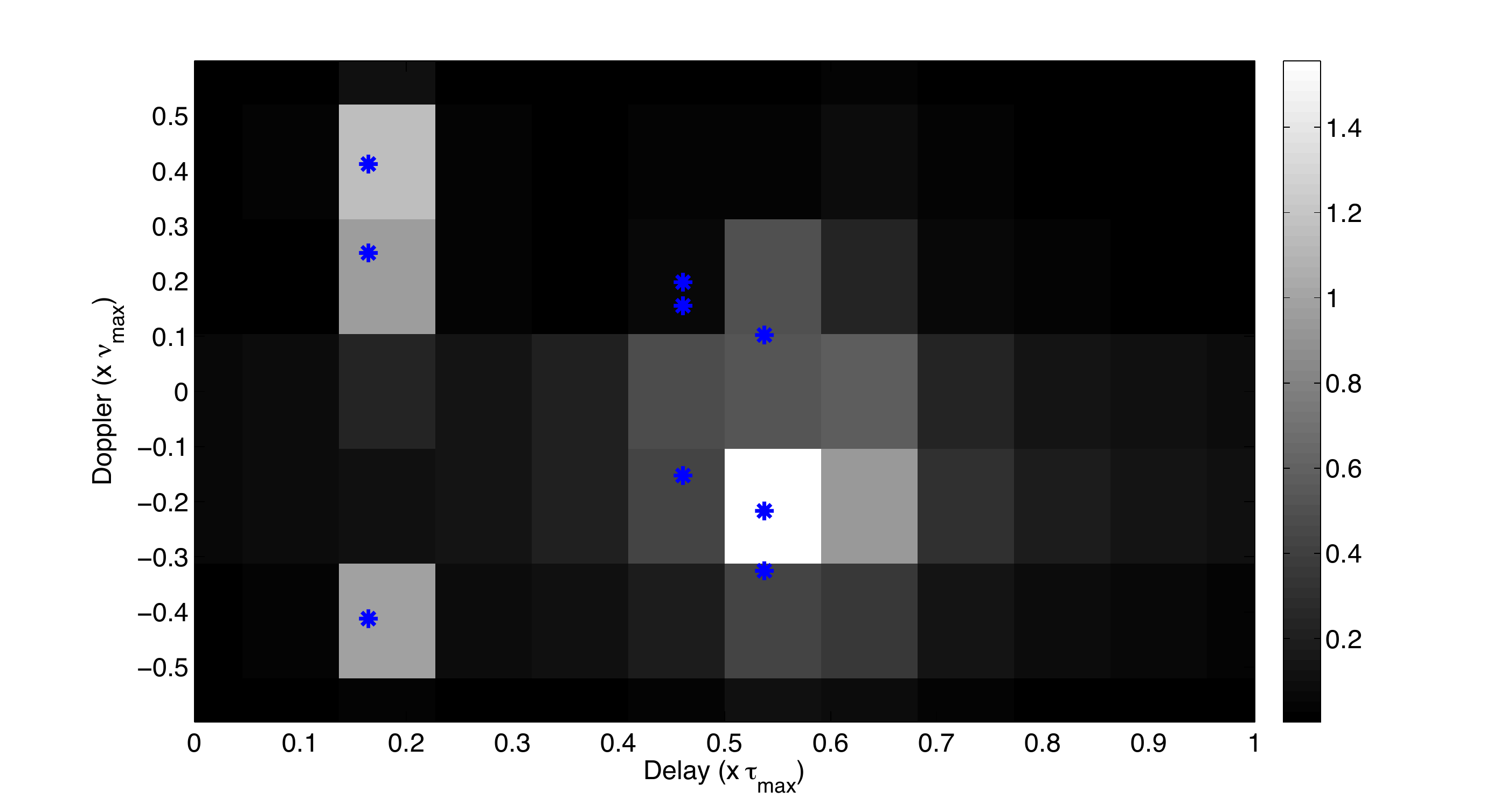}}
\caption{\sl \label{fig:matched_filter}
Comparison between the target-detection performance for the case of
nine targets (represented by $*$) in the delay--Doppler space with $\tau_{max} = 10~\mu\text{s}$, $\nu_{max} = 10 \text{ kHz}$, $W = 1.2$ MHz, and $T = 0.48$ ms. The probing sequence $\{x_n\}$ corresponds to a random binary $(\pm 1)$ sequence with $N = 48$, the pulse $p(t)$ is designed to have a nearly-flat frequency response and the pulse repetition interval is $T = 10~\mu\text{s}$.
Recovery of the
Doppler-delay plane using (a) a union of subspaces approach,
(b) a standard matched filter, and (c) a discretized delay-Doppler plane (taken from~\cite{BGE10}).}
\end{center}
\vspace{-7mm}
\end{figure*}

An example of the identification of nine close targets is illustrated in
Fig.~\ref{fig:matched_filter}(a).
The sampling filter used is a simple LPF. The original and recovered
targets are shown on the Doppler-delay plane. Evidently all the targets
were correctly identified. The result obtained by MF, with the same
TBP, is shown in Fig.~\ref{fig:matched_filter}(b). Clearly, the compressive 
method has superior resolution than the standard MF in this low noise setting.
Thus, the union of subspaces viewpoint not only offers a reduced-rate
sampling method, but allows to increase the resolution in target
identification for a fixed low TBP when the SNR is high enough, which is of 
paramount importance in many practical radar problems.

Previous approaches for identifying the unknown delays and gains involve
sampling the received signal at the Nyquist rate of the pulse
$h(t)$~\cite{bruckstein1985roe,pallas1991ahr,4303294}.
However, prior knowledge of the pulse shape results in a parametric
viewpoint, and we would expect that the rate should be proportional
to the number of degrees of freedom, i.e., the number of paths $L$.

Another approach is to quantize the delay-Doppler plane by
assuming the delays and Doppler shifts lie on a
grid~\cite{MCW05,kay:icassp97,bajwa:allerton08,herman:tsp09}.
After discretization, CS tools for finite representations can be used to capture
the sparsity on the discretized grid. Clearly, this approach has an inherent
resolution limitation. Moreover, in real world scenarios, the targets do not lie
exactly on the grid points. This causes a leakage of their energies in the
quantized space into adjacent grid points~\cite{CSPC10,DB10}.
In contrast, the union of subspaces model avoids the
discretization issue and offers concrete sampling methods that can be
implemented efficiently in hardware (when the SNR is not too poor). 

\subsection{Discussion}

Before concluding, we point out that much of the work in this section 
(Sections~\ref{sec:finiteinfinite}, \ref{sec:infinitefinite} 
and~\ref{sec:infiniteinfinite}) has been 
focused on infinite unions. In each case, the approach we introduced is 
based on a sampling mechanism using analog filters, followed by 
standard array processing tools for recovery in the digital domain.
These techniques allow perfect recovery when no noise is present, and 
generally degrade gracefully in the presence of noise, as has been 
analyzed extensively in the array processing literature. Nonetheless, 
when the SNR is poor, alternative digital recovery methods based on CS 
may be preferable. Thus, we may increase robustness by using the analog 
sampling techniques proposed here in combination with standard CS 
algorithms for recovery in the digital domain. To apply this approach, once 
we have obtained the desired samples, we can view them as the 
measurement vector $y$ in a standard CS system; the CS matrix $\Phi$ is 
now given by a discretized Vandermonde matrix that captures all possible 
frequencies to a desired accuracy; and the vector $x$ is a sparse vector 
with non-zero elements only in those indices corresponding to actual 
frequencies~\cite{TroppLaskaDuarteRombergBaraniuk,MCW05,kay:icassp97,bajwa:allerton08,herman:tsp09}. 
In this way, we combine the benefits of CS with those of analog sampling, 
without requiring discretization in the sampling stage. The discretization now 
only appears in the digital domain during recovery and not in the sampling 
mechanism. If we follow this strategy, then the results developed in 
Section~\ref{sec:csbasics} regarding recovery in the presence of noise are 
relevant here as well.

\section{Conclusions}
\label{sec:conc}

In this review, our aim was to summarize applications of the basic
CS framework that integrate specific constraints
of the problem at hand into the theoretical and algorithmic
formulation of signal recovery, going beyond the original ``random
measurement/sparsity model'' paradigm. Due to constraints given
by the relevant sensing devices, the classes of
measurement matrices available to us are limited. Similarly, some
applications focus on signals that exhibit structure which cannot be
captured by sparsity alone and provide additional information
that can be leveraged during signal recovery. We also considered the
transition between continuous and discrete representations bridged by
analog-to-digital converters. Analog signals are continuous-time by 
nature; in many cases, the application of compressed sensing to 
this larger signal class requires the formulation of new devices, signal 
models, and recovery algorithms. It also necessitates a deeper 
understanding of hardware considerations that must be taken into account 
in theoretical developments. This acquisition framework, in turn, motivates 
the design of new sampling schemes and devices that provide the 
information required for signal recovery in the smallest possible 
representation.

While it is difficult to cover all of the developments in compressive
sensing theory and hardware~\cite{GJBWS07,NK07,BGN08,WJWB08,BW09,CP10,Hansen11},
our aim here was to select few examples that were representative of wider
classes of problems and that offered a balance between a useful theoretical
background and varied applications. We hope that this summary will be
useful to practitioners in signal acquisition and processing that are
interested in leveraging the features of compressive sensing in their
specific applications. We also hope that this review will inspire further
developments in the theory and practice underlying CS: we particularly
envision extending the existing framework to broader signals sets and
inspiring new implementation and design paradigms. With an eye to the
future, more advanced configurations of CS can play a key role in many
new frontiers. Some examples already mentioned
throughout are cognitive radio, optical systems,  medical devices
such as MRI, ultrasound and more. These techniques hold promise
for a complete rethinking of many acquisition systems and 
stretch the limit of current sensing capabilities.

As we have also demonstrated, CS holds promise for increasing resolution
by exploiting signal structure. This can revolutionize many applications such
as radar and microscopy by making efficient use of the available degrees of
freedom in these settings. Consumer electronics, microscopy, civilian and
military surveillance, medical imaging, radar and many other rely on ADCs
and are resolution-limited. Removing the Nyquist barrier in these
applications and increasing resolution can improve the user
experience, increase data transfer, improve imaging quality and reduce
exposure time -- in other words, make a prominent impact on the
analog-digital world surrounding us.

\section{Acknowledgements}

The authors would like to thank their colleagues for many useful comments
and for their collaboration on many topics related to this review. In particular,
they are grateful to Mike Wakin for many discussions on the structure of the
review, to Moshe Mishali, Holger Rauhut, Justin Romberg, Yao Xie, 
and the anonymous reviewers for offering many helpful comments on an 
early draft of the manuscript, and to Pier Luigi Dragotti, Polina Golland, 
Laurent Jacques, Yehia Massoud, Justin Romberg, and Rebecca Willett for 
authorizing the use of their figures. 

\bibliographystyle{IEEEbib}
\bibliography{CSBib}

\begin{thebibliography}{100}

\bibitem{DonohoCS}
D.~L. Donoho,
\newblock ``Compressed sensing,''
\newblock {\em IEEE Trans. Info. Theory}, vol. 52, no. 4, pp. 1289--1306, Sep.
  2006.

\bibitem{CandesUES}
E.~J. Cand\`{e}s and T.~Tao,
\newblock ``Near optimal signal recovery from random projections: Universal
  encoding strategies?,''
\newblock {\em IEEE Trans. Info. Theory}, vol. 52, no. 12, pp. 5406--5425, Dec.
  2006.

\bibitem{BaraniukCS}
R.~G. Baraniuk,
\newblock ``Compressive sensing,''
\newblock {\em {IEEE} Signal Proc. Mag.}, vol. 24, no. 4, pp. 118--120, 124,
  July 2007.

\bibitem{CandesCS}
E.~J. Cand\`{e}s,
\newblock ``Compressive sampling,''
\newblock in {\em Int. Congress of Mathematicians}, Madrid, Spain, 2006,
  vol.~3, pp. 1433--1452.

\bibitem{CandesWakinCS}
E.~J. Cand\`{e}s and M.~B. Wakin,
\newblock ``An introduction to compressive sampling,''
\newblock {\em {IEEE} Signal Proc. Mag.}, vol. 25, no. 2, pp. 21--30, Mar.
  2008.

\bibitem{VMB02}
M.~Vetterli, P.~Marziliano, and T.~Blu,
\newblock ``Sampling signals with finite rate of innovation,''
\newblock {\em {IEEE} Trans. Signal Proc.}, vol. 50, no. 6, pp. 1417--1428,
  2002.

\bibitem{DVB07}
P.~L. Dragotti, M.~Vetterli, and T.~Blu,
\newblock ``Sampling moments and reconstructing signals of finite rate of
  innovation: {S}hannon meets {S}trang-{F}ix,''
\newblock {\em {IEEE} Trans. Signal Proc.}, vol. 55, no. 5, pp. 1741--1757, May
  2007.

\bibitem{BDVMC08}
T.~Blu, P.-L. Dragotti, M.~Vetterli, P.~Marziliano, and L.~Coulot,
\newblock ``Sparse sampling of signal innovations,''
\newblock {\em IEEE Signal Proc. Mag.}, vol. 25, no. 2, pp. 31--40, Mar. 2008.

\bibitem{MEDS09}
M.~Mishali, Y.~C. Eldar, O.~Dounaevsky, and E.~Shoshan,
\newblock ``Xampling: Analog to digital at sub-nyquist rates,''
\newblock {\em {IET} Circuits, Devices and Systems}, vol. 5, no. 1, pp. 8--20,
  Jan. 2011.

\bibitem{MEE10}
M.~Mishali and Y.~C. Eldar,
\newblock ``Xampling: {C}ompressed sensing for analog signals,''
\newblock in {\em Compressed Sensing: {T}heory and Applications}, Y.~C. Eldar
  and G.~Kutyniok, Eds. Cambridge Univ. Press, 2012.

\bibitem{N28}
H.~Nyquist,
\newblock ``Certain topics in telegraph transmission theory,''
\newblock {\em Trans. AIEE}, vol. 47, pp. 617--644, Apr. 1928.

\bibitem{S49}
C.~E. Shannon,
\newblock ``Communications in the presence of noise,''
\newblock {\em Proc. IRE}, vol. 37, pp. 10--21, Jan 1949.

\bibitem{W15}
E.~T. Whittaker,
\newblock ``On the functions which are represented by the expansions of the
  interpolation theory,''
\newblock {\em Proc. Roy. Soc. Edinburgh}, vol. 35, pp. 181--194, 1915.

\bibitem{K33}
V.~A. Kotelnikov,
\newblock ``{On the transmission capacity of the `ether' and of cables in
  electrical communications},''
\newblock in {\em All-Union Conf. Technological Reconstruction of the
  Communications Sector and the Development of Low-current Engineering}, 1933.

\bibitem{LD08}
Y.~M. Lu and M.~N. Do,
\newblock ``A theory for sampling signals from a union of subspaces,''
\newblock {\em {IEEE} Trans. Signal Proc.}, vol. 56, no. 6, pp. 2334--2345,
  2008.

\bibitem{BD09a}
T.~Blumensath and M.~E. Davies,
\newblock ``Sampling theorems for signals from the union of finite-dimensional
  linear subspaces,''
\newblock {\em IEEE Trans. Info. Theory}, vol. 55, no. 4, pp. 1872--1882, Apr.
  2009.

\bibitem{EM09a}
Y.~C. Eldar and M.~Mishali,
\newblock ``{Robust recovery of signals from a structured union of
  subspaces},''
\newblock {\em IEEE Trans. Info. Theory}, vol. 55, no. 11, pp. 5302--5316,
  2009.

\bibitem{E09}
Y.~C. Eldar,
\newblock ``Compressed sensing of analog signals in shift-invariant spaces,''
\newblock {\em IEEE Trans. Signal Proc.}, vol. 57, no. 8, pp. 2986--2997, Aug.
  2009.

\bibitem{GE10}
K.~Gedalyahu and Y.~C. Eldar,
\newblock ``Time delay estimation from low rate samples: A union of subspaces
  approach,''
\newblock {\em {IEEE} Trans. Signal Proc.}, vol. 58, no. 6, pp. 3017--3031,
  June 2010.

\bibitem{BCDH10}
R.~G. Baraniuk, V.~Cevher, M.~F. Duarte, and C.~Hegde,
\newblock ``Model-based compressive sensing,''
\newblock {\em IEEE Trans. Info. Theory}, vol. 56, no. 4, pp. 1982--2001, Apr.
  2010.

\bibitem{GR97}
I.~F. Gorodnitsky and B.~D. Rao,
\newblock ``{Sparse signal reconstruction from limited data using FOCUSS: A
  re-weighted minimum norm algorithm},''
\newblock {\em IEEE Trans. Signal Proc.}, vol. 45, no. 3, pp. 600--616, Mar.
  1997.

\bibitem{DonohoBP}
S.~Chen, D.~L. Donoho, and M.~Saunders,
\newblock ``Atomic decomposition by basis pursuit,''
\newblock {\em SIAM J. Sci. Computing}, vol. 20, no. 1, pp. 33--61, 1998.

\bibitem{BDE09}
A.~M. Bruckstein, D.~L. Donoho, and M.~Elad,
\newblock ``From sparse solutions of systems of equations to sparse modeling of
  signals and images,''
\newblock {\em SIAM Review}, vol. 51, no. 1, pp. 34--81, Feb. 2009.

\bibitem{MZ93}
S.~Mallat and Z.~Zhang,
\newblock ``Matching pursuit with time-frequency dictionaries,''
\newblock {\em {IEEE} Trans. Signal Proc.}, vol. 41, no. 12, pp. 3397--3415,
  Dec. 1993.

\bibitem{Foucart}
S.~Foucart,
\newblock ``A note on guaranteed sparse recovery via $\ell_1$-minimization,''
\newblock {\em Appl. Comput. Harmon. Anal.}, vol. 29, no. 1, pp. 97 -- 103,
  2010.

\bibitem{FoucartGribonval}
S.~Foucart and R.~Gribonval,
\newblock ``Real versus complex null space properties for sparse vector
  recovery,''
\newblock {\em C. R. Acad. Sci. Paris, Ser. I}, vol. 348, no. 15--16, pp.
  863--865, 2010.

\bibitem{CDD09}
A.~Cohen, W.~Dahmen, and R.~A. DeVore,
\newblock ``Compressed sensing and best $k$-term approximation,''
\newblock {\em J. American Math. Society}, vol. 22, no. 1, pp. 211--231, Jan.
  2009.

\bibitem{DonohoOSR}
D.~L. Donoho and M.~Elad,
\newblock ``{Optimally sparse representation in general (nonorthogonal)
  dictionaries via $\ell_1$ minimization},''
\newblock {\em Proc. Nat. Acad. Sci.}, vol. 100, no. 5, pp. 2197--2202, Mar.
  2003.

\bibitem{DH01}
D.L. Donoho and Xiaoming Huo,
\newblock ``Uncertainty principles and ideal atomic decompositions,''
\newblock {\em IEEE Trans. Info. Theory}, vol. 47, no. 7, pp. 2845--2862, Nov.
  2001.

\bibitem{TroppGreed}
J.~A. Tropp,
\newblock ``Greed is good: {A}lgorithmic results for sparse approximation,''
\newblock {\em IEEE Trans. Info. Theory}, vol. 50, no. 10, pp. 2231--2242, Oct.
  2004.

\bibitem{GN03}
R.~Gribonval and M.~Nielsen,
\newblock ``Sparse representations in unions of bases,''
\newblock {\em IEEE Trans. Info. Theory}, vol. 49, no. 12, pp. 3320--3325, Dec.
  2003.

\bibitem{Welch}
L.~R. Welch,
\newblock ``Lower bounds on the maximum cross correlation of signals,''
\newblock {\em IEEE Trans. Info. Theory}, vol. 20, no. 3, pp. 397--399, May
  1974.

\bibitem{StrohmerHeath}
T.~Strohmer and R.~Heath,
\newblock ``Grassmanian frames with applications to coding and communication,''
\newblock {\em Appl. Comput. Harmon. Anal.}, vol. 14, no. 3, pp. 257--275, Nov.
  2003.

\bibitem{Gershgorin}
S.~A. Ger\v{s}gorin,
\newblock ``{\"U}ber die abgrenzung der eigenwerte einer matrix,''
\newblock {\em Izv. Akad. Nauk SSSR Ser. Fiz.-Mat.}, vol. 6, pp. 749--754,
  1931.

\bibitem{HS10}
M.A. Herman and T.~Strohmer,
\newblock ``General deviants: An analysis of perturbations in compressed
  sensing,''
\newblock {\em {IEEE} J. Selected Topics in Signal Proc.}, vol. 4, no. 2, pp.
  342 --349, Apr. 2010.

\bibitem{CSPC10}
Y.~Chi, L.~L. Scharf, A.~Pezeshki, and R.~Calderbank,
\newblock ``Sensitivity to basis mismatch in compressed sensing,''
\newblock {\em {IEEE} Trans. Signal Proc.}, vol. 59, no. 5, pp. 2182--2195, May
  2011.

\bibitem{Treichler}
J.~Treichler, M.~A. Davenport, and R.~G. Baraniuk,
\newblock ``Application of compressive sensing to the design of wideband signal
  acquisition receivers,''
\newblock in {\em U.S./Australia Joint Workshop on Defense Applications of
  Signal Proc. (DASP)}, Lihue, HI, Sep. 2009.

\bibitem{Aeron}
S.~Aeron, V.~Saligrama, and M.~Zhao,
\newblock ``Information theoretic bounds for compressed sensing,''
\newblock {\em IEEE Trans. Info. Theory}, vol. 56, no. 10, pp. 5111--5130, Oct.
  2010.

\bibitem{BenHaimMichaeliEldar}
Z.~Ben-Haim, T.~Michaeli, and Y.~C. Eldar,
\newblock ``Performance bounds and design criteria for estimating finite rate
  of innovation signals,''
\newblock Sep. 2010,
\newblock Preprint, arXiv:1009.2221.

\bibitem{Arias}
E.~Arias-Castro and Y.~C. Eldar,
\newblock ``Noise folding in compressed sensing,''
\newblock {\em {IEEE} Signal Proc. Letters}, 2011,
\newblock To appear.

\bibitem{CaiXuZhang}
T.~T. Cai, G.~Xu, and J.~Zhang,
\newblock ``On recovery of sparse signals via $\ell_1$ minimization,''
\newblock {\em IEEE Trans. Info. Theory}, vol. 55, no. 7, pp. 3388--3397, July
  2009.

\bibitem{herman:tsp09}
M.~A. Herman and T.~Strohmer,
\newblock ``High-resolution radar via compressed sensing,''
\newblock {\em {IEEE} Trans. Signal Proc.}, vol. 57, no. 6, pp. 2275--2284,
  June 2009.

\bibitem{DeVoreDeterministic}
R.~A. DeVore,
\newblock ``Deterministic constructions of compressed sensing matrices,''
\newblock {\em J. Complex.}, vol. 23, no. 4, pp. 918--925, Aug. 2007.

\bibitem{DonohoL1L0}
D.~Donoho,
\newblock ``For most large underdetermined systems of linear equations, the
  minimal $\ell_1$-norm solution is also the sparsest solution,''
\newblock {\em Comm. Pure Appl. Math.}, vol. 59, no. 6, pp. 797--829, 2006.

\bibitem{CandesPlan}
E.~J. Cand\`{e}s and Y.~Plan,
\newblock ``Near-ideal model selection by $\ell_1$ minimization,''
\newblock {\em Ann. Stat.}, vol. 37, no. 5A, pp. 2145--2177, Oct. 2009.

\bibitem{BDDW08}
R.~G. Baraniuk, M.~Davenport, R.~DeVore, and M.~Wakin,
\newblock ``{A simple proof of the restricted isometry property for random
  matrices},''
\newblock {\em Constructive Approximation}, vol. 28, no. 3, pp. 253--263, 2008.

\bibitem{TD09}
D.~L. Donoho and J.~Tanner,
\newblock ``Observed universality of phase transitions in high-dimensional
  geometry, with implications for modern data analysis and signal processing,''
\newblock {\em Phil. Trans. Royal Soc. A}, vol. 367, no. 1906, pp. 4273--4293,
  Nov. 2009.

\bibitem{DPF10}
C.~Dossal, G.~Peyr\'{e}, and J.~Fadili,
\newblock ``A numerical exploration of compressed sampling recovery,''
\newblock {\em Linear Algebra and its Applications}, vol. 432, no. 7, pp. 1663
  -- 1679, Mar. 2010.

\bibitem{LP}
S.~Boyd and L.~Vanderberghe,
\newblock {\em Convex Optimization},
\newblock Cambridge Univ. Press, 2004.

\bibitem{ROF92}
L.~I. Rudin, S.~Osher, and E.~Fatemi,
\newblock ``Nonlinear total variation based noise removal algorithms,''
\newblock {\em Phys. D}, vol. 60, pp. 259--268, Nov. 1992.

\bibitem{TroppWright}
J.A. Tropp and S.J. Wright,
\newblock ``Computational methods for sparse solution of linear inverse
  problems,''
\newblock {\em Proc. IEEE}, vol. 98, no. 6, pp. 948 --958, June 2010.

\bibitem{HauptNowak}
J.~Haupt and R.~Nowak,
\newblock ``Signal reconstruction from noisy random projections,''
\newblock {\em IEEE Trans. Info. Theory}, vol. 52, no. 9, pp. 4036--4048, Sep.
  2006.

\bibitem{JXC08}
S.~Ji, Y.~Xue, and L.~Carin,
\newblock ``Bayesian compressive sensing,''
\newblock {\em {IEEE} Trans. Signal Proc.}, vol. 56, no. 6, pp. 2346--2356,
  June 2008.

\bibitem{CandesDS}
E.~Cand\`{e}s and T.~Tao,
\newblock ``The {D}antzig selector: {S}tatistical estimation when $p$ is much
  larger than $n$,''
\newblock {\em Ann. Stat.}, vol. 35, no. 6, pp. 2313--2351, Dec. 2005.

\bibitem{PatiOMP}
Y.~Pati, R.~Rezaifar, and P.~Krishnaprasad,
\newblock ``Orthogonal matching pursuit: {R}ecursive function approximation
  with applications to wavelet decomposition,''
\newblock in {\em Asilomar Conf. Signals, Systems, and Computers}, Pacific
  Grove, CA, Nov. 1993.

\bibitem{cosamp}
D.~Needell and J.~A. Tropp,
\newblock ``{CoSaMP}: {I}terative signal recovery from incomplete and
  inaccurate samples,''
\newblock {\em Appl. Comput. Harmon. Anal.}, vol. 26, no. 3, pp. 301--321, May
  2008.

\bibitem{DM09}
W.~Dai and O.~Milenkovic,
\newblock ``Subspace pursuit for compressive sensing signal reconstruction,''
\newblock {\em IEEE Trans. Info. Theory}, vol. 55, no. 5, pp. 2230--2249, May
  2009.

\bibitem{IHT}
T.~Blumensath and M.~E. Davies,
\newblock ``Iterative hard thresholding for compressed sensing,''
\newblock {\em Appl. Comput. Harmon. Anal.}, vol. 27, no. 3, pp. 265--274, Nov.
  2008.

\bibitem{DonohoBPIC}
D.~L. Donoho, M.~Elad, and V.~N. Temlyakov,
\newblock ``{Stable recovery of sparse overcomplete representations in the
  presence of noise},''
\newblock {\em IEEE Trans. Info. Theory}, vol. 52, no. 1, pp. 6--18, Jan. 2006.

\bibitem{CandesSSR}
E.~J. Cand\`{e}s, J.~K. Romberg, and T.~Tao,
\newblock ``Stable signal recovery from incomplete and inaccurate
  measurements,''
\newblock {\em Comm. Pure Appl. Math.}, vol. 59, no. 8, pp. 1207--1223, 2006.

\bibitem{BenHaimEldarElad}
Z.~Ben-Haim, Y.~C. Eldar, and M.~Elad,
\newblock ``Coherence-based performance guarantees for estimating a sparse
  vector under random noise,''
\newblock {\em {IEEE} Trans. Signal Proc.}, vol. 58, no. 10, pp. 5030--5043,
  Oct. 2010.

\bibitem{BenHaimEldar}
Z.~Ben-Haim and Y.~C. Eldar,
\newblock ``The {C}ram\'{e}r-{R}ao bound for estimating a sparse parameter
  vector,''
\newblock {\em {IEEE} Trans. Signal Proc.}, vol. 58, no. 6, pp. 3384--3389,
  June 2010.

\bibitem{WakinDavenport}
M.~B. Wakin and M.~A. Davenport,
\newblock ``Analysis of orthogonal matching pursuit using the restricted
  isometry property,''
\newblock {\em IEEE Trans. Info. Theory}, vol. 56, no. 9, pp. 4395--4401, Sep.
  2010.

\bibitem{ZhangOMP}
T.~Zhang,
\newblock ``Sparse recovery with orthogonal matching pursuit under {RIP},''
\newblock May 2010,
\newblock Preprint, arXiv:1005.2449.

\bibitem{T08}
J.~A. Tropp,
\newblock ``{O}n the conditioning of random subdictionaries,''
\newblock {\em {A}ppl. {C}omput. {H}armon. {A}nal.}, vol. 25, pp. 1--24, 2008.

\bibitem{GJBWS07}
M.~E. Gehm, R.~John, D.~Brady, R.~Willett, and T.~J. Schulz,
\newblock ``Single-shot compressive spectral imaging with a dual-disperser
  architecture,''
\newblock {\em Optics Express}, vol. 15, no. 21, pp. 14013--14027, Oct. 2007.

\bibitem{NK07}
M.~A. Neifeld and J.~Ke,
\newblock ``Optical architectures for compressive imaging,''
\newblock {\em Appl. Optics}, vol. 46, no. 22, pp. 5293--5303, July 2007.

\bibitem{WJWB08}
A.~Wagadarikar, R.~John, R.~Willett, and D.~Brady,
\newblock ``Single disperser design for coded aperture snapshot spectral
  imaging,''
\newblock {\em Appl. Optics}, vol. 47, no. 10, pp. B44--B51, Apr. 2008.

\bibitem{CandesRomberg}
E.~J. Cand\`{e}s and J.~K. Romberg,
\newblock ``Sparsity and incoherence in compressive sampling,''
\newblock {\em Inverse Problems}, vol. 23, no. 3, pp. 969--985, 2007.

\bibitem{EldarUncertainty}
Y.~C. Eldar,
\newblock ``Uncertainty relations for shift-invariant analog signals,''
\newblock {\em IEEE Trans. Info. Theory}, vol. 55, no. 12, pp. 5742--5757, Dec.
  2009.

\bibitem{RudelsonVershynin}
M.~Rudelson and R.~Vershynin,
\newblock ``On sparse reconstruction from {F}ourier and {G}aussian
  measurements,''
\newblock {\em Comm. Pure Appl. Math.}, vol. 61, no. 8, pp. 1025--1171, Aug.
  2008.

\bibitem{LustigDonohoSantosPauly}
M.~Lustig, D.~L. Donoho, J.~M. Santos, and J.~M. Pauly,
\newblock ``Compressed sensing {MRI},''
\newblock {\em {IEEE} Signal Proc. Mag.}, vol. 25, no. 2, pp. 72--82, Mar.
  2008.

\bibitem{CandesRUP}
E.~J. Cand\`{e}s, J.~K. Romberg, and T.~Tao,
\newblock ``Robust uncertainty principles: {E}xact signal reconstruction from
  highly incomplete frequency information,''
\newblock {\em IEEE Trans. Info. Theory}, vol. 52, no. 2, pp. 489--509, 2006.

\bibitem{GSES09}
S.~Gazit, A.~Szameit, Y.~C. Eldar, and M.~Segev,
\newblock ``Super-resolution and reconstruction of sparse sub-wavelength
  images,''
\newblock {\em Opt. Express}, vol. 17, pp. 23920--23946, 2009.

\bibitem{SGSES10}
Y.~Shechtman, S.~Gazit, A.~Szameit, Y.~C. Eldar, and M.~Segev,
\newblock ``Super-resolution and reconstruction of sparse images carried by
  incoherent light,''
\newblock {\em Opt. Letters}, vol. 35, pp. 23920--23946, April 2010.

\bibitem{DuarteDavenportTakharLaskaKellyBaraniuk}
M.~F. Duarte, M.~A. Davenport, D.~Takhar, J.~N. Laska, T.~Sun, K.~F. Kelly, and
  R.~G. Baraniuk,
\newblock ``Single pixel imaging via compressive sampling,''
\newblock {\em IEEE Signal Proc. Mag.}, vol. 25, no. 2, pp. 83--91, March 2008.

\bibitem{Noiselets}
R.~Coifman, F.~Geshwind, and Y.~Meyer,
\newblock ``Noiselets,''
\newblock {\em Appl. Comp. Harmonic Analysis}, vol. 10, pp. 27--44, 2001.

\bibitem{Coifman}
R.~A. DeVerse, R.~R. Coifman, A.~C. Coppi, W.~G. Fateley, F.~Geshwind, R.~M.
  Hammaker, S.~Valenti, F.~J. Warner, and G.~L. Davis,
\newblock ``Application of spatial light modulators for new modalities in
  spectrometry and imaging,''
\newblock in {\em Spectral Imaging: Instrumentation, Applications, and Analysis
  II}, San Jose, CA, Jan. 2003, vol. 4959 of {\em Proc. SPIE}, pp. 12--22.

\bibitem{ChanCharanTakharKellyBaraniukMittleman}
W.~L. Chan, K.~Charan, D.~Takhar, K.~F. Kelly, R.~G. Baraniuk, and D.~M.
  Mittleman,
\newblock ``A single-pixel terahertz imaging system based on compressed
  sensing,''
\newblock {\em Appl. Phy. Letters}, vol. 93, no. 12, Sep. 2008.

\bibitem{YeParedesArceWuChenPrather}
P.~Ye, J.~L. Paredes, G.~R. Arce, Y.~Wu, C.~Chen, and D.~W. Prather,
\newblock ``Compressive confocal microscopy,''
\newblock in {\em IEEE Int. Conf. Acoustics, Speech, and Signal Proc.
  (ICASSP)}, Taipei, Taiwan, Apr. 2009, pp. 429--432.

\bibitem{PRLNGSBM}
S.~Pfetsch, T.~Ragheb, J.~Laska, H.~Nejati, A.~Gilbert, M.~Strauss,
  R.~Baraniuk, and Y.~Massoud,
\newblock ``On the feasibility of hardware implementation of sub-nyquist
  random-sampling based analog-to-information conversion,''
\newblock in {\em IEEE Int. Symp. Circuits and Systems (ISCAS)}, Seattle, WA,
  May 2008, pp. 1480--1483.

\bibitem{GST08}
A.C. Gilbert, M.J. Strauss, and J.A. Tropp,
\newblock ``A tutorial on fast fourier sampling,''
\newblock {\em IEEE Signal Proc. Mag.}, vol. 25, no. 2, pp. 57--66, Mar. 2008.

\bibitem{BajwaSayeedNowak}
W.~U. Bajwa, A.~Sayeed, and R.~Nowak,
\newblock ``A restricted isometry property for structurally subsampled unitary
  matrices,''
\newblock in {\em Allerton Conf. Communication, Control, and Computing},
  Monticello, IL, Sep. 2009, pp. 1005--1012.

\bibitem{TroppLaskaDuarteRombergBaraniuk}
J.~A. Tropp, J.~N. Laska, M.~F. Duarte, J.~K. Romberg, and R.~G. Baraniuk,
\newblock ``Beyond {N}yquist: Efficient sampling of sparse bandlimited
  signals,''
\newblock {\em IEEE Trans. Info. Theory}, vol. 56, no. 1, pp. 520--544, Jan.
  2010.

\bibitem{Ragheb}
T.~Ragheb, J.~N. Laska, H.~Nejati, S.~Kirolos, R.~G. Baraniuk, and Y.~Massoud,
\newblock ``A prototype hardware for random demodulation based compressive
  analog-to-digital conversion,''
\newblock in {\em IEEE Midwest Symp. Circuits and Systems}, Knoxville, TN, Aug.
  2008, pp. 37--40.

\bibitem{YuHoyosSadler}
Z.~Yu, S.~Hoyos, and B.~M. Sadler,
\newblock ``Mixed-signal parallel compressed sensing and reception for
  cognitive radio,''
\newblock in {\em IEEE Int. Conf. Acoustics, Speech, and Signal Proc.
  (ICASSP)}, Las Vegas, NV, Apr. 2008, pp. 3861--3864.

\bibitem{DB10}
M.~F. Duarte and R.~G. Baraniuk,
\newblock ``Spectral compressive sensing,''
\newblock Feb. 2010,
\newblock Preprint.

\bibitem{VS10}
G.~C. Valley and T.~J. Shaw,
\newblock ``Compressive sensing recovery of sparse signals with arbitrary
  frequencies,''
\newblock 2010,
\newblock Preprint.

\bibitem{HauptBajwaRazNowak}
J.~D. Haupt, W.~U. Bajwa, G.~Raz, and R.~Nowak,
\newblock ``Toeplitz compressed sensing matrices with applications to sparse
  channel estimation,''
\newblock {\em IEEE Trans. Info. Theory}, vol. 56, no. 11, pp. 5862--5875, June
  2010.

\bibitem{RauhutStructured}
H.~Rauhut,
\newblock ``Compressive sensing and structured random matrices,''
\newblock in {\em Theoretical Foundations and Numerical Methods for Sparse
  Recovery}, H.~Rauhut, Ed., vol.~9 of {\em Radon Series on Computational and
  Applied Mathematics}. De Gruyter, 2010.

\bibitem{RRT10}
H.~Rauhut, J.~K. Romberg, and J.~A. Tropp,
\newblock ``Restricted isometries for partial random circulant matrices,''
\newblock {\em {A}ppl. {C}omput. {H}armon. {A}nal.}, May 2011,
\newblock To appear.

\bibitem{RombergConvolution}
J.~K. Romberg,
\newblock ``Compressive sensing by random convolution,''
\newblock {\em SIAM J. Imaging Science}, vol. 2, no. 4, pp. 1098--1128, Dec.
  2009.

\bibitem{MarciaW_ICASSP08}
R.~F. Marcia and R.~M. Willett,
\newblock ``Compressive coded aperture superresolution image reconstruction,''
\newblock in {\em IEEE Int. Conf. Acoustics, Speech, and Signal Proc.
  (ICASSP)}, Las Vegas, NV, USA, March 2008, pp. 833--836.

\bibitem{MarciaHarmanyWillett}
R.~Marcia, Z.~Harmany, and R.~Willett,
\newblock ``Compressive coded aperture imaging,''
\newblock in {\em Computational Imaging VII}, San Jose, CA, Jan. 2009, vol.
  7246 of {\em Proc. SPIE}.

\bibitem{JacquesVandergheynstBibetMajidzadehSchmidLeblebici}
L.~Jacques, P.~Vandergheynst, A.~Bibet, V.~Majidzadeh, A.~Schmid, and
  Y.~Leblebici,
\newblock ``{CMOS} compressed imaging by random convolution,''
\newblock in {\em IEEE Int. Conf. Acoustics, Speech, and Signal Proc.
  (ICASSP)}, Taipei, Taiwan, Apr. 2009, pp. 1113--1116.

\bibitem{DuarteBaraniuk}
M.~F. Duarte and R.~G. Baraniuk,
\newblock ``Kronecker compressive sensing,''
\newblock 2009,
\newblock Preprint.

\bibitem{RivensonStern}
Y.~Rivenson and A.~Stern,
\newblock ``Compressed imaging with a separable sensing operator,''
\newblock {\em IEEE Signal Proc. Letters}, vol. 16, no. 6, pp. 449--452, June
  2009.

\bibitem{SunKelly}
T.~Sun and K.~F. Kelly,
\newblock ``Compressive sensing hyperspectral imager,''
\newblock in {\em Computational Optical Sensing and Imaging (COSI)}, San Jose,
  CA, Oct. 2009.

\bibitem{RobucciGrayChiuRombergHasler}
R.~Robucci, J.~Gray, L.~K. Chiu, J.~K. Romberg, and P.~Hasler,
\newblock ``Compressive sensing on a {CMOS} separable-transform image sensor,''
\newblock {\em Proc. IEEE}, vol. 98, no. 6, pp. 1089--1101, June 2010.

\bibitem{HornJohnson}
R.~A. Horn and C.~R. Johnson,
\newblock {\em Topics in matrix analysis},
\newblock Cambridge University Press, 1991.

\bibitem{BWDSB06}
D.~Baron, M.~B. Wakin, M.~F. Duarte, S.~Sarvotham, and R.~G. Baraniuk,
\newblock ``Distributed compressed sensing,''
\newblock Tech. {R}ep. TREE-0612, Rice University, Department of Electrical and
  Computer Engineering, Houston, TX, Nov. 2006.

\bibitem{PLM97}
J.~W. Phillips, R.~M. Leahy, and J.~C. Mosher,
\newblock ``{MEG}-based imaging of focal neuronal current sources,''
\newblock {\em IEEE Trans. Medical Imaging}, vol. 16, no. 3, pp. 338--348, June
  1997.

\bibitem{G93}
R.~Gribonval,
\newblock ``Sparse decomposition of stereo signals with matching pursuit and
  application to blind separation of more than two sources from a stereo
  mixture,''
\newblock in {\em IEEE Int. Conf. Acoustics, Speech, and Signal Proc.
  (ICASSP)}, Minneapolis, MN, Apr. 1993.

\bibitem{MCW05}
D.~Malioutov, M.~Cetin, and A.~S. Willsky,
\newblock ``A sparse signal reconstruction perspective for source localization
  with sensor arrays,''
\newblock {\em IEEE Trans. Signal Proc.}, vol. 53, no. 8, pp. 3010--3022, Aug.
  2005.

\bibitem{CR02}
S.~F. Cotter and B.~D. Rao,
\newblock ``Sparse channel estimation via matching pursuit with application to
  equalization,''
\newblock {\em {IEEE} Trans. Communications}, vol. 50, no. 3, pp. 374--377,
  Mar. 2002.

\bibitem{FGF99}
I.~J. Fevrier, S.~B. Gelfand, and M.~P. Fitz,
\newblock ``Reduced complexity decision feedback equalization for multipath
  channels with large delay spreads,''
\newblock {\em {IEEE} Trans. Communications}, vol. 47, no. 6, pp. 927--937,
  June 1999.

\bibitem{Bazerque}
J.~A. Bazerque and G.~B. Giannakis,
\newblock ``Distributed spectrum sensing for cognitive radio networks by
  exploiting sparsity,''
\newblock {\em {IEEE} Trans. Signal Proc.}, vol. 58, no. 3, pp. 1847--1862,
  Mar. 2010.

\bibitem{ME09b}
M.~Mishali and Y.~C. Eldar,
\newblock ``Blind multiband signal reconstruction: {C}ompressed sensing for
  analog signals,''
\newblock {\em IEEE Trans. Signal Proc.}, vol. 57, pp. 993--1009, Mar. 2009.

\bibitem{ME10}
M.~Mishali and Y.~C. Eldar,
\newblock ``From theory to practice: {Sub-Nyquist} sampling of sparse wideband
  analog signals,''
\newblock {\em {IEEE} J. Selected Topics in Signal Proc.}, vol. 4, no. 2, pp.
  375--391, Apr. 2010.

\bibitem{DE10}
M.~E. Davies and Y.~C. Eldar,
\newblock ``Rank awareness in joint sparse recovery,''
\newblock Apr. 2010,
\newblock Preprint, arXiv:1004.4529.

\bibitem{FengPhD}
P.~Feng,
\newblock {\em Universal minimum-rate sampling and spectrum-blind
  reconstruction for multiband signals},
\newblock Ph.D. thesis, University of Illinois, Urbana, IL, 1998.

\bibitem{CH06}
J.~Chen and X.~Huo,
\newblock ``Theoretical results on sparse representations of
  multiple-measurement vectors,''
\newblock {\em IEEE Trans. Signal Proc.}, vol. 54, no. 12, pp. 4634--4643, Dec.
  2006.

\bibitem{ME08a}
M.~Mishali and Y.~C. Eldar,
\newblock ``Reduce and boost: Recovering arbitrary sets of jointly sparse
  vectors,''
\newblock {\em IEEE Trans. Signal Proc.}, vol. 56, no. 10, pp. 4692--4702, Oct.
  2008.

\bibitem{TGS06}
J.~A. {T}ropp, A.~C. {G}ilbert, and M.~J. {S}trauss,
\newblock ``{Algorithms for simultaneous sparse approximation. Part {I}: Greedy
  pursuit},''
\newblock {\em Signal Proc.}, vol. 86, pp. 572--588, Apr. 2006.

\bibitem{T06}
J.~A. Tropp,
\newblock ``{Algorithms for simultaneous sparse approximation. {P}art {II}:
  Convex relaxation},''
\newblock {\em Signal Proc.}, vol. 86, pp. 589--602, Apr. 2006.

\bibitem{FR08}
M.~{F}ornasier and H.~{R}auhut,
\newblock ``{R}ecovery algorithms for vector valued data with joint sparsity
  constraints,''
\newblock {\em {S}{I}{A}{M} {J}. {N}umer. {A}nal.}, vol. 46, no. 2, pp.
  577--613, 2008.

\bibitem{CR05}
S.~F. Cotter, B.~D. Rao, K.~Engan, and K.~Kreutz-Delgado,
\newblock ``Sparse solutions to linear inverse problems with multiple
  measurement vectors,''
\newblock {\em IEEE Trans. Signal Proc.}, vol. 53, no. 7, pp. 2477--2488, July
  2005.

\bibitem{GRSV08}
R.~{G}ribonval, H.~{R}auhut, K.~{S}chnass, and P.~{V}andergheynst,
\newblock ``{A}toms of all channels, unite! {A}verage case analysis of
  multi-channel sparse recovery using greedy algorithms,''
\newblock {\em {J}. {F}ourier {A}nal. {A}ppl.}, vol. 14, no. 5, pp. 655--687,
  2008.

\bibitem{ER10}
Y.~C. Eldar and H.~Rauhut,
\newblock ``Average case analysis of multichannel sparse recovery using convex
  relaxation,''
\newblock {\em IEEE Trans. Info. Theory}, vol. 6, no. 1, pp. 505--519, Jan.
  2010.

\bibitem{EKB10}
Y.~C. Eldar, P.~Kuppinger, and H.~B{\"o}lcskei,
\newblock ``Block-sparse signals: Uncertainty relations and efficient
  recovery,''
\newblock {\em IEEE Trans. Signal Proc.}, pp. 3042--3054, June 2010.

\bibitem{S79}
R.~O. Schmidt,
\newblock ``Multiple emitter location and signal parameter estimation,''
\newblock in {\em RADC Spectral Estimation Workshop}, Rome, NY, Oct. 1979, pp.
  243--258.

\bibitem{FB961}
P.~Feng and Y.~Bresler,
\newblock ``Spectrum-blind minimum-rate sampling and reconstruction of
  multiband signals,''
\newblock in {\em IEEE Int. Conf. Acoustics, Speech, and Signal Proc.
  (ICASSP)}, Atlanta, GA, May. 1996, vol.~3, pp. 1688--1691.

\bibitem{SV07}
K.~Schnass and P.~Vandergheynst,
\newblock ``Average performance analysis for thresholding,''
\newblock {\em IEEE Signal Proc. Letters}, vol. 14, no. 11, pp. 828--831, Nov.
  2007.

\bibitem{OuEEG}
W.~Ou, M.~S. H\"{a}m\"{a}l\"{a}inen, and P.~Golland,
\newblock ``A distributed spatio-temporal {EEG}/{MEG} inverse solver,''
\newblock {\em Neuroimage}, vol. 44, no. 3, pp. 932--946, Feb. 2009.

\bibitem{csmrf}
V.~Cevher, M.~F. Duarte, C.~Hegde, and R.~G. Baraniuk,
\newblock ``Sparse signal recovery using {M}arkov {R}andom {F}ields,''
\newblock in {\em Workshop on Neural Info. Proc. Systems (NIPS)}, Vancouver,
  Canada, Dec. 2008.

\bibitem{Schniter}
P.~Schniter,
\newblock ``Turbo reconstruction of structured sparse signals,''
\newblock in {\em Conf. Info. Sci. and Sys. (CISS)}, Princeton, NJ, Mar. 2010.

\bibitem{BSB10}
D.~Baron, S.~Sarvotham, and R.G. Baraniuk,
\newblock ``Bayesian compressive sensing via belief propagation,''
\newblock {\em {IEEE} Trans. Signal Proc.}, vol. 58, no. 1, pp. 269 --280, Jan.
  2010.

\bibitem{WaveletBCS}
L.~He and L.~Carin,
\newblock ``Exploiting structure in wavelet-based {B}ayesian compressive
  sensing,''
\newblock {\em {IEEE} Trans. Signal Proc.}, vol. 57, no. 9, pp. 3488--3497,
  Sep. 2009.

\bibitem{JDC09}
S.~Ji, D.~Dunson, and L.~Carin,
\newblock ``Multi-task compressive sensing,''
\newblock {\em {IEEE} Trans. Signal Proc.}, vol. 57, no. 1, pp. 92--106, Jan.
  2009.

\bibitem{HCC10}
L.~He, H.~Chen, and L.~Carin,
\newblock ``Tree-structured compressive sensing with variational {B}ayesian
  analysis,''
\newblock {\em {IEEE} Signal Proc. Letters}, vol. 17, no. 3, pp. 233--236, Mar.
  2010.

\bibitem{FEE10}
T.~Faktor, Y.~C. Eldar, and M.~Elad,
\newblock ``Exploiting statistical dependencies in sparse representations for
  signal recovery,''
\newblock Oct. 2010,
\newblock Preprint, arXiv:1010.5734.

\bibitem{JAB09}
R.~Jenatton, J.~Y. Audibert, and F.~Bach,
\newblock ``Structured variable selection with sparsity inducing norms,''
\newblock April 2009,
\newblock Preprint, arXiv:0904.3523.

\bibitem{PV07}
L.~Peotta and P.~Vandergheynst,
\newblock ``Matching pursuit with block incoherent dictionaries,''
\newblock {\em {IEEE} Trans. Signal Proc.}, vol. 55, no. 9, pp. 4549--4557,
  Sep. 2007.

\bibitem{YL06}
M.~Yuan and Y.~Lin,
\newblock ``Model selection and estimation in regression with grouped
  variables,''
\newblock {\em J. Roy. Stat. Soc. Ser. B Stat. Methodol.}, vol. 68, no. 1, pp.
  49--67, 2006.

\bibitem{EM09}
Y.~C. Eldar and T.~Michaeli,
\newblock ``Beyond bandlimited sampling,''
\newblock {\em IEEE Signal Proc. Mag.}, vol. 26, no. 3, pp. 48--68, May 2009.

\bibitem{B08}
F.~R. Bach,
\newblock ``Consistency of the group lasso and multiple kernel learning,''
\newblock {\em J. Mach. Learn. Res.}, vol. 9, pp. 1179--1225, 2008.

\bibitem{NR08}
Y.~Nardi and A.~Rinaldo,
\newblock ``On the asymptotic properties of the group lasso estimator for
  linear models,''
\newblock {\em Electron. J. Statist.}, vol. 2, pp. 605--633, 2008.

\bibitem{SPH09}
M.~Stojnic, F.~Parvaresh, and B.~Hassibi,
\newblock ``On the reconstruction of block-sparse signals with an optimal
  number of measurements,''
\newblock {\em {IEEE} Trans. Signal Proc.}, vol. 57, no. 8, pp. 3075--3085,
  Aug. 2009.

\bibitem{MGB08}
L.~Meier, S.~{van}~{de} Geer, and P.~B{\"u}hlmann,
\newblock ``The group lasso for logistic regression,''
\newblock {\em J. R. Statist. Soc. B}, vol. 70, no. 1, pp. 53--77, 2008.

\bibitem{ES05}
S.~Erickson and C.~Sabatti,
\newblock ``Empirical {B}ayes estimation of a sparse vector of gene expression
  changes,''
\newblock {\em Statistical Applications in Genetics and Molecular Biology},
  vol. 4, no. 1, pp. 22, 2005.

\bibitem{PVMH08}
F.~Parvaresh, H.~Vikalo, S.~Misra, and B.~Hassibi,
\newblock ``Recovering sparse signals using sparse measurement matrices in
  compressed {DNA} microarrays,''
\newblock {\em {IEEE} J. Selected Topics in Signal Proc.}, vol. 2, no. 3, pp.
  275--285, June 2008.

\bibitem{HDC09}
C.~Hegde, M.~F. Duarte, and V.~Cevher,
\newblock ``Compressive sensing recovery of spike trains using structured
  sparsity,''
\newblock in {\em Workshop on Signal Proc. with Adaptive Sparse Structured
  Representations (SPARS)}, Saint Malo, France, Apr. 2009.

\bibitem{CIHB09}
V.~Cevher, P.~Indyk, C.~Hegde, and R.~G. Baraniuk,
\newblock ``Recovery of clustered sparse signals from compressive
  measurements,''
\newblock in {\em Int. Conf. Sampling Theory and Applications (SAMPTA)},
  Marseille, France, May 2009.

\bibitem{DCB09}
M.~F. Duarte, V.~Cevher, and R.~G. Baraniuk,
\newblock ``Model-based compressive sensing for signal ensembles,''
\newblock in {\em Allerton Conf. Communication, Control, and Computing},
  Monticello, IL, Sep. 2009.

\bibitem{BenHaimEldar10}
Z.~Ben-Haim and Y.~C. Eldar,
\newblock ``Near-oracle performance of greedy block-sparse estimation
  techniques from noisy measurements,''
\newblock 2010,
\newblock Preprint, arXiv:1009:0906.

\bibitem{FHT10}
J.~Friedman, T.~Hastie, and R.~Tibshirani,
\newblock ``A note on the group lasso and a sparse group lasso,''
\newblock Jan. 2010,
\newblock Preprint, arXiv:1001.0736.

\bibitem{SRSE10}
P.~Sprechmann, I.~Ramirez, G.~Sapiro, and Y.~C. Eldar,
\newblock ``{C-HiLasso}: A collaborative hierarchical sparse modeling
  framework,''
\newblock {\em {IEEE} Trans. Signal Proc.}, June 2010,
\newblock To appear.

\bibitem{kay:icassp97}
A.~W. Habboosh, R.~J. Vaccaro, and S.~Kay,
\newblock ``An algorithm for detecting closely spaced delay/{D}oppler
  components,''
\newblock in {\em IEEE Int. Conf. Acoustics, Speech, and Signal Proc.
  (ICASSP)}, Munich, Germany, Apr. 1997, pp. 535--538.

\bibitem{bajwa:allerton08}
W.~U. Bajwa, A.~M. Sayeed, and R.~Nowak,
\newblock ``Learning sparse doubly-selective channels,''
\newblock in {\em Allerton Conf. Communication, Control, and Computing},
  Monticello, IL, Sep. 2008, pp. 575--582.

\bibitem{BGE10}
W.~U. Bajwa, K.~Gedalyahu, and Y.~C. Eldar,
\newblock ``Identification of parametric underspread linear systems with
  application to super-resolution radar,''
\newblock {\em {IEEE} Trans. Signal Proc.}, vol. 59, no. 6, pp. 2548--2561,
  Aug. 2011.

\bibitem{U00}
M.~Unser,
\newblock ``Sampling---50 years after {S}hannon,''
\newblock {\em Proc. IEEE}, vol. 88, pp. 569--587, Apr. 2000.

\bibitem{DDR94}
C.~de~Boor, R.~DeVore, and A.~Ron,
\newblock ``{The structure of finitely generated shift-invariant spaces in
  $L_2(\mathbb{R}^d)$},''
\newblock {\em J. Funct. Anal}, vol. 119, no. 1, pp. 37--78, 1994.

\bibitem{RS95b}
A.~Ron and Z.~Shen,
\newblock ``Frames and stable bases for shift-invariant subspaces of
  {L}$_{2}(${R}$^d$),''
\newblock {\em Canadian J. Mathematics}, vol. 47, no. 5, pp. 1051--1094, 1995.

\bibitem{AG01}
A.~Aldroubi and K.~Gr{\"o}chenig,
\newblock ``Non-uniform sampling and reconstruction in shift-invariant
  spaces,''
\newblock {\em SIAM Review}, vol. 43, pp. 585--620, 2001.

\bibitem{LV98}
Y.-P. Lin and P.~P. Vaidyanathan,
\newblock ``Periodically nonuniform sampling of bandpass signals,''
\newblock {\em IEEE Trans. Circuits Syst. II}, vol. 45, no. 3, pp. 340--351,
  Mar. 1998.

\bibitem{HW99}
C.~Herley and P.~W. Wong,
\newblock ``Minimum rate sampling and reconstruction of signals with arbitrary
  frequency support,''
\newblock {\em IEEE Trans. Info. Theory}, vol. 45, no. 5, pp. 1555--1564, July
  1999.

\bibitem{M01}
J.~Mitola~{III},
\newblock ``{Cognitive radio for flexible mobile multimedia communications},''
\newblock {\em Mobile Networks and Applications}, vol. 6, no. 5, pp. 435--441,
  2001.

\bibitem{Landau}
H.~Landau,
\newblock ``Necessary density conditions for sampling and interpolation of
  certain entire functions,''
\newblock {\em Acta Mathematica}, vol. 117, no. 1, pp. 37--52, July 1967.

\bibitem{K53}
A.~Kohlenberg,
\newblock ``Exact interpolation of band-limited functions,''
\newblock {\em J. Appl. Physics}, vol. 24, no. 12, pp. 1432--1436, Dec. 1953.

\bibitem{VB00}
R.~Venkataramani and Y.~Bresler,
\newblock ``Perfect reconstruction formulas and bounds on aliasing error in
  sub-nyquist nonuniform sampling of multiband signals,''
\newblock {\em IEEE Trans. Info. Theory}, vol. 46, no. 6, pp. 2173--2183, Sep.
  2000.

\bibitem{ME09EXRIP}
M.~Mishali and Y.~C. Eldar,
\newblock ``Expected-{RIP}: {C}onditioning of the modulated wideband
  converter,''
\newblock in {\em IEEE Info. Theory Workshop (ITW)}, Taormina, Italy, Oct.
  2009, pp. 343--347.

\bibitem{CMEH10}
Y.~Chen, M.~Mishali, Y.~C. Eldar, and A.~O. Hero~III,
\newblock ``Modulated wideband converter with non-ideal lowpass filters,''
\newblock in {\em IEEE Int. Conf. Acoustics, Speech, and Signal Proc.
  (ICASSP)}, Dallas, TX, Apr. 2010, pp. 3630--3633.

\bibitem{RonenSingleChannel}
R.~Tur, Y.~C. Eldar, and Z.~Friedman,
\newblock ``Innovation rate sampling of pulse streams with application to
  ultrasound imaging,''
\newblock {\em {IEEE} Trans. Signal Proc.}, vol. 59, no. 4, pp. 1827--1842,
  Apr. 2011.

\bibitem{Stoica1997}
P.~Stoica and R.~Moses,
\newblock {\em Introduction to Spectral Analysis},
\newblock Prentice-Hall, Englewood Cliffs, NJ, 1997.

\bibitem{hua-sarkar90-1}
Y.~Hua and T.~K. Sarkar,
\newblock ``Matrix pencil method for estimating parameters of exponentially
  damped/undamped sinusoids in noise,''
\newblock {\em IEEE Trans. Acoust., Speech, Signal Proc.}, vol. 70, pp.
  1272--1281, May 1990.

\bibitem{kung-etal83}
S.~Y. Kung, K.~S. Arun, and B.~D. Rao,
\newblock ``State-space and singular-value decomposition-based approximation
  methods for the harmonic retrieval problem,''
\newblock {\em J. Opt. Soc. Am.}, vol. 73, no. 122, pp. 1799--1811, December
  1983.

\bibitem{kao-arun93-1}
B.~D. Rao and K.~S. Arun,
\newblock ``Model based processing of signals: a state space approach,''
\newblock {\em Proc. IEEE}, vol. 80, no. 2, pp. 283--309, 1992.

\bibitem{kusuma2006multichannel}
J.~Kusuma and V.K. Goyal,
\newblock ``Multichannel sampling of parametric signals with a successive
  approximation property,''
\newblock {\em IEEE Int. Conf. Image Procesing (ICIP)}, pp. 1265 --1268, Oct.
  2006.

\bibitem{UnserFRI2008}
C.~S. Seelamantula and M.~Unser,
\newblock ``A generalized sampling method for finite-rate-of-innovation-signal
  reconstruction,''
\newblock {\em {IEEE} Signal Proc. Letters}, vol. 15, pp. 813--816, 2008.

\bibitem{ronen_kfir}
K.~Gedalyahu, R.~Tur, and Y.~C. Eldar,
\newblock ``Multichannel sampling of pulse streams at the rate of innovation,''
\newblock {\em {IEEE} Trans. Signal Proc.}, vol. 59, no. 4, pp. 1491--1504,
  Apr. 2011.

\bibitem{BD10}
L.~Baboulaz and P.~L. Dragotti,
\newblock ``Exact feature extraction using finite rate of innovation principles
  with an application to image super-resolution,''
\newblock {\em IEEE Trans. Image Proc.}, vol. 18, no. 2, pp. 281--298, Feb.
  2009.

\bibitem{wagner2011xampling}
N.~Wagner, Y.C. Eldar, A.~Feuer, G.~Danin, and Z.~Friedman,
\newblock ``Xampling in ultrasound imaging,''
\newblock in {\em Medical Imaging: Ultrasonic Imaging, Tomography, and
  Therapy}, Lake Buena Vista, FL, 2011, vol. 7968 of {\em Proc. SPIE}.

\bibitem{ESPRIT_Kailath}
R.~Roy and T.~Kailath,
\newblock ``{ESPRIT}-estimation of signal parameters via rotational invariance
  techniques,''
\newblock {\em IEEE Trans. Acoustics, Speech, and Signal Proc.}, vol. 37, no.
  7, pp. 984--995, July 1989.

\bibitem{krim1996tda}
H.~Krim and M.~Viberg,
\newblock ``Two decades of array signal processing research: the parametric
  approach,''
\newblock {\em {IEEE} Signal Proc. Mag.}, vol. 13, no. 4, pp. 67--94, July
  1996.

\bibitem{uwb_char}
M.~Z. Win and R.~A. Scholtz,
\newblock ``Characterization of ultra-wide bandwidth wireless indoor channels:
  {A} communication-theoretic view,''
\newblock {\em {IEEE} J. Selected Areas in Communications}, vol. 20, no. 9, pp.
  1613--1627, Dec 2002.

\bibitem{quazi1981otd}
A.~Quazi,
\newblock ``An overview on the time delay estimate in active and passive
  systems for target localization,''
\newblock {\em IEEE Trans. Acoustics, Speech, and Signal Proc.}, vol. 29, no.
  3, pp. 527--533, 1981.

\bibitem{urick1983pus}
R.~J. Urick,
\newblock {\em {Principles of Underwater Sound}},
\newblock McGraw-Hill New York, 1983.

\bibitem{bruckstein1985roe}
A.~Bruckstein, T.~J. Shan, and T.~Kailath,
\newblock ``The resolution of overlapping echos,''
\newblock {\em IEEE Trans. Acoustics, Speech, and Signal Proc.}, vol. 33, no.
  6, pp. 1357--1367, 1985.

\bibitem{pallas1991ahr}
M.~A. Pallas and G.~Jourdain,
\newblock ``Active high resolution time delay estimation for large {BT}
  signals,''
\newblock {\em {IEEE} Trans. Signal Proc.}, vol. 39, no. 4, pp. 781--788, Apr
  1991.

\bibitem{4303294}
F.-X. Ge, D.~Shen, Y.~Peng, and V.~O.~K. Li,
\newblock ``Super-resolution time delay estimation in multipath environments,''
\newblock {\em IEEE Trans. Circuits Syst. I}, vol. 54, no. 9, pp. 1977--1986,
  Sep. 2007.

\bibitem{BGN08}
L.~Borup, R.~Gribonval, and M.~Nielsen,
\newblock ``Beyond coherence: Recovering structured time-frequency
  representations,''
\newblock {\em Appl. Comp. Harm. Anal.}, vol. 24, no. 1, pp. 120--128, Jan.
  2008.

\bibitem{BW09}
R.~G. Baraniuk and M.~B. Wakin,
\newblock ``Random projections of smooth manifolds,''
\newblock {\em Foundations of Computational Mathematics}, vol. 9, no. 1, pp.
  51--77, Jan. 2009.

\bibitem{CP10}
E.~J. Cand\`{e}s and Y.~Plan,
\newblock ``Matrix completion with noise,''
\newblock {\em Proc. IEEE}, vol. 98, no. 6, pp. 925--936, June 2010.

\bibitem{Hansen11}
A.~Hansen,
\newblock ``Generalized sampling and infinite dimensional compressed sensing,''
\newblock Feb. 2011,
\newblock Preprint.

\end{thebibliography}

\end{document}